\PassOptionsToPackage{unicode}{hyperref}
\PassOptionsToPackage{hyphens}{url}
\documentclass[
]{article}
\usepackage{xcolor}
\usepackage[margin=1in]{geometry}
\usepackage{amsmath,amssymb}
\usepackage{iftex}
\ifPDFTeX
  \usepackage[T1]{fontenc}
  \usepackage[utf8]{inputenc}
  \usepackage{textcomp} 
\else 
  \usepackage{unicode-math} 
  \defaultfontfeatures{Scale=MatchLowercase}
  \defaultfontfeatures[\rmfamily]{Ligatures=TeX,Scale=1}
\fi
\usepackage{lmodern}
\ifPDFTeX\else
\fi
\IfFileExists{upquote.sty}{\usepackage{upquote}}{}
\IfFileExists{microtype.sty}{
  \usepackage[]{microtype}
  \UseMicrotypeSet[protrusion]{basicmath} 
}{}
\makeatletter
\@ifundefined{KOMAClassName}{
  \IfFileExists{parskip.sty}{%
    \usepackage{parskip}
  }{
    \setlength{\parindent}{0pt}
    \setlength{\parskip}{6pt plus 2pt minus 1pt}}
}{
  \KOMAoptions{parskip=half}}
\makeatother
\usepackage{longtable,booktabs,array}
\usepackage{caption}
\usepackage{calc} 
\usepackage{etoolbox}
\makeatletter
\patchcmd\longtable{\par}{\if@noskipsec\mbox{}\fi\par}{}{}
\makeatother
\IfFileExists{footnotehyper.sty}{\usepackage{footnotehyper}}{\usepackage{footnote}}
\makesavenoteenv{longtable}
\usepackage{graphicx}
\makeatletter
\newsavebox\pandoc@box
\newcommand*\pandocbounded[1]{
  \sbox\pandoc@box{#1}%
  \Gscale@div\@tempa{\textheight}{\dimexpr\ht\pandoc@box+\dp\pandoc@box\relax}%
  \Gscale@div\@tempb{\linewidth}{\wd\pandoc@box}%
  \ifdim\@tempb\p@<\@tempa\p@\let\@tempa\@tempb\fi
  \ifdim\@tempa\p@<\p@\scalebox{\@tempa}{\usebox\pandoc@box}%
  \else\usebox{\pandoc@box}%
  \fi%
}
\def\fps@figure{htbp}
\makeatother
\providecommand{\tightlist}{%
  \setlength{\itemsep}{0pt}\setlength{\parskip}{0pt}}
\usepackage{graphicx}
\usepackage{placeins}
\usepackage{bookmark}
\IfFileExists{xurl.sty}{\usepackage{xurl}}{} 
\makeatletter
\@ifundefined{xmpquote}{}{}
\makeatother
\hypersetup{
  hidelinks,
  pdfcreator={LaTeX via pandoc}}

\author{}
\date{\vspace{-2.5em}}

\graphicspath{{figures/}{./}}
\begin{document}

\section{Tournaments determined by three and five
voters}\label{tournaments-determined-by-three-and-five-voters}

\emph{In loving memory of Saif Pathan.}

\textbf{Leonid Chindelevitch\(^{a}\) and Ararat Harutyunyan\(^{b}\)}

\(^{a}\) MRC Centre for Global Infectious Disease Analysis, School of
Public Health, Imperial College London, 90 Wood Lane, London W12 0BZ,
United Kingdom. Email: \texttt{lchindel@ic.ac.uk}. ORCID:
0000-0002-6619-6013.

\(^{b}\) University of Paris-Dauphine, PSL University, CNRS UMR7243,
LAMSADE, Place du Maréchal de Lattre de Tassigny, 75775 Paris Cedex 16,
France. Email: \texttt{ararat.harutyunyan@lamsade.dauphine.fr}.

Corresponding author: Leonid Chindelevitch (\texttt{lchindel@ic.ac.uk}).

\subsection{Abstract}\label{abstract}

Pairwise majority has been studied since Ramon Llull in the 13th century
and Condorcet in the 18th. A tournament records, for each pair of
alternatives, which one a majority of voters prefers. A committee of
\(m\) voters produces only some tournaments; let \(N(m)\) be the
smallest number of alternatives on which some tournament cannot arise
from any \(m\) voters. Only \(N(3) = 8\) is known. We refute three
conjectures. Every minimum feedback arc set of a tournament is a minimal
set of arcs meeting every cyclic triangle, strengthening a theorem of
Milosz, Hamel and Pierrot on Kemeny medians of three voters, and both of
their conjectures fail. The third, going back to a question of
Shepardson and Tovey, is that \(m\) voters produce every tournament of
predictability at least \((m+1)/(2m)\), predictability being the largest
\(\alpha\) for which some electorate produces the tournament with every
majority backed by a fraction \(\alpha\) of the voters. It fails for
three voters: up to isomorphism, exactly 1,013 tournaments on ten
alternatives have predictability \(2/3\) yet arise from no three voters,
none on fewer alternatives does, and failures occur only at the
threshold. For five voters we show \(12 \le N(5) \le 38\) and give the
first explicit tournament of moderate size that no five voters produce,
the Paley tournament on 43 alternatives; its predictability exceeds
\(3/5\), so the conjecture fails even strictly above the threshold. A
companion paper narrows the bounds to \(13 \le N(5) \le 22\). Public
code reproduces every computation.

\textbf{Keywords:} Kemeny median; majority tournament; feedback arc set;
McGarvey number; inducibility; predictability

\subsection{1. Introduction}\label{introduction}

Let \(A\) be a finite set of alternatives. A \emph{profile} is a finite
list of voters, each of whom ranks \(A\) by a linear order. When the
number of voters is odd, the \emph{majority tournament} of the profile
has, for each pair of alternatives \(u\) and \(v\), the arc \(u \to v\)
if more than half of the voters rank \(u\) above \(v\), and the arc
\(v \to u\) otherwise; a \emph{tournament} is a set of arcs with exactly
one arc between every two alternatives. McGarvey (1953) proved that
every tournament on \(A\) is the majority tournament of some profile. So
when the number of voters is unrestricted, majority rule imposes no
structure on collective preference: every pattern of pairwise majorities
occurs, including patterns with cycles. The Condorcet paradox (Condorcet
1785) is the smallest example: three voters ranking
\(a \succ b \succ c\), \(b \succ c \succ a\) and \(c \succ a \succ b\)
produce the cycle \(a \to b \to c \to a\). McGarvey's construction uses
two voters for each pair of alternatives, \(n(n-1)\) voters on \(n\)
alternatives, and the number of voters that suffices for every
tournament on \(n\) alternatives grows as \(\Theta(n/\log n)\) (Stearns
1959; Erdős and Moser 1964). Arbitrary patterns of majority preference
therefore need many voters.

A committee of fixed size is far more constrained. A panel of \(m\)
voters admits only finitely many profiles on a given agenda, so it can
produce only finitely many patterns of pairwise majority, and once the
agenda is long enough most patterns are beyond it. This raises a design
question about small deliberative bodies. On an agenda of \(n\)
alternatives, can a committee of \(m\) members realize every collective
preference, or are there patterns of majority preference that its
members can never produce, however they rank the alternatives? Write
\(N(m)\) for the fewest alternatives at which some pattern escapes every
\(m\)-member committee. For three voters the answer is \(N(3) = 8\):
every pattern on at most seven alternatives is realizable, and on eight
alternatives one is not (Shepardson and Tovey 2009; Eggermont et
al.~2013). For five voters the answer was open, and the smallest pattern
known to escape five voters had over \(6 \times 10^{8}\) alternatives
(Bachmeier et al.~2019).

The same question bears on the Kemeny median. The \emph{Kendall tau
distance} between two linear orders is the number of pairs of
alternatives that they rank in opposite order, and a \emph{Kemeny
median} of a profile is a linear order with the smallest total Kendall
tau distance to the voters' rankings (Kemeny 1959). For a pair with
majority arc \(u \to v\), a median that ranks \(v\) above \(u\)
disagrees with more voters on that pair than one that ranks \(u\) above
\(v\), by the \emph{margin} of the arc: the number of voters ranking
\(u\) above \(v\) minus the number ranking \(v\) above \(u\). A median
therefore reverses a set of majority arcs of least total margin among
the sets whose reversal leaves no directed cycle, that is, a
minimum-weight \emph{feedback arc set} of the majority tournament with
margins as weights. The tournaments and margins that \(m\) voters can
produce are thus exactly the instances of this problem that arise from
Kemeny medians of \(m\) voters. Milosz, Hamel and Pierrot (2018) proved
that a median of three voters reverses only arcs lying on a directed
3-cycle, a triple \(u \to v \to w \to u\). They conjectured that the
same holds for every odd number of voters, and that for every majority
tournament of three voters the fewest arcs meeting every directed
3-cycle are as many as the fewest arcs whose reversal leaves no cycle.
We strengthen their theorem and refute both conjectures: the first fails
for five voters, the second on eleven alternatives.

For five voters, we narrow the gap to \(12 \le N(5) \le 38\) and exhibit
the first explicit pattern of moderate size that five voters cannot
produce, the Paley tournament on \(43\) alternatives. A companion paper
submitted together with this one (Chindelevitch and Harutyunyan 2026)
narrows the bounds further to \(13 \le N(5) \le 22\) with a different
method. Along the way, both papers find that how a majority is carried
matters as well as whether it exists: some patterns are within reach of
three voters only if some pairwise majority is unanimous, and never if
every majority is carried by two votes to one.

The rest of this introduction makes these questions precise.

This paper studies the structure of the classical reduction of the
Kemeny median problem to a weighted feedback arc set problem on majority
tournaments, and of its inverse question -- \emph{which tournaments are
majority tournaments of few voters} -- and refutes three conjectures in
the process. Our starting points are the work of Milosz, Hamel and
Pierrot (2018) on medians of 3 voters, McGarvey's theorem (1953) and its
quantitative refinements (Stearns 1959; Erdős and Moser 1964; Bachmeier
et al.~2019), and the predictability framework of Shepardson and Tovey
(2009). We define all relevant concepts below.

\textbf{Definitions.} We use \(|S|\) for the size of a set \(S\), and
\([n]\) for the set \(\{1, \dots, n\}\) for an integer \(n\).

A \textbf{digraph} \(D = (V, E)\) consists of a finite set \(V\) of
\textbf{vertices} and a set \(E\) of \textbf{arcs}: ordered pairs of
distinct vertices, written \(u \to v\), and shown graphically as an
arrow leaving \(u\) and entering \(v\). The arc \(u \to v\) is said to
be \textbf{incident} on \(u\) and \(v\). The \textbf{order} of
\(D = (V, E)\) is \(n = |V|\) and its \textbf{size} is \(|E|\). All our
digraphs are \textbf{oriented}: at most one of \(u \to v\) and
\(v \to u\) is present for every pair \(\{u, v\}\). A \textbf{dicycle}
(directed cycle) in \(D\) is a sequence of \(k\) arcs connecting
distinct vertices \(v_1, \dots, v_k\) in the order
\(v_1 \to v_2, ..., v_{k-1} \to v_k, v_k \to v_1\). \(D\) is
\textbf{acyclic} if it contains no dicycle. Given a linear order \(\pi\)
of \(V\), an arc \(u \to v\) is \textbf{forward} in \(\pi\) if \(u\)
precedes \(v\) in \(\pi\), and \textbf{backward} otherwise;
\(\mathrm{fwd}_D(\pi)\) denotes the set of arcs of \(D\) forward in
\(\pi\). An acyclic digraph admits a \textbf{topological order}, a
linear order of \(V\) in which every arc is forward. For a vertex \(v\),
the out-neighbourhood \(N^{+}(v)\) (the in-neighbourhood \(N^{-}(v)\))
denote the set of vertices \(u\) which have an arc \(v \to u\)
(\(u \to v\)), respectively. Their sizes \(|N^{+}(v)|\) and
\(|N^{-}(v)|\) are called the \textbf{out-degree} and the
\textbf{in-degree} of \(v\), respectively. \(D\) is \textbf{Eulerian} if
every vertex \(v\) has equal in-degree and out-degree:
\(|N^{+}(v)| = |N^{-}(v)|\).

An \textbf{isomorphism} from a digraph \(D = (V,E)\) to a digraph
\(D' = (V',E')\) of the same order \(n\) is a one-to-one mapping
\(\phi: V \rightarrow V'\) such that \(u \to v\) is an arc in \(E\) if
and only if \(\phi(u) \to \phi(v)\) is an arc in \(E'\); if such a
mapping exists, \(D\) and \(D'\) are called \textbf{isomorphic}. The
\textbf{converse} of a digraph is obtained by reversing the direction of
every arc; \(D\) is \textbf{self-converse} if it is isomorphic to its
converse. A \textbf{subdigraph} of \(D\) is a digraph obtained from
\(D\) by deleting vertices (together with their incident arcs) and/or
arcs; the \textbf{induced subdigraph} \(D[S]\), for \(S \subseteq V\),
is obtained by deleting the vertices outside \(S\), and nothing else.

A \textbf{tournament} \(T = (V, E)\) is an oriented graph obtained from
the complete graph \(K_n\), the undirected graph on \(n\) vertices with
all \(\binom{n}{2}\) edges, by orienting all of its edges. In other
words, it is a digraph with \emph{exactly} one arc between every pair of
vertices. A \textbf{subtournament} is an induced subdigraph of a
tournament (and is itself a tournament). \(T\) is \textbf{regular} if
all out-degrees are equal (so \(n\) is odd), and \textbf{semi-regular}
(equivalently, \textbf{near-regular}) if every out-degree is \(n/2 - 1\)
or \(n/2\) (so \(n\) is even) -- the closest an even-order tournament
can come to regular. Since there are \(C = \binom{n}{2}\) arcs and each
contributes 1 to the total out-degree, the out-degrees of any tournament
sum to \(C\); for even \(n\) it follows that a semi-regular tournament
has exactly \(n/2\) vertices of out-degree \(n/2\) and \(n/2\) of
out-degree \(n/2 - 1\).

\(T\) is \textbf{transitive} if its arcs are exactly the pairs ranked by
some linear order of \(V\) -- equivalently, if it contains no directed
cycle. For tournaments this is further equivalent to containing no
cyclic triangle (Moon 1968). A general digraph, by contrast, can be free
of cyclic triangles while containing longer directed cycles. On three
vertices there are exactly two tournaments up to isomorphism: the
\textbf{transitive triangle} and the \textbf{cyclic triangle} (the
directed 3-cycle). A \textbf{feedback arc set} (FAS) of a digraph is the
set of arcs backward in some linear order of its vertices. The
\textbf{maximum acyclic subgraph} of \(D\) is the largest subdigraph
with no dicycles, and has size
\(\mathrm{MAS}(D) = \max_{\pi} |\mathrm{fwd}_D(\pi)|\), so the minimum
size of a FAS is \(|E| - \mathrm{MAS}(D)\). Every FAS meets every
dicycle (no linear order ranks a dicycle all-forward). In a tournament,
reversing the arcs of a minimum FAS turns it into a transitive (acyclic)
tournament, so a minimum FAS is equivalently a minimum \textbf{reversal
set} (Moon 1968). When the arcs carry weights, the weight of a FAS is
the total weight of its arcs, and a \textbf{minimum-weight FAS}
minimizes this total.

Given \(m\) voters -- linear orders \(\tau_1, \dots, \tau_m\) of the
option set \([n]\) -- the \textbf{Kemeny median problem} (Kemeny 1959;
Young and Levenglick 1978) asks for a linear order \(\mu\) minimizing
\(W(\mu) = \sum_{r} d(\mu, \tau_r)\), where \(d\) is the Kendall tau
distance defined above. The minimizer \(\mu\) is called the
\textbf{median}. The problem of finding a median is NP-hard for every
\(m \ge 3\) (Dwork et al.~2001, Bachmeier et al.~2019, Peters 2026,
Madarasi 2026); its complexity for \(m = 3\) and \(m = 5\) was open
until very recently (Blin et al.~2011; Milosz et al.~2018). For an odd
number \(m\) of voters, the majority tournament defined above has an arc
from \(i\) to \(j\) whenever \(i\) precedes \(j\) in a majority of the
voters (oddness rules out ties, so every pair is decided); its weighted
version assigns each arc the margin \(|L_{ij} - L_{ji}|\), where
\(L_{ij}\) is the number of voters in which \(i\) precedes \(j\). The
Kemeny problem for the profile then reduces to minimum weighted FAS on
the weighted majority tournament (Alon 2002; Charbit et al.~2007;
Kenyon-Mathieu and Schudy 2007):
\(W(\mu) = \mathrm{LB}(R) + (\text{weight of arcs reversed by } \mu)\),
where \(\mathrm{LB}(R) = \sum_{i<j} \min(L_{ij}, L_{ji})\) (Milosz et
al.~2018). As above, a profile is a finite list of voters; a distinct
linear order in a profile is a \textbf{type}; profiles may have more
voters than types.

\(T\) is \textbf{\(m\)-inducible} if there exist \(m\) voters such that
every arc of \(T\) is forward in more than \(m/2\) of them; such a
family of voters \textbf{induces} \(T\) by majority, and is called an
\textbf{inducing profile}. The \textbf{McGarvey number} of \(T\) is the
least \(m\) for which \(T\) is \(m\)-inducible (finite by McGarvey 1953;
Bachmeier et al.~2019 call it the majority dimension). A voter
\textbf{dissents} with an arc if it ranks this arc backward; a profile
induces \(T\) if and only if at most \((m-1)/2\) voters dissent with
each arc. The McGarvey number of a tournament is always odd: if \(T\) is
induced by an even number \(m\) of voters, then every arc is forward for
at least \(m/2 + 1\) of them, so deleting an \emph{arbitrary} voter
leaves at least \(m/2\) supporters among the remaining \(m - 1\) voters,
still a strict majority. Restricting an inducing profile to any vertex
subset \(S\) induces the subtournament \(T[S]\) as every pairwise
majority is preserved, so \(m\)-inducibility is hereditary under
subtournaments.

We further call a digraph \(D\) \textbf{\(m\)-inducible} if some
completion of \(D\) to a tournament -- an orientation of the pairs of
vertices not joined by an arc of \(D\) -- is \(m\)-inducible. In the
majority-\emph{digraph} sense of Bachmeier et al.~(2019), where a pair
not joined by an arc must be an exact tie among the voters, the majority
dimension of any digraph that is not a tournament is \emph{even}
(Bachmeier et al.~2019). \(m\)-inducibility is preserved under taking
converses: reversing every voter of an inducing profile induces the
converse digraph. Shepardson and Tovey (2009) define the
\textbf{predictability} of a tournament as the largest supermajority
threshold at which it is inducible: the maximum \(\alpha\) such that
some finite voter population (equivalently, some assignment of
non-negative weights to linear orders) ranks every arc of \(T\) forward
with total weight fraction at least \(\alpha\). We work with an
equivalent minimax formulation, which we generalize to arbitrary
digraphs for the obstacles at the heart of Section 5. We formally define
it under \textbf{Notation} below, together with the LP duality that
makes the two views agree.

\textbf{Arc classification.} An arc \(a = (u \to v)\) of a tournament is
\textbf{cyclic} if it lies in at least one cyclic triangle, i.e.~if some
\(w\) closes a return path \(v \to w \to u\)
(\(N^{+}(v) \cap N^{-}(u) \ne \varnothing\)); otherwise \(a\) is
\textbf{shortcut-only}: every other vertex \(w\) satisfies \(u \to w\)
or \(w \to v\), so \(a\) appears in triples only as the shortcut
\(u \to w \to v\), \(u \to v\) or as a path arc of a transitive
triangle. In an inducing profile of \(m\) voters, the \textbf{margin} of
an arc is (number of voters ranking it forward) \(-\) (number ranking it
backward); for odd \(m\) this is an odd number in
\(\{1, 3, \dots, m\}\). An arc is \textbf{unanimous} if its margin is
\(m\) (every voter ranks it forward). A profile has \textbf{unit margin}
if \emph{every} arc has margin \(1\), i.e.~every arc has exactly
\((m-1)/2\) voters dissenting with it. As the only possible margins for
\(m = 3\) are \(1\) and \(3\), a profile has unit margin if and only if
it has \emph{no unanimous arc}; for \(m \ge 5\) these notions differ,
e.g.~an arc forward \(4\):\(1\) has margin \(3\), strictly between the
extremes \(1\) and \(5\). We define, for odd \(m\) and odd
\(t \in [m]\), \(\mathcal{I}_{m,t}\) to be the class of tournaments
admitting a \(m\)-voter inducing profile in which \emph{every arc has
margin \(\le t\)}. They satisfy
\(\mathcal{I}_{m,1} \subseteq \mathcal{I}_{m,3} \subseteq \cdots \subseteq \mathcal{I}_{m,m} =
\{\text{$m$-inducible tournaments}\}\), with \(\mathcal{I}_{m,1}\) the
ones inducible with unit margin. Our \(m = 3\) results (Section 5) show
the hierarchy is proper at \(m = 3\):
\(\mathcal{I}_{3,1} \subsetneq \mathcal{I}_{3,3}\) -- some tournaments
are 3-inducible only via a unanimous (margin-\(3\)) arc. We
\textbf{conjecture} that this inclusion remains strict for every odd
\(m\) (Conjecture 8.1).

\textbf{Notation.} \(\mathrm{HS}_3(T)\) is the minimum size of a
\textbf{3-hitting set} of \(T\): a set of arcs meeting every cyclic
triangle, the central object of Milosz, Hamel and Pierrot (2018). Since
a FAS meets every directed cycle, it meets every cyclic triangle, so
\(\mathrm{HS}_3(T) \le \mathrm{FAS}(T)\). \(\mathrm{McG}(T)\) denotes
the \textbf{McGarvey number} of \(T\), and \(N(m)\) the smallest \(n\)
such that some \(n\)-vertex tournament is not \(m\)-inducible. For odd
\(m\), the \textbf{slack} of \(T\) at \(m\) voters is the quantity
\[\mathrm{slack}_m(T) \;:=\; m \cdot \mathrm{MAS}(T) \;-\; \tfrac{m+1}{2}\, C ;\]
it is the total budget available to a \(m\)-voter inducing profile:
summing the majority condition over arcs shows
\(\sum_p \bigl(\mathrm{MAS}(T) - |\mathrm{fwd}_T(\pi_p)|\bigr) \le
\mathrm{slack}_m(T)\) for every inducing profile
\((\pi_1, \dots, \pi_m)\), cf.~(P5) below. In particular
\(\mathrm{slack}_m(T) \ge 0\) is necessary for \(m\)-inducibility, and a
small slack forces all voters close to maximum acyclic orders -- the
effect that drives Section 7.

A tournament that is \emph{not} \(m\)-inducible is
\textbf{vertex-critical} if every vertex-deleted subtournament \(T - x\)
\emph{is} \(m\)-inducible -- i.e.~it is minimal for
non-\(m\)-inducibility (equivalently, it contains no smaller
non-\(m\)-inducible tournament as a subtournament). \(D_n\), \(R_n\) and
\(S_n\) denote the numbers of non-isomorphic tournaments, regular
tournaments (semi-regular for even \(n\)) and self-converse tournaments
on \(n\) vertices. The exact values used in our censuses are collected
in Appendix H; all tournament collections except the Paley family come
from McKay's digraph archive (2026).

\(\Delta(S)\) denotes the probability simplex over a finite set \(S\)
(nonnegative weights summing to 1), and \(\Pi\) the set of all linear
orders of \(V\). For a digraph \(D = (V, E)\), the
\textbf{predictability} is the value
\[\alpha^{*}(D) \;=\; \min_{y \in \Delta(E)} \; \max_{\pi \in \Pi} \sum_{e \in \mathrm{fwd}_D(\pi)} y_e
\;=\; \max_{x \in \Delta(\Pi)} \; \min_{e \in E} \Pr_{\pi \sim x}[\,e \in \mathrm{fwd}_D(\pi)\,]\]
of the associated zero-sum game, which runs as follows (Shepardson and
Tovey 2009): Alice picks an arc \(e\) of \(D\) and, simultaneously, Bob
picks a linear order \(\pi\) of \(V\); Bob wins if his order predicts
the arc correctly, i.e.~if \(e \in \mathrm{fwd}_D(\pi)\). In the
display, \(y\) is a mixed strategy for Alice, \(x\) is one for Bob, and
\(\alpha^{*}(D)\) is Bob's optimal probability of a correct prediction.
The equality of the two sides is von Neumann's minimax theorem --
equivalently, strong LP duality for a finite matrix game (rows = linear
orders, columns = arcs); nothing about the structure of \(D\) is used,
so the definition and the duality are valid for every digraph. For
tournaments, \(\alpha^{*}\) coincides with the predictability of
Shepardson and Tovey (2009): the optimal \(x\) can be taken rational,
and clearing denominators turns it into a finite voter population
achieving the threshold, which is exactly the original definition.

A \textbf{linear program} (LP) optimizes a linear objective over
real-valued variables subject to linear inequalities, and is solvable in
polynomial time; an \textbf{integer linear program} (ILP) additionally
constrains some variables to be integers, and is NP-hard in general. The
primal and dual solutions of a feasible LP have the same value if and
only if they are both optimal; we sometimes refer to these solutions as
certificates. The predictability is the value of an LP, while
\(m\)-inducibility is an integral feasibility question. The refutations
of Conjecture A in Section 5 and Section 7 say precisely that this
LP--ILP gap is non-zero.

\textbf{Group-theoretic notation.} The \textbf{automorphism group}
\(\mathrm{Aut}(D)\) of a digraph \(D\) is the group of automorphisms
from \(D\) to itself; \(D\) is \textbf{rigid} if \(\mathrm{Aut}(D)\)
only contains the identity permutation. For a group \(\Gamma\) of
permutations of a set \(X\), the \textbf{orbit} of \(x \in X\) under
\(\Gamma\) is the set \(\Gamma x = \{\gamma(x) : \gamma \in \Gamma\}\);
the orbits partition \(X\). We use orbits of \(\mathrm{Aut}(T)\) acting
on vertices, on arcs, and on linear orders. Two notations for
permutations of the vertex set are used and must not be confused: a
comma-separated tuple \((a_1, a_2, \dots, a_n)\) lists a \textbf{linear
order} (earliest first), whereas \textbf{cycle notation}
\((a_1\,a_2\,\dots\,a_r)\) denotes the permutation mapping
\(a_1 \to a_2 \to \cdots \to a_r \to a_1\) and fixing every other vertex
(used for automorphisms). \(\langle g_1, \dots, g_r \rangle\) denotes
the group \textbf{generated by} the listed permutations: the smallest
group containing all of them, i.e.~all finite products of the \(g_i\)
and their inverses. For instance
\(\mathrm{Aut}(T^{*}) = \langle (4\,5\,6),
(8\,9\,10) \rangle\) in Section 4 is the group of order 9 whose elements
rotate the triples \(\{4,5,6\}\) and \(\{8,9,10\}\) independently.

\textbf{Lemma 1.1 (symmetrization).} \emph{For every digraph \(D\) there
is an optimal Alice strategy \(y\) (an optimal dual) that is invariant
under \(\mathrm{Aut}(D)\) -- constant on the arc orbits; likewise there
is an \(\mathrm{Aut}(D)\)-invariant optimal Bob strategy \(x\), constant
on the order orbits.}

\emph{Proof.} If \(y\) is optimal and \(\phi \in \mathrm{Aut}(D)\), the
image strategy \(y^{\phi}\) (defined by \(y^{\phi}_{\phi(e)} = y_e\)) is
also optimal: relabelling by \(\phi\) maps \(\mathrm{fwd}_D(\pi)\)
bijectively onto \(\mathrm{fwd}_D(\phi \cdot \pi)\), where
\(\phi \cdot \pi\) is the relabelled order, so \(y\) and \(y^{\phi}\)
have the same worst-case payoff. The payoff is linear in \(y\), so the
set of optimal strategies is convex, and the average of the
\(|\mathrm{Aut}(D)|\) images \(y^{\phi}\) is again optimal -- and
invariant by construction. The same averaging applies to Bob's
strategies \(x\). \(\blacksquare\)

We use Lemma 1.1 frequently: it lets every certificate search be
restricted to \(\mathrm{Aut}\)-invariant strategies, and for
arc-transitive digraphs it guarantees that a uniform dual (one with all
weights equal) is optimal. An \textbf{obstacle} for \(m\)-inducibility
is a subdigraph witnessing \(\alpha^{*} < \frac{m+1}{2m}\) through a
dual predictability certificate (Alice's strategy in the game we defined
above) which is non-zero on its arcs, cf.~(P2) below.

\textbf{Basic facts (used throughout).}

\begin{itemize}
\tightlist
\item
  \textbf{(P1)} \(\alpha^{*}(T) \ge \frac{m+1}{2m}\) is \emph{necessary}
  for \(m\)-inducibility (each arc forward in \(\ge \frac{m+1}{2}\) of
  \(m\) voters; average). For \(m = 3\) the threshold is \(2/3\), for
  \(m = 5\) it is \(3/5\).
\item
  \textbf{(P2) (monotonicity, for general digraphs.)} If \(D'\) is a
  subdigraph of a digraph \(D\), then
  \(\alpha^{*}(D') \ge \alpha^{*}(D)\): restricting each order of an
  optimal distribution for \(D\) to the vertices of \(D'\) preserves the
  forward status of every arc of \(D'\). Dually, a distribution \(y\) on
  \(E(D')\) certifying \(\alpha^{*}(D') < \theta\) extends by zero to
  \(E(D)\) and certifies \(\alpha^{*}(D) \le \alpha^{*}(D') < \theta\)
  -- this is how a subgraph with small \(\alpha^{*}\) (an obstacle)
  bounds its host. Neither direction uses completeness, so monotonicity
  holds for arbitrary digraphs -- as we need, since our obstacle
  certificates are not tournaments.
\item
  \textbf{(P3)} For a cyclic triangle, every linear order makes either 1
  or 2 of its 3 arcs forward -- never 0 or 3; hence any non-transitive
  tournament has \(\alpha^{*} \le 2/3\), and \(\alpha^{*}(T) = 1\) if
  and only if \(T\) is transitive. In particular
  \textbf{\(\alpha^{*} \in (2/3, 1)\) is impossible.}
\item
  \textbf{(P4) (margin rigidity for 3 voters.)} In any inducing profile
  of \(3\) voters, every cyclic triangle receives total forward weight
  at least \(2+2+2 = 6\) (each arc needs a majority of 2), while by (P3)
  each voter contributes at most 2 -- so equality holds throughout. Two
  consequences:

  \begin{itemize}
  \tightlist
  \item
    \textbf{(P4a, margin rigidity on cyclic arcs.)} \emph{Every cyclic
    arc has margin exactly \(1\)} (votes split \(2\):\(1\)), each voter
    is forward on exactly 2 arcs of every cyclic triangle, and on each
    cyclic triangle the map sending an arc to its unique dissenting
    voter is a \emph{bijection} onto the three voters.
  \item
    \textbf{(P4b, unanimity is shortcut-only.)} Consequently every
    unanimous arc is shortcut-only, and a 3-voter profile inducing \(T\)
    has unit margin (no unanimous arc) as soon as it is unanimous on no
    shortcut-only arc: all margin freedom resides on the shortcut-only
    arcs. The special case ``unanimous arcs lie in no directed 3-cycle''
    is the unanimity lemma of Milosz et al.~(2018) (proved there
    directly for weighted 3-voter majority tournaments); we absorb it
    here and cite it as (P4b).
  \end{itemize}
\item
  \textbf{(P5) (unit-margin inducibility uses all the slack; \(m\)
  odd.)} For a \(m\)-voter inducing profile write
  \(\delta_p = \mathrm{MAS}(T) - |\mathrm{fwd}_T(\pi_p)|\). Every arc
  needs \(\ge (m+1)/2\) forward voters, so
  \(\sum_p |\mathrm{fwd}_T(\pi_p)| \ge \frac{m+1}{2} C\),
  i.e.~\(\sum_p \delta_p
  \le \mathrm{slack}_m(T)\) -- with equality if and only if every margin
  is exactly 1.
\end{itemize}

\subsubsection{1.1 Contributions}\label{contributions}

\textbf{1. FAS and 3-cycles (Section 2).} In \emph{any} tournament,
every minimum feedback arc set is a minimal hitting set of the directed
3-cycles (Theorem 2.1). Combined with (P4b) and a weight lemma this
strengthens and recovers the main theorem of Milosz et al.~(2018): for a
3-inducible majority tournament, every minimum-weight FAS is a minimal
hitting set of directed 3-cycles.

\textbf{2. Refuting Conjecture 1 of Milosz et al.~2018 (Section 3).} The
``only 3-cycle edges are ever reversed'' property fails for every odd
\(m \ge 5\): we give minimal counterexamples -- a weak one at
\((m, n) = (7, 4)\), and strong ones (with a unique optimal FAS) at
\((9, 4)\) and \((5, 6)\) -- each with proven minimality.

\textbf{3. Refuting Conjecture 2 of Milosz et al.~2018 (Section 4).}
\(\mathrm{FAS}(T) = \mathrm{HS}_3(T)\) holds for \emph{every} tournament
on \(n \le 10\) vertices (exhaustive), but fails at \(n = 11\): among
the 3-inducible regular tournaments a unique \(T^{*}\) has
\(\mathrm{FAS} = 17 > 16 = \mathrm{HS}_3\) -- refuting the conjecture.
Among the 3-inducible self-converse tournaments on 11 vertices exactly 6
violate the equality.

\textbf{4. Refuting the threshold conjecture at \(m = 3\) (Section 5).}
The conjecture ``\(\alpha^{*}(T) \ge 2/3
\Rightarrow T\) 3-inducible'' (the \(m=3\) case of the question of
Shepardson and Tovey 2009 and Eggermont et al.~2013) is \textbf{false,
and false only on the boundary}: every non-transitive tournament has
\(\alpha^{*} \le 2/3\), so \(\alpha^{*} \in (2/3, 1)\) is impossible and
the strict form is vacuous -- but there are exactly \textbf{1,013}
tournaments on \(n = 10\) vertices (none smaller) with
\(\alpha^{*} = 2/3\) exactly that are not 3-inducible; on \(n = 11\)
there is a \textbf{unique regular} counterexample (a
\(C_{11}\)-circulant we call cA3 -- the counterexample to A(3)), and
exactly 1,548 self-converse ones. All have McGarvey number 5.

\textbf{5. The obstacle landscape at \(n = 9\) (Section 5).} Beyond the
unique 8-vertex obstacle \(G_8\) of Shepardson and Tovey (2009) and
Eggermont et al.~(2013), the non-3-inducible 9-tournaments are
characterized by a complete catalogue of obstacle classes, which we
enumerate exhaustively (all inclusion-minimal optimal-dual supports;
36\% of these tournaments carry several distinct minimal obstacles); we
characterize the unit-margin boundary (254 tournaments 3-inducible only
with some unanimous arc) and prove the forced-arc reversal dichotomy
(Theorem G.2).

\textbf{6. Improved bounds on \(N(k)\) (Section 6).} Every tournament on
\(\le 11\) vertices is 5-inducible (exhaustive; at \(n = 11\) every
non-3-inducible tournament even admits a unit-margin five-voter
profile), so \(12 \le N(5)\). A completion-uniqueness lemma -- for
tournaments of prescribed out-degree sequence, \(m - 1\) of the \(m\)
voters already determine the induced tournament (Proposition 6.1) --
makes it far more efficient to count profiles against \emph{regular} and
\emph{near-regular} tournaments than against all tournaments: it gives
\(N(5) \le 38\) non-constructively (Theorem 6.2, sharpening the
classical bound \(N(5) \le 41\) of Bachmeier et al.~2019), and Theorem
6.3 extends the sharpening to every odd \(m \le 21\), improving all ten
entries of Table 1 of Bachmeier et al.~(2019).

\textbf{7. Refuting the threshold conjecture at \(m = 5\), strictly
(Section 7).} The Paley tournament \(P = \mathrm{Paley}(43)\) has
\(\alpha^{*} = \mathrm{MAS}/C = 543/903 = 181/301 > 3/5\), yet is
\textbf{not 5-inducible}. Unlike the \(m = 3\) case, this refutes even
the strict form. The proof combines a counting argument (two voters are
forced into the ``level \(\le 1\)'' shell of near-maximum acyclic orders
-- the \emph{shell} being the orders whose number of forward arcs falls
short of the maximum, \(\mathrm{MAS}\), by at most one, that shortfall
being the order's \emph{level}), a \textbf{co-backing lemma}: each
voter's \emph{double-back set} -- the set of cyclic triangles in which
that voter ranks two of the three arcs backward -- is disjoint from
every other voter's, and an exhaustive screen of all over 1.6 billion
level-\(\le 1\) orders showing no two have disjoint double-back sets.
Paley(43) is the first \emph{explicit} non-5-inducible tournament whose
order is close to the counting bound (previous explicit constructions
required \(\sim 6 \times 10^{8}\) vertices).

\textbf{8. The strict hierarchy conjecture (Section 8).} We state one
conjecture, that the margin hierarchy is strict for every odd \(m\), and
leave the rest as open problems.

\subsubsection{1.2 Related work}\label{related-work}

McGarvey (1953) proved every tournament of order \(n\) is a majority of
at most \(n(n-1)\) voters; Stearns (1959) improved this to \(n + 2\),
and Erdős--Moser (1964) with Stearns' lower bound determines the optimal
order, \(\Theta(n / \log n)\). On the algorithmic side, minimum FAS on
tournaments is NP-hard (Charbit et al.~2007) and admits a PTAS
(Kenyon-Mathieu and Schudy 2007), while the Kemeny ranking problem is
\(\Theta_2^p\)-complete (Hemaspaandra et al.~2005). Alon (2002) and Mala
(1999) show almost all tournaments have predictability \(\to 1/2\).
Gilboa (1990) asked for the smallest tournament with predictability
\(< 2/3\) and gave a 54-vertex example; Shepardson--Tovey (2009)
answered with the unique-minimal 8-vertex graph \(G_8\) at
\(\alpha^{*} = 13/20\) (so \(N(3) \le 8\)) and reported the \(n = 9\)
record \(64/99\) without proving minimality. Eggermont, Hurkens and
Woeginger (2013) confirmed by exhaustive ILP that all 7-vertex
tournaments are 3-inducible and all 9-vertex tournaments are
5-inducible, with exhaustive censuses of the non-3-inducible tournaments
at \(n = 8\) (96 of them) and \(n = 9\) (Appendix G).

Milosz, Hamel and Pierrot (2018) prove the 3-cycle theorem for medians
of 3 voters and state the two conjectures that we refute in Sections 3
and 4; Blin, Crochemore, Hamel and Vialette (2011) prove the adjacency
and unanimity conditions on medians of odd \(m\). Bachmeier et
al.~(2019) study the McGarvey number (majority dimension) of a digraph,
characterize the digraphs of majority dimension 2 and 3, prove the
hardness of Kemeny medians for odd \(m \geq 7\) voters, and give
non-constructive bounds such as \(N(5) \le 41\), which Theorem 6.3
improves for every odd \(m \le 21\). The best known explicit
non-5-inducible tournament derives from a dominating-set theorem of
Fidler (2011), and requires \(\sim 6 \times 10^{8}\) vertices; we
improve this to just \(43\) vertices in Section 7.

The first author was initially motivated to study the work of Milosz,
Hamel and Pierrot (2018) by previous work on the median of 3
permutations in a different model inspired by genomics, for which we
gave a fast polynomial-time algorithm when generalized permutations
(represented by orthogonal matrices) were allowed as outputs
(Chindelevitch et al.~2018), and which has recently been shown to be
NP-hard to do when standard permutations are required to be returned (de
Moraes and Meidanis 2026). Moving from genomes to voters brought in the
structural study of tournaments, and thus the collaboration with the
second author.

The rest of this paper is organized as follows. Section 2 proves that
every minimum feedback arc set of a tournament is a minimal hitting set
of its cyclic triangles. Sections 3 and 4 refute the two conjectures of
Milosz, Hamel and Pierrot (2018), the first for every odd number of
voters from five on and the second at eleven alternatives. Section 5
turns to predictability and shows that the three-voter threshold fails
to guarantee inducibility exactly on its boundary. Section 6 shows that
every tournament on eleven vertices is 5-inducible and bounds \(N(m)\)
for odd \(m\) from above by counting, and Section 7 proves that the
Paley tournament on 43 vertices is not 5-inducible. Section 8 states one
conjecture and the open problems, and Section 9 describes the code and
data that reproduce every computation. Appendix A proves Lemma 2.2,
Appendix B gives the integer programming formulations, Appendices C and
D list the inducing profiles and certificates behind Sections 4 and 5,
Appendix E describes the maximum acyclic subgraph algorithm used in
Section 7, Appendix F tabulates the counts behind the Paley(43) proof,
Appendix G gives the nine-vertex obstacle catalogue and the forced-arc
dichotomy, and Appendix H lists the census sizes.

\subsection{2. Every minimum FAS is a minimal hitting set of
3-cycles}\label{every-minimum-fas-is-a-minimal-hitting-set-of-3-cycles}

This section begins with the main theorem of Milosz, Hamel and Pierrot:

\textbf{Theorem (Milosz--Hamel--Pierrot 2018).} \emph{If \(T\) is the
majority tournament of 3 voters, every minimum-weight feedback arc set
of \(T\) is a minimal hitting set of the directed 3-cycles of \(T\).}

We strengthen it in two directions: the unweighted statement holds in
\emph{any} tournament, with no voting hypothesis at all (Theorem 2.1),
and the original theorem is then recovered from it (Theorem 2.3).

\textbf{Theorem 2.1.} \emph{In any tournament \(T\), every minimum
feedback arc set is a minimal hitting set of all directed 3-cycles.}

\emph{Proof.} Let \(F\) be a minimum FAS and suppose some strict subset
\(G \subsetneq F\) already hits all 3-cycles; pick
\(e = (i \to j) \in F \setminus G\). Since \(T - F\) is acyclic, fix a
topological order \(\pi\) of \(T - F\); every arc of \(T - F\) is
forward and every arc of \(F\) is backward with respect to \(\pi\)
(otherwise moving it across would contradict minimality). Let
\(S = \{x : \pi(j) < \pi(x) < \pi(i)\}\), the vertices strictly between
\(j\) and \(i\). \(S \ne
\varnothing\) (otherwise reversing \(e\) would yield a smaller feedback
arc set), and for each \(x \in S\) at least one of
\(\{i \to x,\ x \to j\}\) lies in \(F\): if neither exists then
\(x \to i\) and \(j \to x\) are arcs of \(T - F\), giving the 3-cycle
\(i \to j \to x \to i\) whose only backward arc is \(e \notin G\),
contradicting that \(G\) hits all 3-cycles. Now let \(m = |S|\), let
\(c\) count the \(x \in S\) with \((i \to x) \in F\), and let \(c'\)
count the \(x \in S\) with \((x \to j) \in F\); by the previous
sentence, \(c + c' \ge m\). Moving \(i\) to just before \(j\) changes
the backward-arc count by \(\Delta^i = m - 2c - 1\); moving \(j\) to
just after \(i\) changes it by \(\Delta^j = m - 2c' - 1\). Then
\[\Delta^i + \Delta^j \;=\; 2(m - c - c' - 1) \;<\; 0,\] so one of the
two moves strictly decreases the backward-arc count, contradicting the
minimality of \(F\). \(\blacksquare\)

For 3-voter majority tournaments this yields the main theorem of Milosz
et al.~(2018) via (P4b) (unanimous arcs lie in no 3-cycle) and one
further lemma:

\textbf{Lemma 2.2.} \emph{Let \(T\) be a 3-voter majority tournament. No
minimum-weight FAS of \(T\) contains a unanimous (weight-3) arc.} (Proof
in Appendix A.)

\textbf{Theorem 2.3 (reduction to the unweighted case).} \emph{If \(T\)
is the majority tournament of 3 voters, every minimum-weight feedback
arc set of \(T\) is a minimal hitting set of all directed 3-cycles of
\(T\).}

\emph{Proof.} By Lemma 2.2, all arcs of a minimum-weight FAS \(F\) have
weight 1, so \(F\) is a minimum-cardinality FAS of the unweighted
tournament \(T'\); by Theorem 2.1 it is a minimal 3-cycle hitting set of
\(T'\); by (P4a) the 3-cycles of \(T\) and \(T'\) coincide. In
particular, any arc not in a directed 3-cycle is oriented consistently
with every Kemeny median of the three voters. \(\blacksquare\)

Milosz, Hamel and Pierrot (2018) conjectured two extensions of their
theorem, one to more voters and one to the sizes of the two sets
involved; the next two sections refute both.

\subsection{3. Refuting Conjecture 1 of
Milosz--Hamel--Pierrot}\label{refuting-conjecture-1-of-miloszhamelpierrot}

In this section we show that the first conjecture, which extends the
three-voter theorem of Section 2 to every odd number of voters, fails
from five voters on. The conjecture reads as follows.

\textbf{Conjecture 1 (Milosz et al.~2018).} \emph{For any odd \(m\) and
any median \(\pi^{*}\) of \(m\) voters, an arc \((i, j)\) of the
majority tournament that is not contained in any directed 3-cycle
satisfies \(i \prec_{\pi^{*}} j\). Equivalently, no minimum-weight FAS
of the majority tournament reverses the arc \((i,j)\).}

Milosz, Hamel and Pierrot were led to this conjecture by their
three-voter theorem, recovered here as Theorem 2.3: for \(m = 3\), a
median reverses only arcs lying on some directed 3-cycle, leaving every
arc that is in no 3-cycle (one facing no local cyclic conflict) in its
majority direction. It is natural to ask whether this locality survives
as voters are added; Conjecture 1 asserts that it does, for every odd
\(m\).

By Theorem 2.3, the conjecture holds for \(m = 3\). For every odd
\(m \ge 5\), the conjecture fails; we give minimal counterexamples. Both
use the tournament \(T_1\) on \([4]\): the 4-cycle
\(1 \to 2 \to 3 \to 4 \to 1\) plus diagonals \(1 \to 3\), \(4 \to 2\)
(\(T_1\) is Figure 1 of Milosz et al.~2018 minus its cycle-free source).
Its 3-cycles are \(1 \to 3 \to 4 \to 1\) and \(2 \to 3 \to 4 \to 2\);
arc \((1,2)\) lies in no 3-cycle.

\textbf{Counterexample 3.1 (\(m = 7\) weak, \(m = 9\) strong;
\(n = 4\)).} Take the three voter types
\[\rho = (2,3,4,1), \qquad \sigma = (1,3,4,2), \qquad \tau = (4,1,2,3).\]
With multiplicities \((3, 2, 2)\) -- seven voters -- the minimum-weight
FASs are \(\{(3,4)\}\) and \(\{(1,2), (1,3), (4,2)\}\), both of weight
3: \emph{a} minimum-weight FAS reverses \((1,2)\), which is in no
3-cycle (Figure 1a). With multiplicities \((4, 3, 2)\) -- nine voters --
the tie is broken and \(\{(1,2), (1,3), (4,2)\}\) becomes the
\textbf{unique} minimum-weight FAS: every median reverses an arc in no
3-cycle (Figure 1b).

\begin{figure}[t]
\centering
\begin{minipage}[t]{0.48\textwidth}\centering
\includegraphics[width=0.92\linewidth]{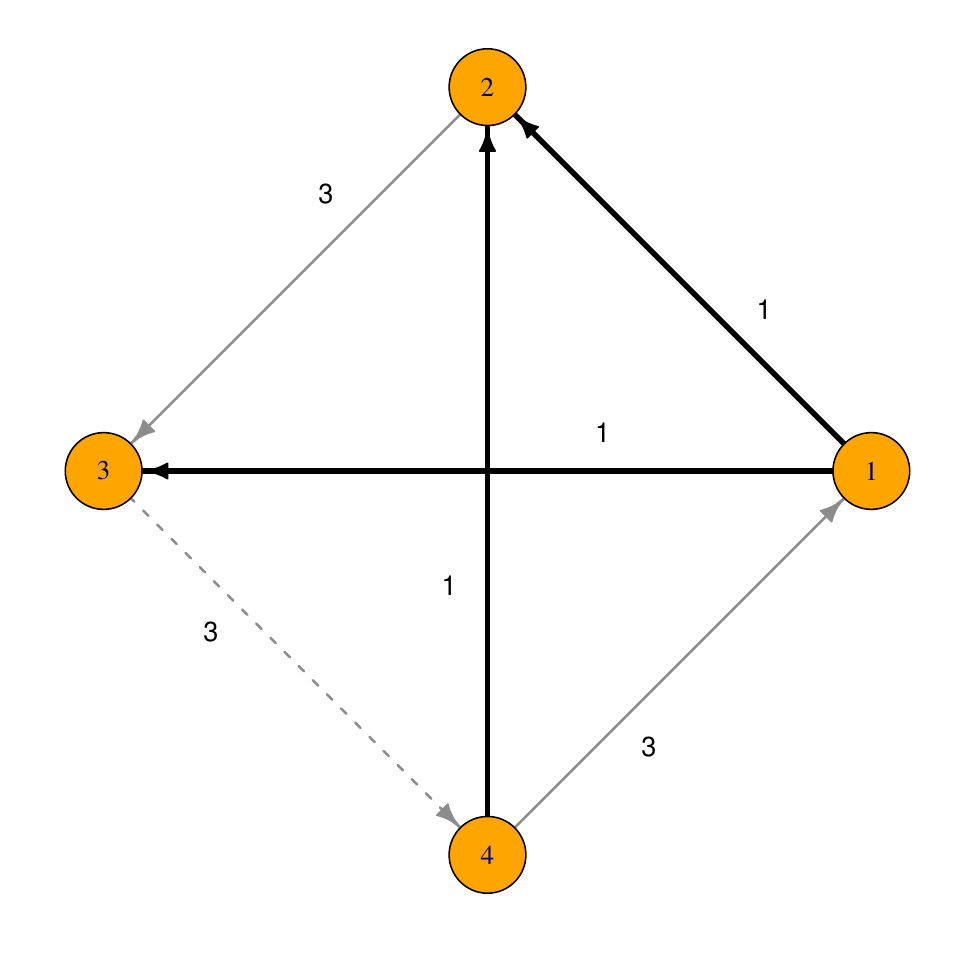}\par (a)
\end{minipage}\hfill
\begin{minipage}[t]{0.48\textwidth}\centering
\includegraphics[width=0.92\linewidth]{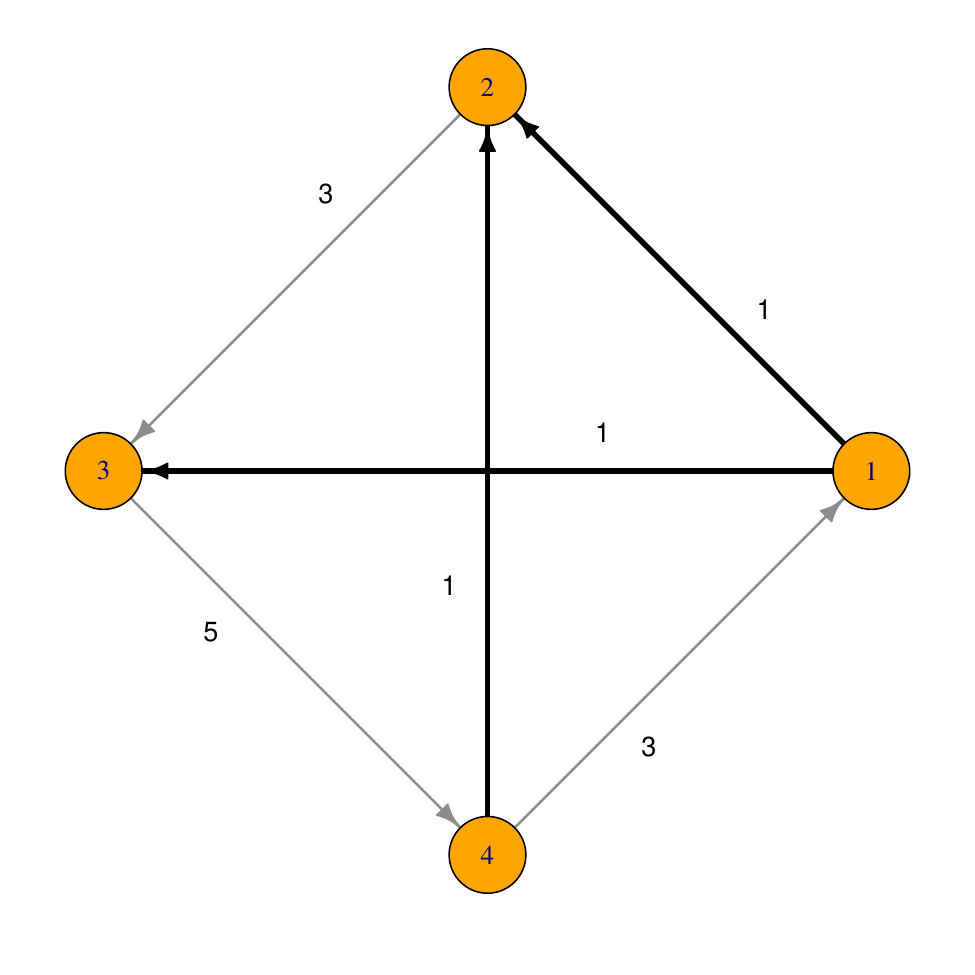}\par (b)
\end{minipage}
\caption{The weighted majority tournament of Counterexample 3.1 (the tournament $T_1$; arc
labels are the majority margins). (a) Seven voters, multiplicities $(3,2,2)$: the two
minimum-weight feedback arc sets are $\{(3,4)\}$ (dashed) and $\{(1,2),(1,3),(4,2)\}$ (thick
black); the latter reverses the arc $(1,2)$, which lies in no directed 3-cycle. (b) Nine
voters, multiplicities $(4,3,2)$: the thick black arcs now form the unique minimum-weight FAS.}
\end{figure}

\textbf{Counterexample 3.2 (strong, \(m = 5\), \(n = 6\)).} There is a
unique (up to isomorphism) tournament on six vertices for which the
5-voter profile
\(\pi = (2,5,3,6,1,4) \times 2\ (\text{multiplicity } 2),\ \rho = (1,4,3,6,2,5),\ \sigma = (4,5,3,6,1,2),\
\tau = (6,1,4,2,5,3)\) makes
\(F_e = \{(1,2), (4,2), (4,3), (4,5), (6,2)\}\) the unique
minimum-weight FAS, where \(e = (4,2)\) lies in no 3-cycle. (Figure 2.)
This tournament is semi-regular, self-converse and has
\(|\mathrm{Aut}| = 3\).

\begin{figure}[t]
\centering
\includegraphics[width=0.52\textwidth]{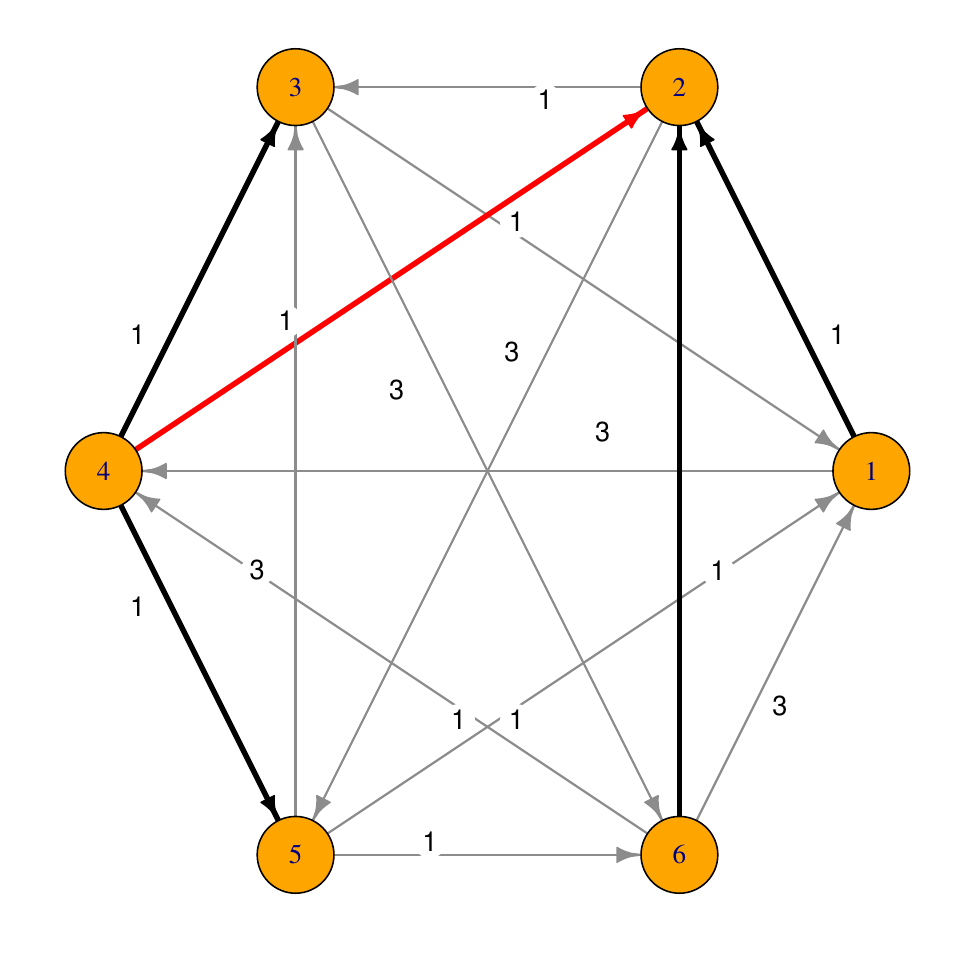}
\caption{The weighted majority tournament of Counterexample 3.2 ($m = 5$, $n = 6$; arc labels
are the margins). The unique minimum-weight feedback arc set $F_e$ (thick) reverses
$e = (4,2)$ (red), which lies in no directed 3-cycle.}
\end{figure}

\textbf{Minimality (all claims ILP-verified; formulations in Appendix
B).} Conjecture 1 holds for \(n = 3\). At \(n = 4\), \(m = 7\)
(respectively \(m = 9\)) is the smallest odd number of voters admitting
a weak (respectively strong) counterexample, and two voter types never
suffice, regardless of multiplicities. At \(m = 5\), the \(n = 6\)
tournament above is the unique order-6 refuting tournament, and none of
the three candidate FASs achieves strict minimality with only 3 voter
types.

\textbf{Remark 3.3 (extension by an even number of voters).} Every
counterexample in this section and in Section 4 extends to any larger
odd number of voters: appending to a profile an arbitrary linear order
\(\pi\) \emph{together with its reverse} does not change any pairwise
margin (the pair splits every ordered pair 1:1), and adds the same
constant to the score of every candidate median -- for any \(\mu\),
\(d(\mu, \pi) + d(\mu, \bar{\pi}) = \binom{n}{2}\) -- so the weighted
majority tournament, the set of medians, and the set of minimum-weight
FASs are all preserved. Thus a counterexample with \(m\) voters yields
counterexamples with \(m + 2, m + 4, \dots\) voters, and the conjectured
property fails for every odd \(m \ge 5\) (Counterexample 3.2 establishes
the case \(m = 5\), and the extension then yields every larger odd
\(m\)).

The first conjecture thus fails as soon as the number of voters exceeds
three. The second concerns three voters only, and compares the sizes of
the two kinds of arc set; the next section shows that it holds up to ten
alternatives and fails at eleven.

\subsection{\texorpdfstring{4. Refuting Conjecture 2:
\(\mathrm{FAS} = \mathrm{HS}_3\) breaks for 3-inducible tournaments at
\(n = 11\)}{4. Refuting Conjecture 2: \textbackslash mathrm\{FAS\} = \textbackslash mathrm\{HS\}\_3 breaks for 3-inducible tournaments at n = 11}}\label{refuting-conjecture-2-mathrmfas-mathrmhs_3-breaks-for-3-inducible-tournaments-at-n-11}

In this section we test the second conjecture against every tournament
on up to ten vertices and against the regular and self-converse
3-inducible tournaments on eleven. It states the following.

\textbf{Conjecture 2 (Milosz et al.~2018).} \emph{For any 3-inducible
tournament \(T\), \(\mathrm{FAS}(T) =
\mathrm{HS}_3(T)\).}

Every feedback arc set meets each directed 3-cycle, so it is in
particular a 3-cycle hitting set and
\(\mathrm{FAS}(T) \ge \mathrm{HS}_3(T)\) for every tournament. Theorem
2.1 sharpens this to the fact that a minimum feedback arc set is always
a \emph{minimal} 3-cycle hitting set. Conjecture 2 is the natural
strengthening that on 3-inducible tournaments this minimal hitting set
is in fact a \emph{minimum} one. In other words, a set hitting all
directed 3-cycles as cheaply as possible already meets every longer
directed cycle, so that \(\mathrm{FAS}\) and \(\mathrm{HS}_3\) coincide.
Milosz, Hamel and Pierrot verified the equality on small tournaments and
a collection of specific examples, and conjectured it in general.

\textbf{Theorem 4.1 (computational).} \emph{For every tournament on
\(n \le 10\) vertices (3-inducible or not),
\(\mathrm{FAS}(T) = \mathrm{HS}_3(T)\).}

\textbf{Counterexample 4.2.} \emph{There is a 3-inducible regular
tournament \(T^{*}\) on 11 vertices with
\(\mathrm{FAS}(T^{*}) = 17 > 16 = \mathrm{HS}_3(T^{*})\).}

\(T^{*}\) is the unique such tournament among the \emph{3-inducible}
regular tournaments on 11 vertices -- the class the conjecture concerns;
we did not test the equality on the non-3-inducible ones. It is the
majority tournament of
\[\rho = (4,5,6,11,8,9,10,2,7,1,3),\quad \sigma = (10,8,9,1,2,3,7,5,6,4,11),\quad
\tau = (3,7,11,1,2,6,4,5,9,10,8),\] with
\(\mathrm{Aut}(T^{*}) = \langle (4\,5\,6), (8\,9\,10) \rangle\). Each
voter keeps both orbits \(\{4,5,6\}\) and \(\{8,9,10\}\) consecutive,
and on each orbit the three voters take its three rotations, so the
profile follows \(\mathrm{Aut}(T^{*})\) as far as it can: no 3-voter
profile of \(T^{*}\) is invariant under a non-trivial automorphism. We
certify the three facts that make it a counterexample (Appendix
B.2--B.3):

\begin{itemize}
\tightlist
\item
  \textbf{\(T^{*}\) is 3-inducible} -- it is by construction the
  majority tournament of the three voters \(\rho, \sigma, \tau\) above,
  and it is regular.
\item
  \textbf{\(\mathrm{HS}_3(T^{*}) = 16\)} -- an explicit set of 16 arcs
  meets \emph{every} directed 3-cycle, and an ILP certifies that no
  15-arc set does; the minimum hitting sets number 9 and form a single
  \(\mathrm{Aut}(T^{*})\)-orbit (Figure 3a).
\item
  \textbf{\(\mathrm{FAS}(T^{*}) = 17\)} -- the linear order
  \((4,5,6,11,8,9,10,1,2,3,7)\) (chosen so that the two
  \(\mathrm{Aut}(T^{*})\)-triples appear consecutively) leaves exactly
  17 arcs backward, an FAS of size 17; no order does better, so
  \(\mathrm{MAS}(T^{*}) = 55 - 17 = 38\) (Figure 3b).
\end{itemize}

\begin{figure}[t]
\centering
\begin{minipage}[t]{0.48\textwidth}\centering
\includegraphics[width=0.92\linewidth]{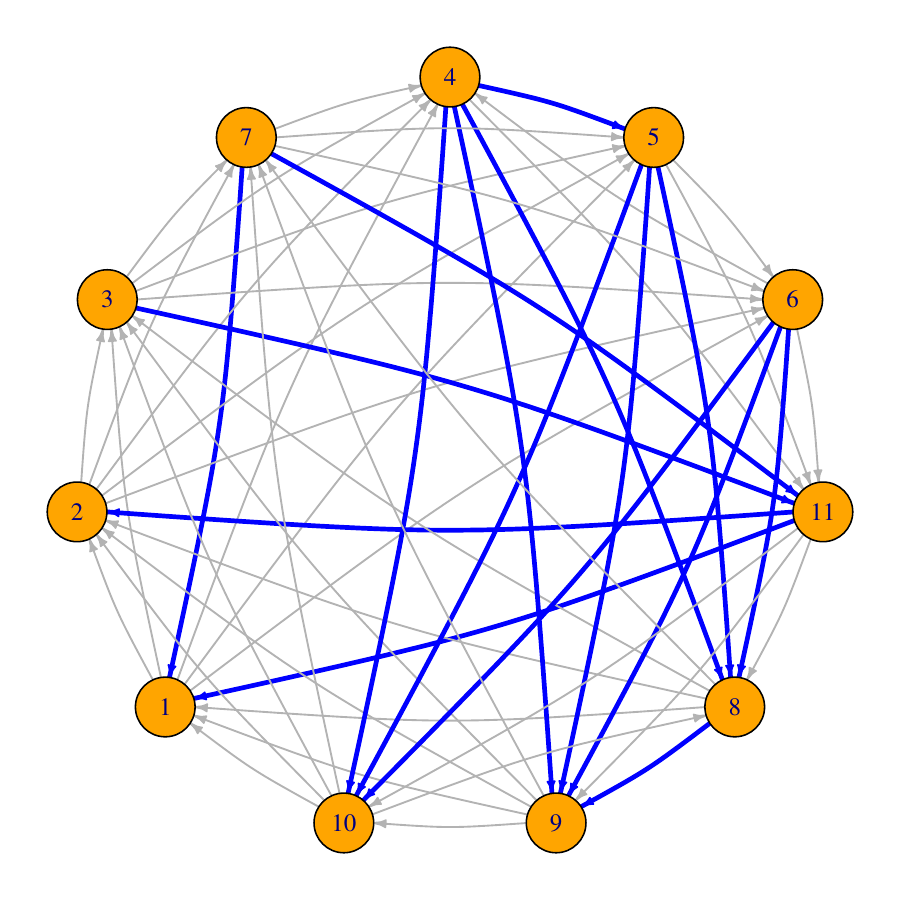}\par (a)
\end{minipage}\hfill
\begin{minipage}[t]{0.48\textwidth}\centering
\includegraphics[width=0.92\linewidth]{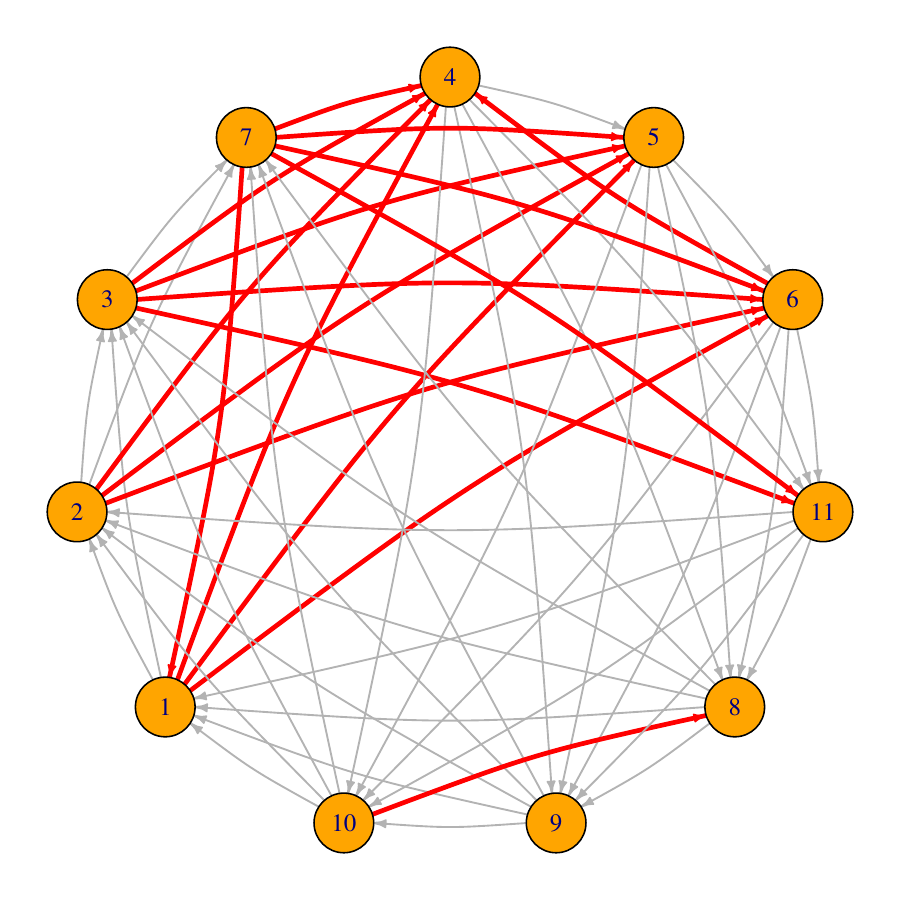}\par (b)
\end{minipage}
\caption{The tournament $T^{*}$ of Counterexample 4.2, drawn with the vertices on a circle in
the witness order $(4,5,6,11,8,9,10,1,2,3,7)$, clockwise from the top. (a) A minimum 3-cycle
hitting set (blue): 16 arcs meeting every one of the 55 cyclic triangles. (b) A minimum
feedback arc set (red): the 17 arcs backward in the witness order.}
\end{figure}

Since \(\mathrm{HS}_3(T^{*}) = 16 < 17 = \mathrm{FAS}(T^{*})\) on a
3-inducible tournament, Conjecture 2 fails at \(n = 11\), the smallest
possible order by Theorem 4.1. Among the \emph{3-inducible}
self-converse tournaments on 11 vertices exactly \textbf{6} violate the
equality, one of them being \(T^{*}\) itself, which is regular as well
as self-converse (Figure 4, where panel (f) is \(T^{*}\); inducing
profiles in Appendix C).

\begin{figure}[p]
\centering
\begin{minipage}[t]{0.38\textwidth}\centering
\includegraphics[width=\linewidth,trim=20 20 20 20,clip]{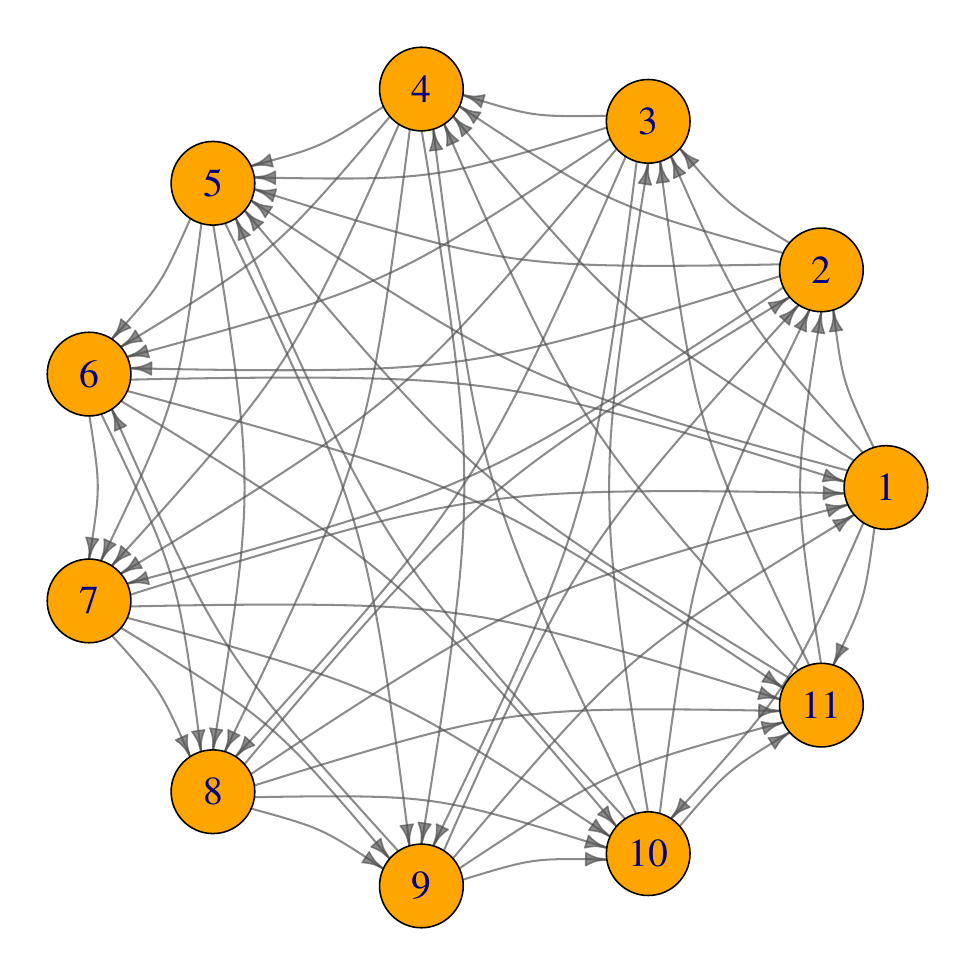}\par (a)
\end{minipage}\hfill
\begin{minipage}[t]{0.38\textwidth}\centering
\includegraphics[width=\linewidth,trim=20 20 20 20,clip]{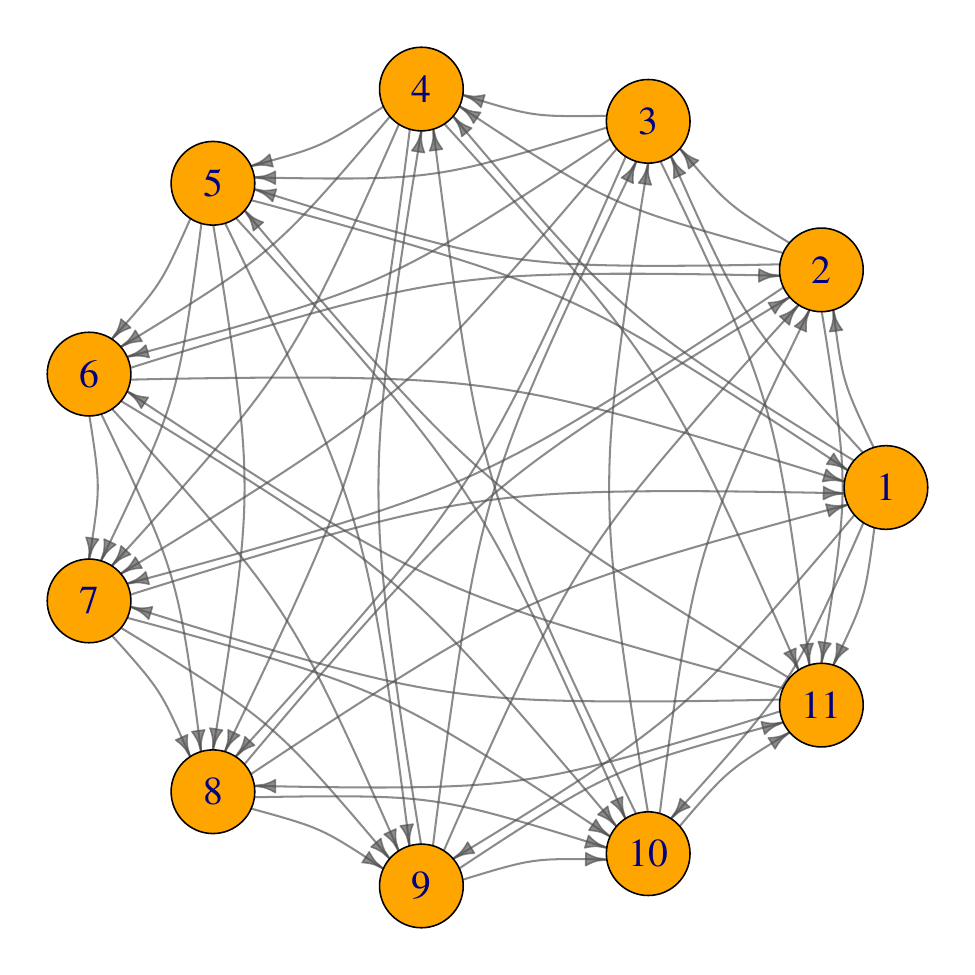}\par (b)
\end{minipage}\par\vspace{1ex}
\begin{minipage}[t]{0.38\textwidth}\centering
\includegraphics[width=\linewidth,trim=20 20 20 20,clip]{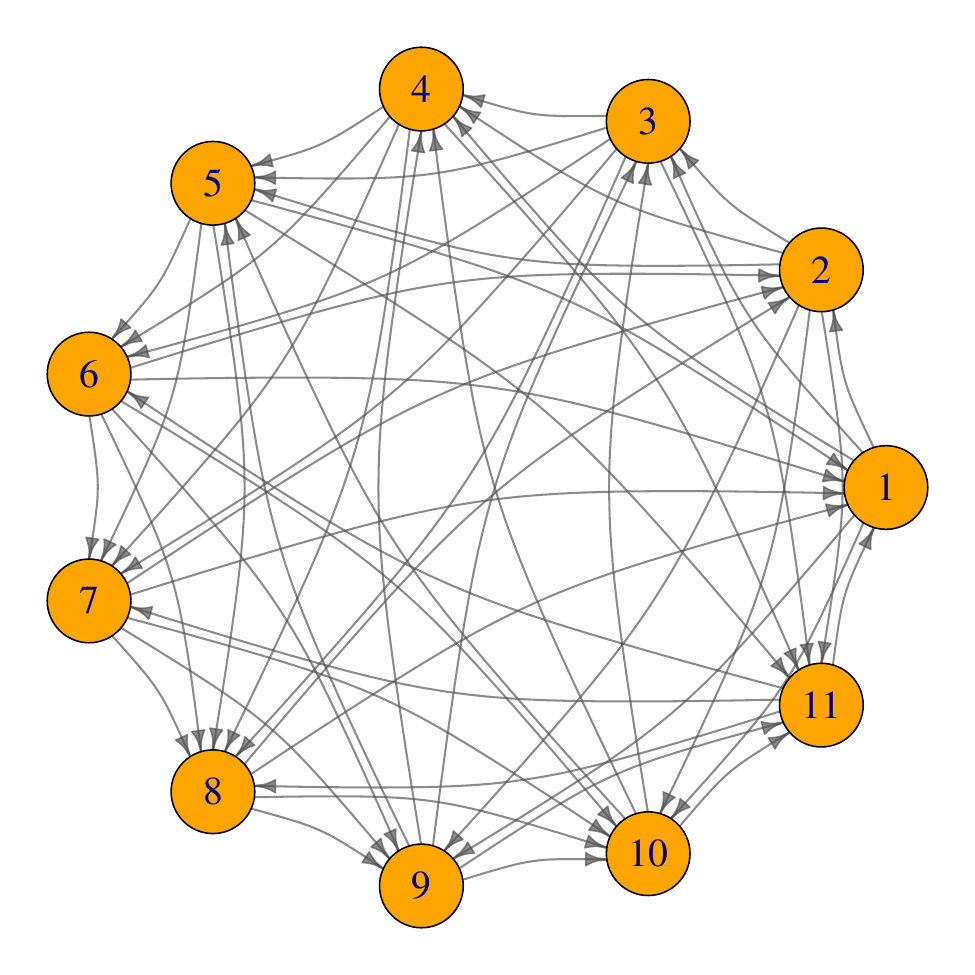}\par (c)
\end{minipage}\hfill
\begin{minipage}[t]{0.38\textwidth}\centering
\includegraphics[width=\linewidth,trim=20 20 20 20,clip]{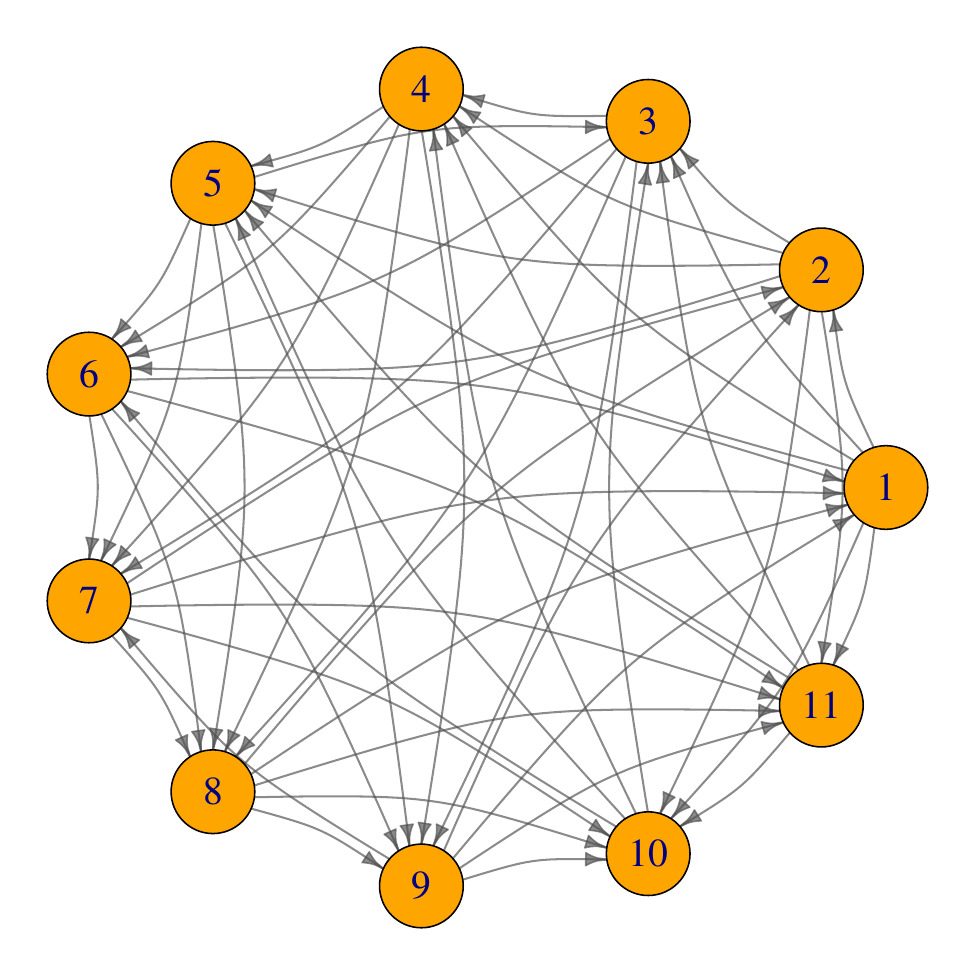}\par (d)
\end{minipage}\par\vspace{1ex}
\begin{minipage}[t]{0.38\textwidth}\centering
\includegraphics[width=\linewidth,trim=20 20 20 20,clip]{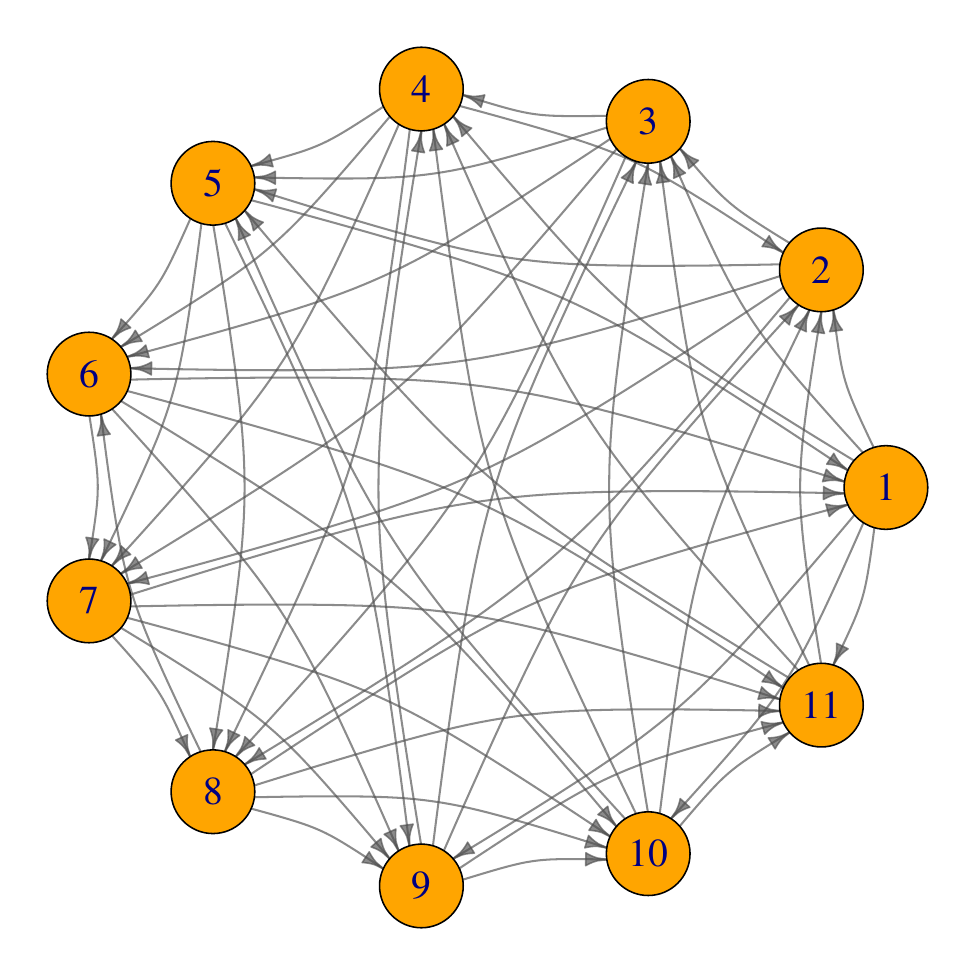}\par (e)
\end{minipage}\hfill
\begin{minipage}[t]{0.38\textwidth}\centering
\includegraphics[width=\linewidth,trim=20 20 20 20,clip]{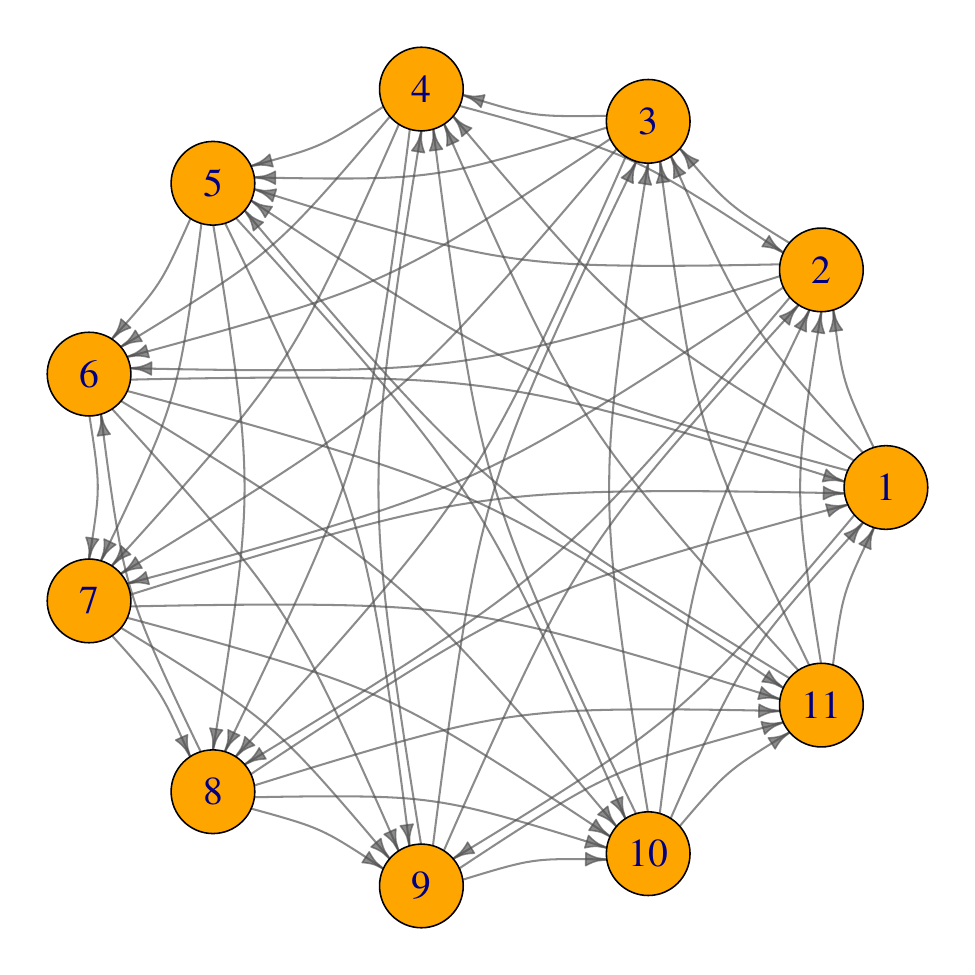}\par (f)
\end{minipage}
\caption{The six 3-inducible self-converse tournaments on 11 vertices violating Conjecture 2
(panels (a)--(f)); panel (f) is isomorphic to $T^{*}$ of Figure 3, the only regular one of the six. Each panel places the vertices in circular order; every arc also bows to
the left of its direction of travel.}
\end{figure}

Both conjectures of Milosz, Hamel and Pierrot therefore fail, and the
counterexamples are rare: the equality holds on every tournament up to
ten vertices and fails for only a handful of the eleven-vertex
tournaments we examined.

\subsection{\texorpdfstring{5. The threshold conjecture fails at
\(m = 3\) exactly on the
boundary}{5. The threshold conjecture fails at m = 3 exactly on the boundary}}\label{the-threshold-conjecture-fails-at-m-3-exactly-on-the-boundary}

Having settled the questions of Milosz, Hamel and Pierrot, we turn for
the rest of the paper to the inverse question: when does high
predictability guarantee inducibility by few voters?

\textbf{Conjecture A.} Shepardson and Tovey (2009) introduced
predictability and proved that \(\alpha^{*} \ge \frac{m+1}{2m}\) is
necessary for \(m\)-inducibility. They explicitly asked whether the
converse also holds (Shepardson and Tovey 2009, p.~502), a question
repeated in Eggermont et al.~(2013). We denote by \textbf{A(\(m\))} the
converse statement: \(\alpha^{*}(T) \ge \frac{m+1}{2m}\) implies \(T\)
is \(m\)-inducible.

\textbf{A(\(m\)) as a no-integrality-gap statement.} The two sides of
A(\(m\)) are the integer and fractional versions of a single covering
program, implicit in Shepardson and Tovey (2009): assign a weight
\(z_\pi \ge 0\) to every linear order \(\pi\) of \(V(T)\), and require
\[\sum_\pi z_\pi = m, \qquad \sum_{\pi\ \text{forwards}\ (u \to v)} z_\pi \;\ge\; \tfrac{m+1}{2}
\quad \forall (u \to v) \in E(T).\] An integral solution is precisely a
\(m\)-voter inducing profile, so integer feasibility is
\(m\)-inducibility; dividing \(z\) by \(m\) shows that LP feasibility is
exactly \(\alpha^{*}(T) \ge \frac{m+1}{2m}\). The necessity direction of
Shepardson and Tovey (2009) is trivial as every integral solution is a
fractional one, and A(\(m\)) is precisely the assertion that this
program has \textbf{no integrality gap}. Theorem 5.1 and Counterexample
5.2 below refute A(3) by exhibiting gap instances, and Section 7 does
the same for A(5).

By (P3), \(\alpha^{*} \in (2/3, 1)\) is impossible, so for \(m = 3\) the
\emph{strict} form is vacuously true and the boundary
\(\alpha^{*} = 2/3\) is the only place A(3) can fail. It does:

\textbf{Theorem 5.1 (refutation of A(3)).} \emph{Exactly 1,013
tournaments on 10 vertices with \(\alpha^{*} = 2/3\) are not
3-inducible, and none on \(\le 9\) vertices. All have McGarvey number
5.}

To our knowledge, this is the first refutation of the sufficiency
direction of the predictability threshold, left open by Shepardson and
Tovey (2009) and Eggermont et al.~(2013).

The census is exhaustive over all \(D_{10}\) ten-vertex tournaments
(McKay 2026); it uses a hereditary pre-filter and computes
\(\alpha^{*}\) in exact rational arithmetic. The 1,013 counterexamples
display no exceptional structure: 1,009 of them are rigid, none is
regular, and 90 distinct out-degree sequences occur among them. By
monotonicity (P2) and the absence of counterexamples on \(\le 9\)
vertices (Appendix G.2), no 10-vertex counterexample contains a smaller
one.

\textbf{Counterexample 5.2 (the regular witness cA3).} \emph{Let \(T\)
be the \textbf{circulant tournament} on \(\mathbb{Z}_{11}\) with
\textbf{connection set} \(S = \{1,2,3,4,6\}\): the vertices are the
residues \(0, 1, \dots, 10\), with an arc \(u \to v\) whenever
\((v - u) \bmod 11 \in S\). Then: (i) \(\alpha^{*}(T) = 2/3\) exactly;
(ii) \(T\) is not 3-inducible; (iii) \(T\) is 5-inducible, so
\(\mathrm{McG}(T) = 5\); (iv) \(T\) is vertex-critical; (v) \(T\) is the
unique counterexample among the \(R_{11}\) regular tournaments on 11
vertices.}

\emph{Verification.} \textbf{(i)} Upper bound: \(T\) is regular, hence
not transitive, so \(\alpha^{*}(T) \le 2/3\) by (P3). Lower bound: a
\textbf{6-voter profile}, given in Appendix D, makes all 55 arcs forward
in exactly 4 of the 6 voters, every arc exactly at \(4/6 = 2/3\) (see
Appendix D). \textbf{(ii)} Non-3-inducibility is verified in three
independent ways: the inducibility ILP of Appendix B in CPLEX; a
different encoding in OR-Tools CP-SAT; and a solver-free enumeration
argument (reproducibility package, Section 9). \textbf{(iii)} Deleting
any voter of the 6-voter profile used in (i) retains a majority for all
the arcs by shifting the balance of \(4\):\(2\) to either \(4\):\(1\) or
\(3\):\(2\), making it a 5-inducing profile; with (ii),
\(\mathrm{McG}(T) = 5\). \textbf{(iv)} It suffices to check that no
vertex deletion of \(T\) matches one of the 1,013 exceptions of Theorem
5.1, since those are the minimum-order counterexamples; we do so by
putting each vertex-deleted tournament into canonical form via the
igraph library (Antonov et al.~2023). \textbf{(v)} The exhaustive sweep
of all \(R_{11}\) regular tournaments (see Appendix G.4) finds exactly
49 with \(\alpha^{*} = 2/3\), of which 48 are 3-inducible and exactly
one, \(T\), is not. \(\blacksquare\)

Figure 5 draws cA3 as a circulant tournament; the residues \(0..10\) are
placed around a circle, with each arc coloured by its step
\((v - u) \bmod 11 \in S = \{1,2,3,4,6\}\). The five step-classes are
the orbits of \(\mathrm{Aut}(\mathrm{cA3}) = C_{11}\) on the arcs, so
the tournament is invariant under the rotation \(x \mapsto x + 1\); this
\(C_{11}\) symmetry pins every arc to the same tight value \(2/3\)
(Lemma 1.1) and underlies the balanced 6-order certificate (Appendix D).

\begin{figure}[t]
\centering
\includegraphics[width=0.6\textwidth]{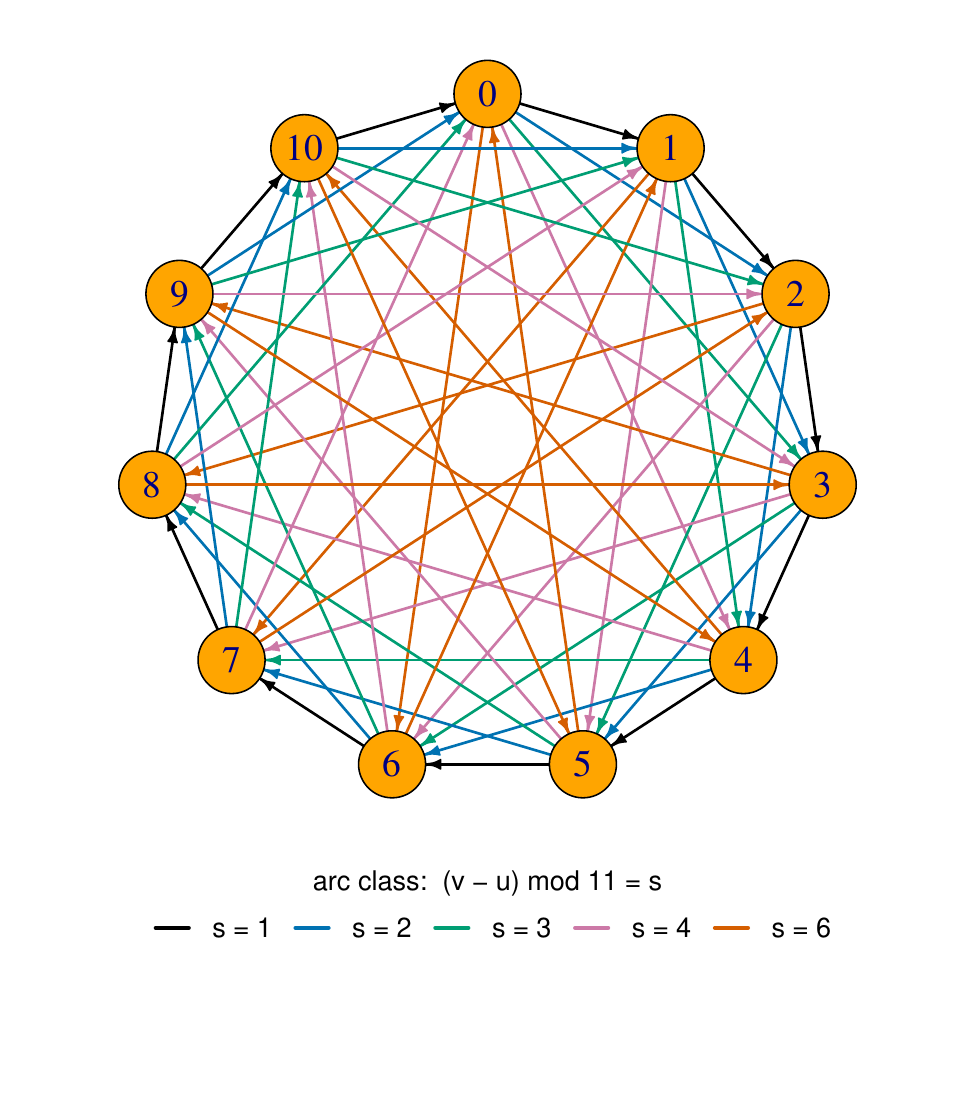}
\caption{The counterexample cA3 (Counterexample 5.2): the circulant tournament on
$\mathbb{Z}_{11}$ with connection set $S = \{1,2,3,4,6\}$, i.e.\ an arc $u \to v$ whenever
$(v - u) \bmod 11 \in S$. Each arc is coloured by its step $s = (v - u) \bmod 11$. This is
the unique regular tournament on $11$ vertices with $\alpha^{*} = 2/3$ that is not 3-inducible.}
\end{figure}

\textbf{The self-converse census.} A complete analysis of the \(S_{11}\)
self-converse tournaments on 11 vertices yields additional examples,
exactly one of which -- which we call \textbf{cA6} -- requires a minimum
of \textbf{9} voters to achieve the predictability \(2/3\). It is shown
in Appendix D, along with the full census.

\textbf{The obstacle landscape at \(n = 9\).} Behind the census lie the
\emph{obstacles} that pin non-3-inducibility below the threshold -- the
supports of the optimal duals of the covering program above (P2).
Eggermont et al.~(2013) found that all 96 non-3-inducible tournaments at
\(n = 8\) share the single 20-arc obstacle, which we call \(G_8\) (see
Figure 6). It has \(\alpha^{*} = 13/20\), is Eulerian and self-converse,
and every order forwards at most 13 of its 20 arcs.

At \(n = 9\) there are many more obstacles: all but 1,054 of the 17,674
non-3-inducible 9-tournaments (Eggermont et al.~2013) contain \(G_8\),
and those 1,054 generate exactly \textbf{40 further obstacles}, with the
values and frequencies of Table 1. Together with \(G_8\), these
\textbf{41 obstacles characterize non-3-inducibility at \(n = 9\)}: a
9-vertex tournament is non-3-inducible iff it contains one of them
(Appendix G). Each value has the form \(\frac{2t-1}{3t-1}\), for
\(t \in \{7, 8, 9, 10, 11, 12, 14, 15, 19\}\) (\(t = 7\) being
\(\alpha^{*}(G_8) = 13/20\)); all 41 lie strictly below \(2/3\), which
is why, consistent with Theorem 5.1, no tournament on \(\le 9\) vertices
with \(\alpha^{*} = 2/3\) can fail to be 3-inducible.

\begin{longtable}[]{@{}
  >{\raggedright\arraybackslash}p{(\linewidth - 16\tabcolsep) * \real{0.1351}}
  >{\raggedright\arraybackslash}p{(\linewidth - 16\tabcolsep) * \real{0.1081}}
  >{\raggedright\arraybackslash}p{(\linewidth - 16\tabcolsep) * \real{0.1081}}
  >{\raggedright\arraybackslash}p{(\linewidth - 16\tabcolsep) * \real{0.1081}}
  >{\raggedright\arraybackslash}p{(\linewidth - 16\tabcolsep) * \real{0.1081}}
  >{\raggedright\arraybackslash}p{(\linewidth - 16\tabcolsep) * \real{0.1081}}
  >{\raggedright\arraybackslash}p{(\linewidth - 16\tabcolsep) * \real{0.1081}}
  >{\raggedright\arraybackslash}p{(\linewidth - 16\tabcolsep) * \real{0.1081}}
  >{\raggedright\arraybackslash}p{(\linewidth - 16\tabcolsep) * \real{0.1081}}@{}}
\caption{The eight predictability values of the 40 obstacle classes not
containing \(G_8\), with their frequencies among the 1,054 such
non-3-inducible 9-tournaments.}\tabularnewline*
\toprule\noalign{}
\begin{minipage}[b]{\linewidth}\raggedright
exact \(\alpha^{*}\)
\end{minipage} & \begin{minipage}[b]{\linewidth}\raggedright
15/23
\end{minipage} & \begin{minipage}[b]{\linewidth}\raggedright
17/26
\end{minipage} & \begin{minipage}[b]{\linewidth}\raggedright
19/29
\end{minipage} & \begin{minipage}[b]{\linewidth}\raggedright
21/32
\end{minipage} & \begin{minipage}[b]{\linewidth}\raggedright
23/35
\end{minipage} & \begin{minipage}[b]{\linewidth}\raggedright
27/41
\end{minipage} & \begin{minipage}[b]{\linewidth}\raggedright
29/44
\end{minipage} & \begin{minipage}[b]{\linewidth}\raggedright
37/56
\end{minipage} \\*
\midrule\noalign{}
\endfirsthead
\toprule\noalign{}
\begin{minipage}[b]{\linewidth}\raggedright
exact \(\alpha^{*}\)
\end{minipage} & \begin{minipage}[b]{\linewidth}\raggedright
15/23
\end{minipage} & \begin{minipage}[b]{\linewidth}\raggedright
17/26
\end{minipage} & \begin{minipage}[b]{\linewidth}\raggedright
19/29
\end{minipage} & \begin{minipage}[b]{\linewidth}\raggedright
21/32
\end{minipage} & \begin{minipage}[b]{\linewidth}\raggedright
23/35
\end{minipage} & \begin{minipage}[b]{\linewidth}\raggedright
27/41
\end{minipage} & \begin{minipage}[b]{\linewidth}\raggedright
29/44
\end{minipage} & \begin{minipage}[b]{\linewidth}\raggedright
37/56
\end{minipage} \\*
\midrule\noalign{}
\endhead
\bottomrule\noalign{}
\endlastfoot
frequency & 529 & 398 & 49 & 36 & 20 & 12 & 6 & 4 \\*
\end{longtable}

One of the simplest new obstacles, which we call \(G_9\), shares some
similarity with \(G_8\): both have an acyclic core of 4 vertices, a
triangle-free group of vertices, and a single 4-cycle joining them
(Figure 6).

\begin{figure}[t]
\centering
\begin{minipage}[t]{0.48\textwidth}\centering
\includegraphics[width=0.96\linewidth]{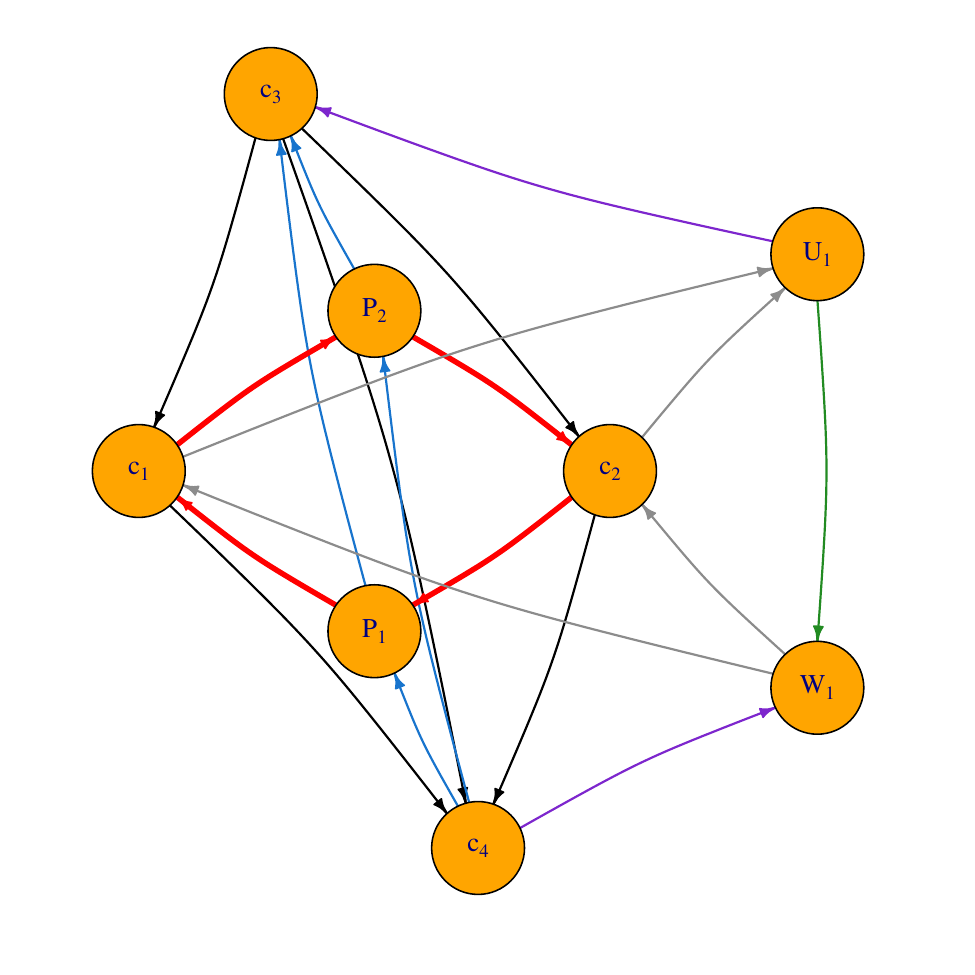}\par (a) $G_8$
\end{minipage}\hfill
\begin{minipage}[t]{0.48\textwidth}\centering
\includegraphics[width=0.96\linewidth]{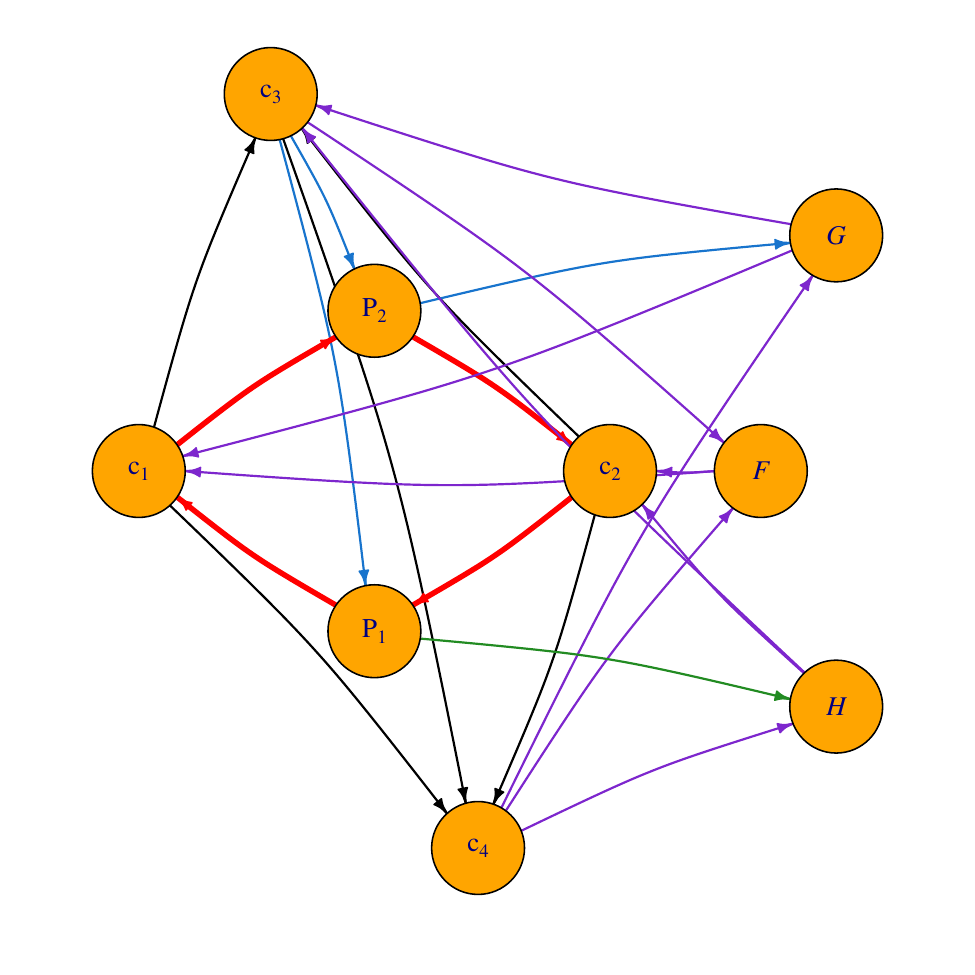}\par (b) $G_9$
\end{minipage}
\caption{$G_8$ (a) and $G_9$ (b) side by side: an acyclic core of 4 vertices, a triangle-free
group of vertices, and a single 4-cycle (thick red) joining them.
$\alpha^{*}(G_8) = 13/20$, $\alpha^{*}(G_9) = 15/23$.}
\end{figure}

\textbf{The unit-margin boundary.} A second phenomenon at \(n = 9\):
exactly \textbf{254} tournaments are 3-inducible but \emph{not} with all
margins equal to 1 -- every inducing profile makes some arc unanimous.
These witness the strict inclusion
\(\mathcal{I}_{3,1} \subsetneq \mathcal{I}_{3,3}\) (Section 1), the
starting level of the margin hierarchy we conjecture strict for every
odd \(m\) (Conjecture 8.1). Call an arc \textbf{forced} if every 3-voter
inducing profile makes it unanimous. Then:

\textbf{Lemma 5.4 (forced arcs gain at least three 3-cycles).} \emph{Let
\(e = (u \to v)\) be a forced arc of a 3-inducible tournament \(T\), and
let \(P = \{w : u \to w \to v\}\). Then \(|P| \ge 3\): reversing a
forced arc always creates at least three cyclic triangles.} (Proof in
Appendix G.4.)

\textbf{Interpretation.} cA3 and the 1,013 census counterexamples show
that predictability and inducibility need not coincide: in each case the
LP attains its maximum possible value \(2/3\), yet 3-inducibility fails
for a global reason that \(\alpha^{*}\) does not detect.

At three voters the threshold conjecture thus fails, but only on its
boundary, where the predictability equals \(2/3\) exactly. At five
voters values strictly above the threshold are possible, and the next
two sections show that the conjecture fails strictly there.

\FloatBarrier

\subsection{\texorpdfstring{6. Improved bounds on
\(N(k)\)}{6. Improved bounds on N(k)}}\label{improved-bounds-on-nk}

For \(m = 5\) the threshold is \(3/5\) and, unlike \(m = 3\), values
strictly above it are possible -- so A(5) has a non-vacuous strict form.
To refute it we need a non-5-inducible tournament with
\(\alpha^{*} > 3/5\); before constructing an explicit example (Section
7), we first bound the orders at which non-5-inducible tournaments can
occur. For the lower bound, our exhaustive census (contribution 6) shows
that every tournament on at most 11 vertices is 5-inducible, so
\(N(5) \ge 12\). Through \(n = 10\) this is the ILP census of Appendix
B.1. At \(n = 11\) no ILP was needed: of the
\(D_{11} = 903{,}753{,}248\) tournaments (McKay 2026),
\(382{,}625{,}248\) are 3-inducible, and every one of the remaining
\(521{,}128{,}000\) admits a \emph{unit-margin} five-voter profile,
found by a dissent-pair constraint search and re-checked by an
independent verifier, so all non-3-inducible tournaments on \(n \le 11\)
vertices in fact lie in \(\mathcal{I}_{5,1}\). This section's counting
arguments bound \(N(k)\) from above, improving the known bounds for
every odd \(m \le 21\).

Comparing counts of profiles and tournaments gives \(N(5) \le 41\)
non-constructively (Eggermont et al.~2013; Bachmeier et al.~2019).
Restricting the count to \emph{regular} tournaments tightens this to
\(N(5) \le 39\), and to \emph{near-regular} tournaments to
\(N(5) \le 38\), because four voters already determine a tournament of
prescribed out-degrees:

\textbf{Proposition 6.1 (algorithmic completion uniqueness).} \emph{Fix
odd \(m\), linear orders \(O_1, \dots, O_{m-1}\) on \([n]\), and a
target out-degree sequence \(d \colon [n] \to
\mathbb{Z}_{\ge 0}\). There is at most one tournament with out-degree
sequence \(d\) realized by the addition of a \(m\)-th voter to the
profile; if it exists, it can be constructed in \(O(n^2)\) time.}

\emph{Proof.} Among the pairs \(\{x, y\}\), those split with margin
\(\ge 2\) by \(O_1, \dots, O_{m-1}\) are majority-decided regardless of
\(O_m\); the evenly split (\(\frac{m-1}{2}\)--\(\frac{m-1}{2}\)) pairs
form an undirected graph \(D\), and on them the majority agrees with
\(O_m\). The completed tournament is therefore determined by the
orientation \(R\) that \(O_m\) induces on \(D\) -- an acyclic
orientation, being the restriction of a linear order -- and conversely
every acyclic orientation of \(D\) is induced by some \(O_m\) (extend it
to a topological order). Prescribing \(d\) determines, at each vertex
\(x\), the number \(b(x)\) of \(D\)-edges that \(R\) must orient out of
\(x\). The following algorithm constructs the unique candidate:

\begin{enumerate}
\def\labelenumi{\arabic{enumi}.}
\tightlist
\item
  Set \(b(x) := d(x) - |\{\text{majority-decided arcs leaving } x\}|\)
  for every vertex. If some \(b(x) < 0\), or \(\sum_x b(x) \ne |E(D)|\),
  halt: \textbf{no completion exists}.
\item
  While \(D\) has surviving vertices: if no surviving vertex has
  \(b(x) = 0\), halt: \textbf{no completion exists}. Otherwise pick such
  an \(x\), orient every surviving \(D\)-edge \(\{x, y\}\) from \(y\) to
  \(x\), decrement \(b(y)\) for each such edge (halting likewise if some
  \(b(y)\) drops below 0), and delete \(x\).
\item
  Output the tournament formed by the majority-decided arcs together
  with the constructed orientation \(R\); a witness \(O_m\) is the
  reverse of the deletion order, which is a topological order of \(R\)
  (this is exactly Kahn's (1962) topological sorting algorithm).
\end{enumerate}

Every step is forced: in any acyclic orientation of the surviving graph
there is a sink, and a sink uses none of its out-budget, so a vertex
with \(b(x) = 0\) exists whenever a valid orientation does; conversely a
vertex with \(b(x) = 0\) can orient no surviving edge outward, so all of
its edges are forced inward, and deleting it updates its neighbours'
budgets correctly. Hence if the algorithm succeeds, its output is the
unique orientation realizing \(d\), and if it halts early, no completion
exists. The output is acyclic: every arc points from a later-deleted to
an earlier-deleted vertex. Each vertex is processed once and each edge
oriented once, for \(O(n^2)\) time in total. \(\blacksquare\)

For even \(n\) there are no regular tournaments, but the
\emph{near-regular} ones substitute for them (recall from Section 1 that
exactly \(n/2\) vertices carry each of the two out-degrees). Deleting
any vertex \(v\) from a regular tournament on \(n+1\) vertices leaves a
near-regular tournament on \(n\): the \(\frac n2\) in-neighbours of
\(v\) each lose an out-arc, dropping to out-degree \(\frac n2 - 1\),
while the \(\frac n2\) out-neighbours keep out-degree \(\frac n2\).
Conversely, each near-regular tournament on \(n\) arises from a unique
such regular tournament: re-add \(v\) beating exactly the \(\frac n2\)
vertices of out-degree \(\frac n2 - 1\). Hence this is a bijection and
\[\#\{\text{near-regular tournaments on } n\} \;=\; R^{L}_{n+1}. \tag{$\ast$}\]
Here and below, \(R^{L}_{\nu}\) denotes the number of \emph{labelled}
regular tournaments on \(\nu\) vertices -- the counting argument takes
place in the labelled world -- while \(R_{n}\) without the superscript
always denotes the unlabelled (isomorphism) count of Section 1.

\textbf{Theorem 6.2.} \emph{Some near-regular tournament on 38 vertices
is not 5-inducible; hence \(N(5) \le 38\).}

\emph{Proof.} Dropping one voter from a realization shows every
5-inducible tournament on \([n]\) with a prescribed out-degree sequence
is the unique (Proposition 6.1) completion of some 4-multiset of orders;
distinct such inducible tournaments determine distinct (4-multiset,
out-degree sequence) pairs. Fixing the \emph{precise} out-degree
sequence is exactly what makes the completion unique, so the count is
one per fixed sequence: the near-regular sequences on even \(n\) number
\(\binom{n}{n/2}\) (the defining choice is which half of the vertices
have out-degree \(n/2\)), so
\[\log_2 \#\{\text{5-inducible near-regular on } n\} \;\le\;
\log_2\Bigl[\binom{n}{n/2}\binom{n! + 3}{4}\Bigr] \;<\; 624.7 \quad (n = 38).\]
Equivalently, by vertex-relabelling symmetry each of the
\(\binom{n}{n/2}\) sequences is realized by
\(R^{L}_{n+1}/\binom{n}{n/2}\) tournaments, which must exceed
\(\binom{n! + 3}{4}\) for a witness. By \((\ast)\) their total is
\(R^{L}_{39}\), and Schrijver's (1983) lower bound on Eulerian
orientations at \(K_{39}\) (a regular tournament is an Eulerian
orientation of \(K_{n+1}\)) gives
\[\log_2 \#\{\text{near-regular on } 38\} \;=\; \log_2 R^{L}_{39} \;\ge\;
\log_2\Bigl[2^{-\binom{39}{2}} \binom{38}{19}^{39}\Bigr] \;>\; 625.5 .\]
Since \(625.5 > 624.7\), some near-regular 38-tournament is not
5-inducible. (The plain regular count already gives the weaker
\(N(5) \le 39\):
\(\log_2 R^{L}_{39} > 625.5 > 610.8 > \log_2 \binom{39! + 3}{4}\), so
all but a \(2^{-14.8}\) fraction of regular 39-tournaments are
non-5-inducible.) \(\blacksquare\)

\emph{Remark.} The same proof with \(m = 3\) shows a pair admits at most
one regular 3-voter majority completion, so all but a small fraction of
regular tournaments on \(\ge 17\) vertices are not 3-inducible,
consistent with, and far weaker than, \(N(3) = 8\) (Shepardson and Tovey
2009).

The argument is not specific to five voters: Proposition 6.1 already
holds for every odd \(m\), and the near-regular bijection \((\ast)\) is
\(m\)-free. Counting both classes gives:

\textbf{Theorem 6.3 (all odd \(m\)).} \emph{For odd \(m \ge 3\): (a) at
most \(\binom{n! + m - 2}{m - 1}\) regular tournaments on odd \([n]\)
are \(m\)-inducible, so a non-\(m\)-inducible regular tournament exists
once \(R^{L}_{n} > \binom{n! + m - 2}{m - 1}\); (b) at most
\(\binom{n}{n/2}\binom{n! + m - 2}{m - 1}\) near-regular tournaments on
even \([n]\) are \(m\)-inducible, so a non-\(m\)-inducible near-regular
tournament exists once
\(R^{L}_{n+1} > \binom{n}{n/2}\binom{n! + m - 2}{m - 1}\), using in both
cases Schrijver's bound
\(R^{L}_{\nu} \ge 2^{-\binom{\nu}{2}}\binom{\nu-1}{(\nu-1)/2}^{\nu}\)
(odd \(\nu\)).}

Together with the exact multiset count of profiles, this extends Table 1
of Bachmeier et al.~(2019). Their counting lemma bounds the majoritarian
expressiveness \(n^{\mathcal{T}}(k)\), the largest \(n\) such that every
\(n\)-vertex tournament is \(m\)-inducible, via the estimate
\(\binom{n! + k - 1}{k} \le (2 \cdot n!)^k / k!\). Using the exact
multiset count in place of the Stirling estimate (the ``exact'' row)
already sharpens three entries (the \(m = 5\) case noted in a footnote
of Bachmeier et al.~2019); the regular and near-regular bounds of
Theorem 6.3 then sharpen all ten. The 10 best bounds are equally split
between \emph{regular} tournaments of odd size and \emph{near-regular}
tournaments of even size. The smallest non-\(m\)-inducible tournament
has \(N(m) = n^{\mathcal{T}}(m) + 1\) vertices, so that
\(n^{\mathcal{T}}(5) \le 37\) is exactly the bound \(N(5) \le 38\) of
Theorem 6.2.

\begin{longtable}[]{@{}
  >{\raggedright\arraybackslash}p{(\linewidth - 20\tabcolsep) * \real{0.3243}}
  >{\raggedright\arraybackslash}p{(\linewidth - 20\tabcolsep) * \real{0.0676}}
  >{\raggedright\arraybackslash}p{(\linewidth - 20\tabcolsep) * \real{0.0676}}
  >{\raggedright\arraybackslash}p{(\linewidth - 20\tabcolsep) * \real{0.0676}}
  >{\raggedright\arraybackslash}p{(\linewidth - 20\tabcolsep) * \real{0.0676}}
  >{\raggedright\arraybackslash}p{(\linewidth - 20\tabcolsep) * \real{0.0676}}
  >{\raggedright\arraybackslash}p{(\linewidth - 20\tabcolsep) * \real{0.0676}}
  >{\raggedright\arraybackslash}p{(\linewidth - 20\tabcolsep) * \real{0.0676}}
  >{\raggedright\arraybackslash}p{(\linewidth - 20\tabcolsep) * \real{0.0676}}
  >{\raggedright\arraybackslash}p{(\linewidth - 20\tabcolsep) * \real{0.0676}}
  >{\raggedright\arraybackslash}p{(\linewidth - 20\tabcolsep) * \real{0.0676}}@{}}
\caption{Upper bounds on the majoritarian expressiveness
\(n^{\mathcal{T}}(k)\), the largest \(n\) such that every \(n\)-vertex
tournament is \(m\)-inducible. Columns are indexed by the odd voter
count \(m\); bold marks a strict improvement on Table 1 of Bachmeier et
al.~(2019).}\tabularnewline*
\toprule\noalign{}
\begin{minipage}[b]{\linewidth}\raggedright
\end{minipage} & \begin{minipage}[b]{\linewidth}\raggedright
3
\end{minipage} & \begin{minipage}[b]{\linewidth}\raggedright
5
\end{minipage} & \begin{minipage}[b]{\linewidth}\raggedright
7
\end{minipage} & \begin{minipage}[b]{\linewidth}\raggedright
9
\end{minipage} & \begin{minipage}[b]{\linewidth}\raggedright
11
\end{minipage} & \begin{minipage}[b]{\linewidth}\raggedright
13
\end{minipage} & \begin{minipage}[b]{\linewidth}\raggedright
15
\end{minipage} & \begin{minipage}[b]{\linewidth}\raggedright
17
\end{minipage} & \begin{minipage}[b]{\linewidth}\raggedright
19
\end{minipage} & \begin{minipage}[b]{\linewidth}\raggedright
21
\end{minipage} \\*
\midrule\noalign{}
\endfirsthead
\toprule\noalign{}
\begin{minipage}[b]{\linewidth}\raggedright
\end{minipage} & \begin{minipage}[b]{\linewidth}\raggedright
3
\end{minipage} & \begin{minipage}[b]{\linewidth}\raggedright
5
\end{minipage} & \begin{minipage}[b]{\linewidth}\raggedright
7
\end{minipage} & \begin{minipage}[b]{\linewidth}\raggedright
9
\end{minipage} & \begin{minipage}[b]{\linewidth}\raggedright
11
\end{minipage} & \begin{minipage}[b]{\linewidth}\raggedright
13
\end{minipage} & \begin{minipage}[b]{\linewidth}\raggedright
15
\end{minipage} & \begin{minipage}[b]{\linewidth}\raggedright
17
\end{minipage} & \begin{minipage}[b]{\linewidth}\raggedright
19
\end{minipage} & \begin{minipage}[b]{\linewidth}\raggedright
21
\end{minipage} \\*
\midrule\noalign{}
\endhead
\bottomrule\noalign{}
\endlastfoot
Bachmeier et al.~(2019) & 18 & 41 & 66 & 93 & 122 & 152 & 183 & 216 &
249 & 282 \\*
exact & 18 & \textbf{40} & \textbf{65} & 93 & 122 & 152 & 183 &
\textbf{215} & 249 & 282 \\*
Thm 6.3 & \textbf{16} & \textbf{37} & \textbf{62} & \textbf{89} &
\textbf{117} & \textbf{147} & \textbf{178} & \textbf{210} & \textbf{243}
& \textbf{276} \\*
\end{longtable}

Each threshold is sharp for the class attaining it: the count wins at
the witness size and falls short one vertex lower.

These bounds show that a non-5-inducible tournament exists on 38
vertices, but they do not exhibit one. The next section gives an
explicit example.

\subsection{7. Paley(43) is not the majority of five
voters}\label{paley43-is-not-the-majority-of-five-voters}

Section 6 established the existence of a non-5-inducible tournament on
38 vertices by counting. We now construct an explicit example of
comparable size, which also refutes the strict form of A(5).

We briefly recall the definition of the Paley tournament. Let \(q\) be a
prime with \(q \equiv 3 \pmod 4\). A nonzero element \(a\) of the field
\(\mathrm{GF}(q)\) is a \textbf{quadratic residue} if \(a = b^2\) for
some nonzero \(b\); exactly \((q-1)/2\) of the \(q - 1\) nonzero
elements are quadratic residues, and since \(q \equiv 3 \pmod 4\), the
element \(-1\) is \emph{not} one -- so for every \(a \ne 0\) exactly one
of \(a, -a\) is a quadratic residue. The \textbf{Paley tournament}
\(\mathrm{Paley}(q)\) has vertex set \(\mathrm{GF}(q)\) and an arc
\(u \to v\) whenever \(v - u\) is a quadratic residue; this property of
\(-1\) is exactly what makes every pair of vertices get one arc,
i.e.~what makes this a tournament. For \(q = 43\), the field is
\(\mathbb{Z}_{43}\) and the arithmetic is mod 43; for background on
Paley tournaments see Bachmeier et al.~(2019).

\textbf{Theorem 7.1.} \emph{\(P = \mathrm{Paley}(43)\) satisfies
\(\alpha^{*}(P) = \mathrm{MAS}/C = 543/903
= 181/301 > 3/5\), yet \(P\) is not 5-inducible. Hence
\(\alpha^{*} > \frac{m+1}{2m}\) does not imply \(m\)-inducibility, and
\(N(5) \le 43\) with an explicit witness.}

The identity \(\alpha^{*} = \mathrm{MAS}/C\) holds because \(P\) is
arc-transitive: \(\mathrm{Aut}(P)\) acts transitively on the arcs, so by
Lemma 1.1 the uniform dual is optimal. Throughout, \(q = 43\),
\(C = \binom{q}{2} = 903\), the number of cyclic triangles is
\(N_{\Delta} = (q^3 - q)/24 = 3311\), and
\(|\mathrm{Aut}(P)| = q(q-1)/2 = 903\).

\subsubsection{7.1 MAS and the forced
shell}\label{mas-and-the-forced-shell}

\(\mathrm{MAS}(P) = 543\) (\textbf{MAS-orders} are the orders attaining
it), computed exactly by a subset-DP modulo \(\mathrm{Aut}(P)\) with a
meet-in-the-middle optimization (Appendix E). The slack (Section 1)
\[\mathrm{slack}_5(P) \;=\; 5 \cdot \mathrm{MAS} - 3C \;=\; 6\] is
small: \(\mathrm{slack}_5 = 5C\,(\alpha^{*} - 3/5)\), so it grows in
proportion to \(C\) whenever the predictability stands off from the
\(3/5\) threshold, yet for \(P\) it is only \(6\), the smallest value
anywhere in the Paley family (Table 3, whose other members range from
\(7\) to \(47\)). Equivalently, \(\alpha^{*}(P) = 181/301\) exceeds
\(3/5\) by \(2/1505 \approx 0.0013\). For an inducing profile
\((O_1, \dots, O_5)\), let \(f_i = |\mathrm{fwd}_P(O_i)|\) and relabel
the voters so that \(f_1 \ge f_2 \ge \cdots \ge f_5\). Summing the
majority condition over arcs gives \(\sum_{i=1}^{5} f_i \ge 3C = 2709\);
since \(f_1 \le \mathrm{MAS} =
543\), the other four sum to at least \(2709 - 543 = 2166\), and the
largest of them satisfies \(4 f_2 \ge 2166\), i.e.~\(f_2 \ge 542\).
Since \(f_1 \ge f_2 \ge 542\) and \(f_1 \le \mathrm{MAS} = 543\), the
two largest forward-counts each miss \(\mathrm{MAS}\) by at most one.
Hence \textbf{at least two voters lie at level \(\le 1\)} (the
\textbf{level} of an order \(O\) being
\(\delta(O) = \mathrm{MAS} - |\mathrm{fwd}_P(O)|\)). The slack leaves
essentially no room for disagreement: the five levels sum to at most
\(\mathrm{slack}_5(P) = 6\), since
\(\sum_{i} \delta(O_i) = 5\,\mathrm{MAS} - \sum_i f_i
\le 5 \cdot 543 - 3C = 2715 - 2709 = 6\).

\subsubsection{7.2 Co-backing}\label{co-backing}

\textbf{Lemma 7.2 (co-backing).} \emph{In any inducing profile of 5
voters, every cyclic triangle is double-backed (2 of its 3 arcs are
directed backward) by at most one voter.}

\emph{Proof.} Each arc of the triangle is ranked backward by \(\le 2\)
of the 5 voters, so there are at most six backward incidences; each
voter ranks \(\ge 1\) arc of a cyclic triangle backward (a linear order
cannot agree with all three arcs of a cycle); if \(d\) voters
double-back, incidences \(= 5 + d \le 6\) -- the \(5\) counts the one
forced backward arc from each of the five voters, the \(d\) the extra
backward arc contributed by each double-backing voter -- so \(d \le 1\).
\(\blacksquare\)

Hence the five voters' \textbf{double-back sets}
\(\mathrm{DB}(O) = \{\)cyclic triangles with exactly 1 forward arc in
\(O\}\) are pairwise disjoint; in particular the two level-\(\le 1\)
voters of Section 7.1 form a DB-disjoint pair of level-\(\le 1\) orders.
\textbf{If no such pair exists, \(P\) is not 5-inducible} -- a finite
criterion.

\subsubsection{7.3 The screen}\label{the-screen}

By Section 7.1--7.2 an inducing profile would contain a pair of
level-\(\le 1\) orders with disjoint double-back sets; the screen
decides this finite criterion by direct enumeration. The level-\(\le 1\)
shell has about 1.8 million \(\mathrm{Aut}\)-orbits, and over 1.6
billion orders (see Appendix F). Two optimizations make the exhaustive
search practical:

\begin{itemize}
\tightlist
\item
  \textbf{Razor.} We fix the window \(\mathcal{W} = [23]\) and restrict
  attention to the 538 cyclic triangles inside \(\mathcal{W}\). The
  restriction of \(\mathrm{DB}(O)\) to these triangles depends only on
  how \(O\) ranks \(\mathcal{W}\), so the shell collapses to a few
  million distinct restrictions; DB-disjointness implies disjointness of
  the restrictions, so only the pairs with disjoint restrictions (about
  five million; Appendix F) need the exact check.
\item
  \textbf{One-side automorphism reduction.}
  \(\mathrm{DB}(\sigma O) = \sigma(\mathrm{DB}(O))\) for
  \(\sigma \in \mathrm{Aut}(P)\) (proved using only the fact that
  \(\sigma\) permutes vertices and preserves arcs in Appendix F), so
  overlap sizes are \(\mathrm{Aut}\)-invariant and it suffices to test
  canonical orbit representatives against the full pool; a factor of 903
  reduction on one side.
\end{itemize}

The exact pass examined all of the roughly 4.4 billion (representative,
pool-order) candidate pairs: \textbf{none is DB-disjoint} (minimum
overlap among candidates: 68 triangles). The verification checks -- an
exhaustive level-0 run over all \(\approx 1.6 \times 10^{14}\) MAS-order
pairs, a positive-detection coverage test, and an exhaustive cross-check
of the equivariance over all 903 automorphisms -- are itemized with the
complete search statistics in Appendix F.

\subsubsection{7.4 Conclusion of the
proof}\label{conclusion-of-the-proof}

An inducing profile of 5 voters would force (Section 7.1 + Section 7.2)
a DB-disjoint pair of level-\(\le 1\) orders; the screen (Section 7.3)
shows none exists. \(\blacksquare\)

\begin{longtable}[]{@{}
  >{\raggedright\arraybackslash}p{(\linewidth - 12\tabcolsep) * \real{0.0678}}
  >{\raggedright\arraybackslash}p{(\linewidth - 12\tabcolsep) * \real{0.0847}}
  >{\raggedright\arraybackslash}p{(\linewidth - 12\tabcolsep) * \real{0.1017}}
  >{\raggedright\arraybackslash}p{(\linewidth - 12\tabcolsep) * \real{0.1356}}
  >{\raggedright\arraybackslash}p{(\linewidth - 12\tabcolsep) * \real{0.1356}}
  >{\raggedright\arraybackslash}p{(\linewidth - 12\tabcolsep) * \real{0.1186}}
  >{\raggedright\arraybackslash}p{(\linewidth - 12\tabcolsep) * \real{0.3559}}@{}}
\caption{The Paley family: size, MAS, predictability, slack, status;
``companion'' marks a result of the companion paper (Chindelevitch and
Harutyunyan 2026).}\tabularnewline*
\toprule\noalign{}
\begin{minipage}[b]{\linewidth}\raggedright
\(q\)
\end{minipage} & \begin{minipage}[b]{\linewidth}\raggedright
\(C\)
\end{minipage} & \begin{minipage}[b]{\linewidth}\raggedright
\(\mathrm{MAS}\)
\end{minipage} & \begin{minipage}[b]{\linewidth}\raggedright
\(\alpha^{*}\)
\end{minipage} & \begin{minipage}[b]{\linewidth}\raggedright
\(\approx\)
\end{minipage} & \begin{minipage}[b]{\linewidth}\raggedright
\(\mathrm{slack}_5\)
\end{minipage} & \begin{minipage}[b]{\linewidth}\raggedright
5-inducible?
\end{minipage} \\*
\midrule\noalign{}
\endfirsthead
\toprule\noalign{}
\begin{minipage}[b]{\linewidth}\raggedright
\(q\)
\end{minipage} & \begin{minipage}[b]{\linewidth}\raggedright
\(C\)
\end{minipage} & \begin{minipage}[b]{\linewidth}\raggedright
\(\mathrm{MAS}\)
\end{minipage} & \begin{minipage}[b]{\linewidth}\raggedright
\(\alpha^{*}\)
\end{minipage} & \begin{minipage}[b]{\linewidth}\raggedright
\(\approx\)
\end{minipage} & \begin{minipage}[b]{\linewidth}\raggedright
\(\mathrm{slack}_5\)
\end{minipage} & \begin{minipage}[b]{\linewidth}\raggedright
5-inducible?
\end{minipage} \\*
\midrule\noalign{}
\endhead
\bottomrule\noalign{}
\endlastfoot
7 & 21 & 14 & \(2/3\) & 0.6667 & 7 & yes, McG 3 \\*
11 & 55 & 35 & \(7/11\) & 0.6364 & 10 & yes, McG 5 \\*
19 & 171 & 107 & \(107/171\) & 0.6257 & 22 & yes, McG 5 \\*
23 & 253 & 161 & \(7/11\) & 0.6364 & 46 & no (companion) \\*
27 & 351 & 216 & \(8/13\) & 0.6154 & 27 & no (companion) \\*
31 & 465 & 285 & \(19/31\) & 0.6129 & 30 & no (companion) \\*
43 & 903 & 543 & \(181/301\) & 0.6013 & \textbf{6} & \textbf{no (Thm
7.1)} \\*
47 & 1081 & 658 & \(14/23\) & 0.6087 & 47 & open \\*
\end{longtable}

\textbf{Remark (why Paley is a convenient family).} The necessity (P1)
plus arc-transitivity in combination with Lemma 1.1 makes
\(\mathrm{MAS}/C\) the exact predictability, and Paley(43) exceeds the
\(3/5\) threshold by only \(2/1505\) -- the slack of 6 that forces two
voters into the level-\(\le 1\) shell. Table 3 gives the full picture
across the family. MAS values are exact for \(q \le 47\) (Appendix E).
Two lessons are visible. First, \(\alpha^{*}\) is \emph{not monotone} in
\(q\) (it coincides at \(q = 11\) and \(q = 23\), and the \(q = 47\)
value exceeds the \(q = 43\) value), so no Paley tournament is certified
below the \(3/5\) threshold, and extrapolation is unreliable. Second,
the slack \(\mathrm{slack}_5\) -- which controls the depth of the shell
that the Section 7 argument must screen -- is \emph{exceptionally} small
at \(q = 43\) and at no other known member: the proof of Theorem 7.1
exploits a genuine numerical coincidence. Paley(11) and Paley(19) are
5-inducible with explicit certificates, so non-5-inducibility is not
generic to the family. The slack argument of Section 7 does not reach
Paley(23), Paley(27) or Paley(31); the companion paper (Chindelevitch
and Harutyunyan 2026) decides all three by a different method, and
Paley(47) remains open.

\textbf{Remark (why five voters is the current frontier).} Theorem 7.1
refutes the sufficiency of the predictability threshold at \(m = 5\)
\emph{strictly}; the analogous question for every odd \(m \ge 7\) --
does \(\alpha^{*} > \frac{m+1}{2m}\) imply \(m\)-inducibility? --
remains open, for the following reason. Since \(m\)-inducibility is
upward-closed in \(m\) (adding to a profile any order together with its
reverse does not change any margin, so \(m\)-inducible \(\Rightarrow\)
\((m+2)\)-inducible), every candidate witness for \(m = 7\) lies among
the non-5-inducible tournaments -- objects of which, prior to this work,
no explicit example of reasonable size was known at all, and of which,
before the companion paper, \(\mathrm{Paley}(43)\) was the only explicit
specimen below \(\sim 6 \times 10^{8}\) vertices. \(\mathrm{Paley}(43)\)
does satisfy the \(m = 7\) threshold (\(181/301 > 4/7\)), so it is a
perfectly good \emph{candidate} for the \(m = 7\) refutation; what is
missing is the certification, and both the arithmetic and the structural
pillars of the proof of Theorem 7.1 collapse at once when \(m = 7\).
Refuting the threshold at \(m = 7\) therefore appears to require either
a genuinely new certification method or a new explicit family; Theorem
6.3 guarantees non-7-inducible tournaments exist on \(63\) vertices, but
non-constructively.

Explicit witnesses were previously astronomical, and rest on domination:
\emph{every} majority tournament of 5 voters has a dominating set of
size at most 12 -- a theorem of Fidler (2011), valid for tournaments of
every order and sharpening, at 5 voters, the \(O(m \log m)\) bound of
Alon et al.~(see Bachmeier et al.~2019) -- so any tournament with no
dominating set of size 12 is a witness. The smallest known such
tournament, a quadratic residue tournament obtained via the
Graham--Spencer bound, has \(\sim 6 \times 10^{8}\) vertices (Bachmeier
et al.~2019). Paley(43) is thus the first explicit non-5-inducible
tournament whose order approaches the counting bound:
\[12 \;\le\; N(5) \;\le\; 38 \ \text{(non-constructive; Theorem 6.2)}, \qquad N(5) \le 43 \ \text{(explicit)}.\]

The companion paper (Chindelevitch and Harutyunyan 2026) improves both
ends to \(13 \le N(5) \le 22\) with explicit 22-vertex tournaments,
using a dedicated search instead of the counting argument of this
section. The final section collects what the refutations leave open.

\subsection{8. Discussion and open
problems}\label{discussion-and-open-problems}

In this section we summarize what the paper has refuted and established,
state one conjecture, and list the open problems.

\textbf{What is refuted.} Three conjectures have been refuted in this
paper: Milosz et al.'s Conjectures 1 and 2 (Sections 3 and 4), and the
threshold conjecture A(\(m\)) at both \(m = 3\) (boundary; Section 5)
and \(m = 5\) (strict; Section 7).

On the positive side, our work has established that
\(\mathrm{FAS} = \mathrm{HS}_3\) up to \(n = 10\) (Theorem 4.1), that
A(3) restricted to \(\alpha^{*} > 2/3\) holds (vacuously), and that
empirically, violations are \emph{rare} among tournaments. Namely, there
is a unique regular witness at \(n = 11\) in both the FAS/HS\(_3\) and
the A(3) questions, and only a handful of others among the large number
of candidates that we examined. Additionally, we have exhibited the
first explicit tournament of moderate size that is provably not
inducible by 5 voters.

\subsubsection{8.1 Conjecture}\label{conjecture}

\textbf{Conjecture 8.1 (strictness of the margin hierarchy).} \emph{For
every odd \(m\) and every odd \(t\) with \(1 \le t \le m - 2\):
\(\mathcal{I}_{m,t} \subsetneq \mathcal{I}_{m,t+2}\).}

Section 1 defined the classes \(\mathcal{I}_{m,t}\) (tournaments
\(m\)-inducible with every arc of margin \(\le t\)), nested
\(\mathcal{I}_{m,1} \subseteq \mathcal{I}_{m,3} \subseteq \cdots
\subseteq \mathcal{I}_{m,m}\). We establish strictness only at
\(m = 3\), namely \(\mathcal{I}_{3,1} \subsetneq \mathcal{I}_{3,3}\) --
the 254 tournaments of Section 5, which are 3-inducible but only via a
unanimous (margin-\(3\)) arc. The first open case is \(m = 5\), where
the conjecture asserts two strict inclusions concretely:

\begin{itemize}
\tightlist
\item
  \(\mathcal{I}_{5,3} \subsetneq \mathcal{I}_{5,5}\): some tournament is
  5-inducible but \emph{only} with a \textbf{unanimous} arc -- every
  5-voter inducing profile has some arc split \(5\):\(0\) (margin
  \(5\));
\item
  \(\mathcal{I}_{5,1} \subsetneq \mathcal{I}_{5,3}\): some tournament is
  5-inducible with \emph{no} unanimous arc (only \(4\):\(1\) and
  \(3\):\(2\) splits) but \emph{not} with all margins exactly \(1\) --
  every such profile needs some arc split \(4\):\(1\) (margin \(3\)).
\end{itemize}

The companion paper (Chindelevitch and Harutyunyan 2026) establishes the
second: Paley(19) is 5-inducible with every margin at most \(3\), but
not with unit margin.

\subsubsection{8.2 Open problems}\label{open-problems}

\begin{enumerate}
\def\labelenumi{\arabic{enumi}.}
\tightlist
\item
  \textbf{The exact value of \(N(5)\).} This paper leaves the gap
  \(12 \le N(5) \le 38\). The companion paper (Chindelevitch and
  Harutyunyan 2026) narrows it to \(13 \le N(5) \le 22\) and shows that
  Paley(23), Paley(27) and Paley(31) are not 5-inducible; the exact
  value remains open. Paley(23) had resisted about six weeks of SAT
  computation in Bachmeier et al.~(2019). Before that search, we
  verified that all seven of its order-20 subtournaments (the deletion
  classes of triples: \((q-2)/3\) classes for \(q \equiv 2 \bmod 3\) by
  a Burnside count; see Burnside 1911) are 5-inducible.
\item
  \textbf{Vertex-deleted Paley(43).} The companion paper also shows that
  \(P - v\) (order 42) is not 5-inducible, so Paley(43) is not
  vertex-critical. Whether \(P - 2v\) (order 41, unique up to
  isomorphism since \(\mathrm{Aut}(P)\) acts transitively on unordered
  pairs of vertices) is 5-inducible is open. The screen of Section 7.3
  extends to it: the deleted set can be chosen \emph{a posteriori} (a
  pair of orders is dangerous if and only if its DB-overlap has a
  \(\le k\)-vertex transversal -- an \(\mathrm{Aut}\)-invariant
  property, so the factor-903 reduction survives deletion), and the
  budget region sharpens to
  \(\delta_1 + 4 \delta_2 \le 5\,\mathrm{MAS} - 3C\), where
  \(\delta_1, \delta_2\) are the levels of the two best-agreeing voters
  (yielding 45 for \(P - 2v\)).
\item
  \textbf{Predictability vs.~McGarvey number.} All our A(3)
  counterexamples have McGarvey number \(5\); is the McGarvey number
  bounded above by 5 on \(\{\alpha^{*} = 2/3\}\)? The strike-a-voter
  argument (any \(2/3\)-tight \(6\)-order certificate yields a
  majority-5 induction) explains the observed bound except at the single
  exceptional tournament cA6 (Appendix D), which needs 9 voters at
  threshold \(2/3\) yet still 5 for a simple majority. We expect this to
  hold in general, but advance it only tentatively.
\end{enumerate}

\subsection{9. Reproducibility}\label{reproducibility}

The code and data behind every computational result in this paper
(Sections 3--8 and the appendices) are at
\url{https://github.com/Leonardini/Tournaments}, organized by
section, with a manifest mapping each claim to its verifier. The
Paley(43) result of Section 7 is reproduced end to end by a single
script. Every census count can be regenerated, every reported optimum is
certified by an integer or linear program and checked in exact rational
arithmetic, and every inducing profile and certificate cited in the text
and the appendices is included.

\subsection{Declaration of competing
interest}\label{declaration-of-competing-interest}

The authors declare that they have no known competing financial
interests or personal relationships that could have appeared to
influence the work reported in this paper.

\subsection{Funding}\label{funding}

LC acknowledges funding from the MRC Centre for Global Infectious
Disease Analysis (reference MR/X020258/1), funded by the UK Medical
Research Council (MRC). This UK funded award is carried out in the frame
of the Global Health EDCTP3 Joint Undertaking.

\subsection{CRediT authorship contribution
statement}\label{credit-authorship-contribution-statement}

\textbf{Leonid Chindelevitch:} Conceptualization, Methodology, Software,
Formal analysis, Investigation, Writing -- original draft, Writing --
review and editing. \textbf{Ararat Harutyunyan:} Conceptualization,
Formal analysis, Writing -- review and editing.

\subsection{Data availability}\label{data-availability}

The code and data supporting the findings of this study, together with a
manifest mapping every claim to its verifier, are maintained at
\url{https://github.com/Leonardini/Tournaments} and archived at
Zenodo, DOI:
\href{https://doi.org/10.5281/zenodo.22040443}{10.5281/zenodo.22040443},
as a tagged snapshot corresponding to this manuscript, described in
Section 9.

\subsection{Declaration of generative AI and AI-assisted technologies in
the manuscript preparation
process}\label{declaration-of-generative-ai-and-ai-assisted-technologies-in-the-manuscript-preparation-process}

During the preparation of this work the authors used Claude (Anthropic)
for exploratory reasoning, implementation assistance, and drafting and
editing. After using this tool, the authors reviewed and edited the
content as needed and take full responsibility for the content of the
published article.

\subsection{Acknowledgements}\label{acknowledgements}

This work was granted access to the HPC resources of IDRIS under the
allocation 2025-AD010616623R1 made by the Grand Équipement National de
Calcul Intensif (GENCI).

\subsection{References}\label{references}

Alon N (2002) Voting paradoxes and digraphs realizations. \emph{Advances
in Applied Mathematics} 29(1):126--135.
\url{doi:10.1016/S0196-8858(02)00007-6}

Antonov M, Csárdi G, Horvát S, Müller K, Nepusz T, Noom D, Salmon M,
Traag V, Welles BF, Zanini F (2023) igraph enables fast and robust
network analysis across programming languages. arXiv:2311.10260

Bachmeier G, Brandt F, Geist C, Harrenstein P, Kardel K, Peters D,
Seedig HG (2019) \(k\)-Majority digraphs and the hardness of voting with
a constant number of voters. \emph{Journal of Computer and System
Sciences} 105:130--157. \url{doi:10.1016/j.jcss.2019.04.005}

Blin G, Crochemore M, Hamel S, Vialette S (2011) Medians of an odd
number of permutations. \emph{Pure Mathematics and Applications}
21(2):161--175

Burnside W (1911) \emph{Theory of Groups of Finite Order}, 2nd edn.
Cambridge University Press

Charbit P, Thomassé S, Yeo A (2007) The minimum feedback arc set problem
is NP-hard for tournaments. \emph{Combinatorics, Probability and
Computing} 16(1):1--4. \url{doi:10.1017/S0963548306007887}

Chindelevitch L, Harutyunyan A (2026) Tournaments not inducible by five
voters. Submitted to \emph{Mathematical Social Sciences} jointly with
this paper

Chindelevitch L, Zanetti JPP, Meidanis J (2018) On the rank-distance
median of 3 permutations. \emph{BMC Bioinformatics} 19(Suppl 6):142.
\url{doi:10.1186/s12859-018-2131-4}

Condorcet, Marquis de (1785) \emph{Essai sur l'application de l'analyse
à la probabilité des décisions rendues à la pluralité des voix}.
Imprimerie Royale, Paris

de Moraes V, Meidanis J (2026) Time complexity and relaxation gap for
the rank median of three genomes. In: \emph{Comparative Genomics}, LNBI
16569, Springer, pp 3--28. \url{doi:10.1007/978-3-032-26891-4_1}

Dwork C, Kumar R, Naor M, Sivakumar D (2001) Rank aggregation methods
for the Web. In: \emph{Proceedings of the 10th International Conference
on World Wide Web (WWW10)}, ACM, pp 613--622. \url{doi:10.1145/371920.372165}

Eggermont C, Hurkens C, Woeginger GJ (2013) Realizing small tournaments
through few permutations. \emph{Acta Cybernetica} 21(2):267--271.
\url{doi:10.14232/actacyb.21.2.2013.4}

Erdős P, Moser L (1964) On the representation of directed graphs as
unions of orderings. \emph{Publications of the Mathematical Institute of
the Hungarian Academy of Sciences} 9(1--2):125--132

Fidler D (2011) A recurrence for bounds on dominating sets in
\(k\)-majority tournaments. \emph{The Electronic Journal of
Combinatorics} 18(1):\#P166. \url{doi:10.37236/653}

Gilboa I (1990) A necessary but insufficient condition for the
stochastic binary choice problem. \emph{Journal of Mathematical
Psychology} 34(4):371--392. \url{doi:10.1016/0022-2496(90)90019-6}

Grötschel M, Jünger M, Reinelt G (1985) Facets of the linear ordering
polytope. \emph{Mathematical Programming} 33(1):43--60.
\url{doi:10.1007/BF01582010}

Held M, Karp RM (1962) A dynamic programming approach to sequencing
problems. \emph{Journal of the Society for Industrial and Applied
Mathematics} 10(1):196--210. \url{doi:10.1137/0110015}

Hemaspaandra E, Spakowski H, Vogel J (2005) The complexity of Kemeny
elections. \emph{Theoretical Computer Science} 349(3):382--391.
\url{doi:10.1016/j.tcs.2005.08.031}

Kahn AB (1962) Topological sorting of large networks.
\emph{Communications of the ACM} 5(11):558--562.
\url{doi:10.1145/368996.369025}

Kemeny JG (1959) Mathematics without numbers. \emph{Daedalus}
88(4):577--591

Kenyon-Mathieu C, Schudy W (2007) How to rank with few errors: a PTAS
for weighted feedback arc set on tournaments. In: \emph{Proceedings of
the 39th Annual ACM Symposium on Theory of Computing (STOC)}, ACM, pp
95--103. \url{doi:10.1145/1250790.1250806}

Madarasi P (2026) The complexity of Kemeny aggregation with three
rankings. arXiv:2607.28588

Mala J (1999) On \(\lambda\)-majority voting paradoxes.
\emph{Mathematical Social Sciences} 37(1):39--44.
\url{doi:10.1016/S0165-4896(98)00012-2}

McGarvey DC (1953) A theorem on the construction of voting paradoxes.
\emph{Econometrica} 21(4):608--610. \url{doi:10.2307/1907926}

McKay BD (2026) Combinatorial data: digraphs (tournament catalogues).
Accessed 10 August 2026.
\url{https://users.cecs.anu.edu.au/~bdm/data/digraphs.html}

Milosz R, Hamel S, Pierrot A (2018) Median of 3 permutations, 3-cycles
and 3-hitting set problem. In: \emph{Proceedings of IWOCA 2018}, LNCS
10979, Springer, pp 224--236. \url{doi:10.1007/978-3-319-94667-2_19}

Moon JW (1968) \emph{Topics on Tournaments}. Holt, Rinehart and Winston

Peters D (2026) Kemeny rank aggregation is NP-hard for three voters.
arXiv:2607.25540

Schrijver A (1983) Bounds on the number of Eulerian orientations.
\emph{Combinatorica} 3(3--4):375--380. \url{doi:10.1007/BF02579193}

Shepardson D, Tovey CA (2009) Smallest tournaments not realizable by
\(2/3\)-majority voting. \emph{Social Choice and Welfare}
33(3):495--503. \url{doi:10.1007/s00355-009-0375-7}

Stearns R (1959) The voting problem. \emph{American Mathematical
Monthly} 66(9):761--763. \url{doi:10.1080/00029890.1959.11989405}

Wolsey LA (2020) \emph{Integer Programming}, 2nd edn. Wiley.
\url{doi:10.1002/9781119606475}

Young HP, Levenglick A (1978) A consistent extension of Condorcet's
election principle. \emph{SIAM Journal on Applied Mathematics}
35(2):285--300. \url{doi:10.1137/0135023}

\FloatBarrier

\setcounter{table}{0}\renewcommand{\thetable}{A.\arabic{table}}\setcounter{figure}{0}\renewcommand{\thefigure}{A.\arabic{figure}}

\subsection{Appendix A. Proof of Lemma
2.2}\label{appendix-a.-proof-of-lemma-2.2}

\textbf{Lemma 2.2.} \emph{Let \(T\) be a 3-voter majority tournament. No
minimum-weight feedback arc set of \(T\) contains a unanimous (weight-3)
arc.}

\emph{Proof.} Let \((i \to j)\) be a unanimous arc (weight 3) and
suppose for contradiction that some minimum-weight feedback arc set
\(F\) contains it. Let \(\pi\) be a topological order of \(T - F\), so
that every arc of \(F\) is decreasing with respect to \(\pi\); in
particular \(\pi(j) < \pi(i)\). Let \(\pi'\) be obtained from \(\pi\) by
swapping the positions of \(i\) and \(j\). For each voter \(\tau_r\)
define \(\mathrm{cost}_{\tau_r}(\sigma)\) as the number of arcs that are
backward with respect to \(\sigma\) and supported by \(\tau_r\); the
total weight of the feedback arc set induced by an order \(\sigma\) is
\[\mathrm{cost}(\sigma) \;=\; \sum_{r=1}^{3} \mathrm{cost}_{\tau_r}(\sigma).\]
Let us fix \(\tau_r\). Since \(i \prec_{\tau_r} j\) and
\(\pi(j) < \pi(i)\), the arc \((i, j)\) contributes 1 to
\(\mathrm{cost}_{\tau_r}(\pi)\) and 0 to
\(\mathrm{cost}_{\tau_r}(\pi')\), a change of \(-1\). The only other
arcs whose backward status can change under the swap are those joining
\(i\) or \(j\) to a vertex \(k\) lying strictly between \(j\) and \(i\)
in \(\pi\), and for each such \(k\) only the pairs \((i, k)\) and
\((k, j)\) are affected. If a vertex \(k\) existed for which both
affected arcs were backward with respect to \(\tau_r\), then
\(j \prec_{\tau_r} k \prec_{\tau_r} i\), contradicting
\(i \prec_{\tau_r} j\) by transitivity. Hence at most one of the two
arcs at each such \(k\) can be backward in \(\tau_r\), so swapping \(i\)
and \(j\) cannot increase the number of backward arcs supported by
\(\tau_r\): the total contribution of each such \(k\) to the change is
non-positive. All other arcs are unaffected by the swap, and therefore
\[\mathrm{cost}_{\tau_r}(\pi') \;<\; \mathrm{cost}_{\tau_r}(\pi).\]
Summing over \(r\) gives \(\mathrm{cost}(\pi') < \mathrm{cost}(\pi)\),
contradicting the minimality of \(F\). \(\blacksquare\)

\setcounter{table}{0}\renewcommand{\thetable}{B.\arabic{table}}\setcounter{figure}{0}\renewcommand{\thefigure}{B.\arabic{figure}}

\subsection{Appendix B. Integer linear programming
formulations}\label{appendix-b.-integer-linear-programming-formulations}

Every optimum reported in this paper is certified by an integer linear
program (ILP) or a linear program (LP): unlike a heuristic,
branch-and-cut (Wolsey 2020) returns \emph{certified} optimality or
infeasibility, not merely a good solution. We use IBM CPLEX 22.1
throughout, with OR-Tools CP-SAT as an independent second solver where
stated (a different engine \emph{and} a different encoding), and every
optimal value reported in this paper is verified in exact rational
arithmetic (we do not repeat this qualifier below). The formulations:

\textbf{B.1 (\(m\)-inducibility).} \emph{Decides whether \(T\) is
induced by \(m\) voters.} Binary variables \(b^p_{uv}\) for each voter
\(p \in [m]\) and ordered pair \((u, v)\), meaning ``\(u\) precedes
\(v\) in voter \(p\)'' -- an ILP in \(O(m n^2)\) binary variables and
\(O(m n^3)\) constraints:
\[(1)\ \ b^p_{uv} + b^p_{vu} = 1, \qquad (2)\ \ b^p_{uv} + b^p_{vw} + b^p_{wu} \le 2 \ \ \forall\ u,v,w,
\qquad (3)\ \ \sum_{p=1}^{m} b^p_{uv} \ge \tfrac{m+1}{2} \ \ \forall\ (u \to v) \in E(T).\]
Families (1)--(2) cut out exactly the linear ordering polytope's integer
points (Grötschel et al.~1985) (the encoding of Eggermont et al.~2013)
-- so each voter's variables encode a genuine linear order -- while (3)
enforces the majority; feasibility is thus equivalent to
\(m\)-inducibility. Used in Sections 3--4 (minimality of the
counterexamples, the \(n \le 10\) census), Section 5 (the \(n = 10\) and
self-converse censuses, with the CP-SAT position-variable encoding as
cross-check), and the \(n \le 10\) layers of the exhaustive
5-inducibility census of Section 6 (the \(n = 11\) layer needed no ILP;
see Section 6).

\textbf{B.2 (minimum FAS).} \emph{Certifies the minimum feedback arc set
of \(T\) -- equivalently \(\mathrm{MAS}(T)\).} One voter's order
variables \(b_{uv}\) as in B.1 (the case \(m = 1\)); minimize
\(\sum_{(u \to v) \in E(T)} w_{uv} (1 - b_{uv})\) with \(w \equiv 1\)
(unweighted) or \(w\) = margins (weighted). Used in Section 3 and
Section 4.

\textbf{B.3 (minimum 3-cycle hitting set).} \emph{Certifies the fewest
arcs meeting every directed 3-cycle.} Binary \(x_e\) per arc;
\(\sum_{e \in t} x_e \ge 1\) for every directed 3-cycle \(t\); minimize
\(\sum_e x_e\). Used in Section 4 (with B.2, the two ILPs of Theorem
4.1).

\textbf{B.4 (minimum set cover).} \emph{Certifies the fewest obstacle
classes whose containment covers a given family of tournaments.} Given
the containment matrix \(M\) (\(M_{ij} = 1\) if and only if tournament
\(i\) contains obstacle \(j\)), binary \(y_j\);
\(\sum_j M_{ij} y_j \ge 1\) for all \(i\); minimize \(\sum_j y_j\).
Produces the 26-class set cover of the non-3-inducible tournaments with
\(n = 9\) (Appendix G.1) and the 25-class cover of the tournaments with
\(n = 9\) that are 3-inducible, but not with unit margin (Appendix G.2).

\textbf{B.5 (supermajority certificates).} (i) \(m\)-voter
\(2/3\)-supermajority feasibility: this extends B.1, raising family
(3)'s right-hand side to \(\tfrac{2m}{3}\) (\(3 \mid m\)) -- deciding
the 6-vs-9-voter dichotomy of Section 5 and Appendix D directly, with no
order pool. (ii) Minimum-cardinality balanced certificate over a pool
\(\mathcal{P}\) of orders: integer weights \(w_O \ge 0\),
\(\sum_O w_O = m\),
\(\sum_{O:\, e \in \mathrm{fwd}_T(O)} w_O = \tfrac{2m}{3}\) for every
arc \(e\); minimize \(m\) (or the support size). Produces the 6-order
certificate of Counterexample 5.2 (with column generation + \(C_{11}\)
enrichment supplying the pool \(\mathcal{P}\)).

\textbf{B.6 (the predictability LP).} \emph{Computes \(\alpha^{*}(T)\),
the predictability value of Section 5.} The LP of Section 5 is not
integer; it is solved by column generation (cutting planes in the dual)
(Wolsey 2020), with the exact weighted-MAS DP from Appendix E.1 as the
separation routine. The Paley-family computations additionally exploit
automorphisms (averaging over the orbits of arcs under
\(\mathrm{Aut}\)), cf.~Section 7.

\textbf{B.7 (enumerating all minimal obstacle supports).}
\emph{Enumerates every inclusion-minimal obstacle carried by a
non-3-inducible tournament.} The inclusion-minimal supports of the
inducibility LP's optimal duals are exactly the obstacles that certify
non-inducibility; enumerating \emph{all} of them, not just the first one
a solver happens to return, makes the obstacle catalogue of Appendix G
complete. For a tournament with dual-LP value \(\alpha^{*}\): variables
\(y_e\) (non-negative, or free in the exact-coverage variant), binaries
\(z_e\); \(\sum_e y_e = 1\); support linking
\(\varepsilon z_e \le |y_e| \le z_e\); the dual-optimality rows
\(\sum_{e \in \mathrm{fwd}_T(\pi)} y_e \le \alpha^{*}\) generated lazily
(separation by the exact weighted-MAS DP from Appendix E.1); minimize
\(\sum_e z_e\). After each optimal support \(S\) is found, we add a
no-good cut \(\sum_{e \in S} z_e \le |S| - 1\), which excludes \(S\) and
all its supersets. The solutions therefore form an antichain, and are
exactly the inclusion-minimal supports. We iterate until infeasibility
is reached. The forced-arc test of Theorem G.2 is a variant of
formulation B.1 with the margin of one arc pinned (to 3, or to 1 for
avoidability), and is decided by CP-SAT.

\setcounter{table}{0}\renewcommand{\thetable}{C.\arabic{table}}\setcounter{figure}{0}\renewcommand{\thefigure}{C.\arabic{figure}}

\subsection{\texorpdfstring{Appendix C. The six self-converse violators
of Conjecture 2
(\(n = 11\))}{Appendix C. The six self-converse violators of Conjecture 2 (n = 11)}}\label{appendix-c.-the-six-self-converse-violators-of-conjecture-2-n-11}

The following 3-voter profiles induce the six self-converse violators of
Conjecture 2 (the tournaments of Figure 4). Each row is one tournament;
the three columns \(\rho, \sigma, \tau\) are its voters' orders, each
written as the sequence of the eleven vertices from first to last.

\begin{longtable}[]{@{}
  >{\raggedright\arraybackslash}p{(\linewidth - 6\tabcolsep) * \real{0.2500}}
  >{\raggedright\arraybackslash}p{(\linewidth - 6\tabcolsep) * \real{0.2500}}
  >{\raggedright\arraybackslash}p{(\linewidth - 6\tabcolsep) * \real{0.2500}}
  >{\raggedright\arraybackslash}p{(\linewidth - 6\tabcolsep) * \real{0.2500}}@{}}
\caption{Inducing profiles for the six self-converse violators of
Conjecture 2 (one row per tournament; the tournaments of Figure
4).}\tabularnewline*
\toprule\noalign{}
\begin{minipage}[b]{\linewidth}\raggedright
\end{minipage} & \begin{minipage}[b]{\linewidth}\raggedright
\(\rho\)
\end{minipage} & \begin{minipage}[b]{\linewidth}\raggedright
\(\sigma\)
\end{minipage} & \begin{minipage}[b]{\linewidth}\raggedright
\(\tau\)
\end{minipage} \\*
\midrule\noalign{}
\endfirsthead
\toprule\noalign{}
\begin{minipage}[b]{\linewidth}\raggedright
\end{minipage} & \begin{minipage}[b]{\linewidth}\raggedright
\(\rho\)
\end{minipage} & \begin{minipage}[b]{\linewidth}\raggedright
\(\sigma\)
\end{minipage} & \begin{minipage}[b]{\linewidth}\raggedright
\(\tau\)
\end{minipage} \\*
\midrule\noalign{}
\endhead
\bottomrule\noalign{}
\endlastfoot
A & (1,11,10,2,3,4,5,6,7,8,9) & (4,8,9,2,6,7,1,10,11,3,5) &
(5,3,7,9,6,8,10,11,1,2,4) \\*
B & (1,10,3,4,11,6,8,2,5,7,9) & (5,6,7,8,1,9,10,2,3,4,11) &
(9,2,11,4,3,7,5,8,6,10,1) \\*
C & (1,9,2,10,3,4,5,11,7,8,6) & (3,6,7,8,2,9,10,5,11,1,4) &
(4,11,5,6,8,7,1,10,2,9,3) \\*
D & (1,2,11,10,5,3,4,6,9,7,8) & (3,6,8,9,7,11,1,10,2,4,5) &
(4,5,7,8,9,2,10,6,1,11,3) \\*
E & (1,9,11,10,3,4,2,5,8,6,7) & (2,3,5,6,7,8,9,1,10,11,4) &
(4,7,8,6,10,11,5,1,9,2,3) \\*
F & (2,3,4,5,8,6,7,9,10,11,1) & (6,11,1,5,9,10,3,4,2,7,8) &
(7,8,10,1,9,11,4,2,3,5,6) \\*
\end{longtable}

\setcounter{table}{0}\renewcommand{\thetable}{D.\arabic{table}}\setcounter{figure}{0}\renewcommand{\thefigure}{D.\arabic{figure}}

\subsection{Appendix D. A(3) counterexample
data}\label{appendix-d.-a3-counterexample-data}

This appendix records the explicit certificate data behind the A(3)
counterexamples of Section 5: cA3's minimum \(2/3\)-certificate and its
symmetry, the self-converse census at \(n = 11\), and the exceptional
9-voter tournament cA6.

cA3 is the circulant on \(\mathbb{Z}_{11}\) with connection set
\(S = \{1,2,3,4,6\}\) (Counterexample 5.2); in McKay's catalogue it is
\texttt{regulartournaments11} index 1068, under the relabelling stated
in Section 5. The six orders of its minimum \(2/3\)-certificate (each of
weight \(1/6\); residue labels, earliest first; each lists
\(\mathbb{Z}_{11}\) in arithmetic progression of the stated step):

\begin{longtable}[]{@{}lll@{}}
\caption{The six orders of cA3's minimum \(2/3\)-certificate (each of
weight \(1/6\); residue labels, earliest first).}\tabularnewline*
\toprule\noalign{}
step & order & backward arcs \\*
\midrule\noalign{}
\endfirsthead
\toprule\noalign{}
step & order & backward arcs \\*
\midrule\noalign{}
\endhead
\bottomrule\noalign{}
\endlastfoot
1 & (10, 0, 1, 2, 3, 4, 5, 6, 7, 8, 9) & 16 \\*
1 & (4, 5, 6, 7, 8, 9, 10, 0, 1, 2, 3) & 16 \\*
2 & (7, 9, 0, 2, 4, 6, 8, 10, 1, 3, 5) & 19 \\*
2 & (8, 10, 1, 3, 5, 7, 9, 0, 2, 4, 6) & 19 \\*
3 & (2, 5, 8, 0, 3, 6, 9, 1, 4, 7, 10) & 20 \\*
3 & (3, 6, 9, 1, 4, 7, 10, 2, 5, 8, 0) & 20 \\*
\end{longtable}

The six orders have further structure: each lists the residues in
arithmetic progression with a common step \(d\) -- two orders of step
\(d = 1\) (16 backward arcs each), two of step \(d = 2\) (19 each), and
two of step \(d = 3\) (20 each). Here
\(\mathrm{Aut}(\mathrm{cA3}) = C_{11}\), acting by translation
\(x \mapsto x + 1\), which shifts the starting point of a progression
and preserves its step, so the six orders come from three
\(\mathrm{Aut}\)-orbits of linear orders, with equal backward count per
orbit. Every arc is backward in exactly \(2\) orders, as \(2 \cdot 55 =
110 = 2(16 + 19 + 20)\).

The certificate is \textbf{minimum-cardinality}: integer-optimality over
a \(C_{11}\)-symmetric order pool is confirmed by CPLEX (Appendix B,
B.5); structurally, every arc lies on a cyclic triangle, so by (P3)
every arc must be tight in any \(2/3\)-certificate, whence for \(m\)
orders of equal weight the total backward count \(55m/3\) forces
\(3 \mid m\). However, \(m = 3\) is impossible by Counterexample 5.2
(ii). It is also \textbf{orbit-minimal}: the support of \emph{any}
\(2/3\)-certificate must meet at least three \(\mathrm{Aut}\)-orbits of
orders (see the reproducibility package, Section 9.)

\textbf{The self-converse census at \(n = 11\).} Among all \(S_{11}\)
self-converse 11-vertex tournaments, exactly \textbf{1,548} are
counterexamples (\(\alpha^{*} = 2/3\) and not 3-inducible); 1,388 are
vertex-critical, and the counterexample counts among vertex-deletions
are distributed \(\{1: 32,\ 2: 120,\ 3: 7,\ 4: 1\}\) among the rest. All
but one are inducible by 6 voters at the \(2/3\) threshold; the unique
exception is \textbf{cA6} below. Thirteen representative counterexamples
with certificates are given in the reproducibility package (Section 9).
The exceptional tournament \textbf{cA6} has an integer-infeasible
\(m = 6\) at threshold \(2/3\); its minimum at the \(2/3\) threshold is
\textbf{9 voters}, attained by 8 types, one used twice (a 6- or 7-type
certificate is likewise infeasible):

\begin{longtable}[]{@{}ll@{}}
\caption{A minimum profile for cA6 at \(\alpha^{*} = 2/3\): 9 voters, 8
types.}\tabularnewline*
\toprule\noalign{}
multiplicity & order (earliest first, 1-based) \\*
\midrule\noalign{}
\endfirsthead
\toprule\noalign{}
multiplicity & order (earliest first, 1-based) \\*
\midrule\noalign{}
\endhead
\bottomrule\noalign{}
\endlastfoot
2 & 5 2 9 4 6 1 10 7 8 3 11 \\*
1 & 1 7 8 3 11 5 2 9 4 6 10 \\*
1 & 3 4 7 5 6 8 9 1 2 11 10 \\*
1 & 3 6 7 9 1 4 11 5 2 8 10 \\*
1 & 1 4 11 5 2 8 10 3 6 7 9 \\*
1 & 6 8 9 1 2 11 10 3 4 7 5 \\*
1 & 1 2 11 10 3 4 7 5 6 8 9 \\*
1 & 2 8 10 3 6 7 9 1 4 11 5 \\*
\end{longtable}

Because cA6 is not 6-voter \(2/3\)-inducible, the strike-a-voter
shortcut does not apply to it (unlike cA3 and every other counterexample
in the census, whose majority-of-5 profiles come free from their 6-order
certificates and are therefore not tabulated); its 5-inducibility was
checked directly. An explicit majority-of-5 inducing profile is given
below:

\begin{longtable}[]{@{}ll@{}}
\caption{An explicit majority-of-5 inducing profile for
cA6.}\tabularnewline*
\toprule\noalign{}
voter & order (earliest first, 1-based) \\*
\midrule\noalign{}
\endfirsthead
\toprule\noalign{}
voter & order (earliest first, 1-based) \\*
\midrule\noalign{}
\endhead
\bottomrule\noalign{}
\endlastfoot
1 & 1 3 4 7 5 2 6 9 8 11 10 \\*
2 & 2 6 7 11 9 1 4 8 5 10 3 \\*
3 & 10 3 5 6 1 2 7 8 9 11 4 \\*
4 & 4 11 5 6 8 10 7 9 1 2 3 \\*
5 & 2 8 3 9 4 1 10 11 7 5 6 \\*
\end{longtable}

The \(n = 10\) census pipeline (hereditary pre-filter via the 17,674
nine-vertex non-3-inducibles, ILP on survivors, exact rational
\(\alpha^{*}\)): 2,804,603 pre-filtered; 6,902,592 inducible; 25,861
vertex-critical non-3-inducible; exactly 1,013 at \(\alpha^{*} = 2/3\).
A rigid self-converse witness from McKay's catalogue (2026) and its
6-voter profile is in the reproducibility package (Section 9).

\setcounter{table}{0}\renewcommand{\thetable}{E.\arabic{table}}\setcounter{figure}{0}\renewcommand{\thefigure}{E.\arabic{figure}}

\subsection{Appendix E. The MAS algorithm: orbit subset-DP,
meet-in-the-middle, and the negation/converse join
key}\label{appendix-e.-the-mas-algorithm-orbit-subset-dp-meet-in-the-middle-and-the-negationconverse-join-key}

This appendix defines the dynamic program that certifies
\(\mathrm{MAS}(\mathrm{Paley}(43)) = 543\) (Section 7.1) and enumerates
the level-\(\le 1\) shell (Section 7.3).

\textbf{E.1 Base recurrence (Held--Karp 1962).} For a vertex subset
\(S \subseteq V\) let \(g(S) =
\mathrm{MAS}(T[S])\), the maximum number of internally-forward arcs over
all linear orders of the sub-tournament on \(S\). Placing the last
vertex \(v\) of the order gains exactly its in-arcs from the rest of
\(S\), yielding the recurrence:
\[g(\varnothing) = 0, \qquad g(S) \;=\; \max_{v \in S}\ \Bigl[\, g(S \setminus \{v\}) \;+\;
\bigl|\{u \in S \setminus \{v\} : u \to v\}\bigr| \,\Bigr],\] and
\(\mathrm{MAS}(T) = g(V)\). The same recurrence computes the
\textbf{maximum-weight acyclic subgraph} for arbitrary arc weights \(w\)
-- the separation oracle invoked by the certificate LPs of Appendix
B.6--B.7 -- the only change being that the in-arc count
\(\bigl|\{u \in S \setminus \{v\} : u \to v\}\bigr|\) becomes the in-arc
weight \(\sum_{u \in S \setminus \{v\},\ u \to v} w_{uv}\); the
unweighted DP above is the case \(w \equiv 1\). Here, this would produce
a table with \(2^{43} \approx 8.8 \times 10^{12}\) states.

\textbf{E.2 Automorphism quotient.} \(g\) is \(\mathrm{Aut}\)-invariant
because \(T[\sigma S] \cong T[S]\) for \(\sigma \in \mathrm{Aut}(T)\),
so the DP may be run over orbit representatives of subsets modulo
\(\mathrm{Aut} = \{x \mapsto ax + b : a \in \mathrm{QR}\}\), leaving
almost \(903\) times fewer states, making the run feasible:
\(4.9 \times 10^{9}\) states to reach the middle layer
\(|S| = \lfloor q/2 \rfloor\) of E.3, which we refer to as the
\textbf{equator}. The canonical representative of an orbit is chosen as
the lexicographically minimum bit-mask over the 903 images; at each
level, the child gain is computed \emph{before} canonicalizing the child
(in the parent's labelling), which keeps the quotient correct.

\textbf{E.3 Meet-in-the-middle at the equator.} Any order that ranks all
of \(S\) before all of \(V \setminus S\) has forward count
\(g(S) + \mathrm{cross}(S) + g(V \setminus S)\), where
\(\mathrm{cross}(S)\) counts arcs from \(S\) to \(V \setminus S\); every
order induces such a split at position \(h = \lfloor q/2 \rfloor = 21\)
and attains equality at its own split, so
\[\mathrm{MAS}(T) \;=\; \max_{|S| = h}\ \bigl[\, g(S) + \mathrm{cross}(S) + g(V \setminus S) \,\bigr].\]
This allows us to halve the DP depth: layer tables are built only to
layer 21 and joined at the equator, with layer 22 streamed from layer-21
parents. Because \(\mathrm{MAS}\) is easily seen to be
converse-invariant, a single DP table serves both the prefix \(S\) and
the suffix \(V \setminus S\).

\textbf{E.4 The join: the negation/complement commutes with the quotient
by \(\mathrm{Aut}\).} Reading \(g(V \setminus S)\) (a 22-set) off the
21-layer table requires mapping complements into 21-orbit keys. Let
\(N(S) = -S\), \(C(S) = V \setminus S\), \(\varphi = N \circ C\). Then
(i) \(C\) commutes with every bijection of \(V\); (ii)
\(N \circ \sigma_{a,b} = \sigma_{a,-b} \circ N\) -- negation normalizes
\(\mathrm{Aut}\), flipping only the translation (it lies \emph{outside}
\(\mathrm{Aut}\) because \(-1\) is a non-residue for
\(q \equiv 3 \bmod 4\)); hence (iii)
\(\varphi \circ \sigma_{a,b} = \sigma_{a,-b} \circ \varphi\).

Consequently \(\mathrm{canon}(\varphi(S))\) is constant on the
\(\mathrm{Aut}\)-orbit of \(S\) (the key is well-defined); \(\varphi\)
is an involution between 21-sets and 22-sets; and distinct 21-orbits
probe distinct keys -- a perfect 1:1 join. Negation rather than plain
complement is forced: the suffix side of an order is covered through its
reversed-negated twin (\(x \mapsto -x\) is the converse isomorphism), so
the suffix lookup must be at \(\mathrm{canon}(N(T))\). The DP is correct
in both directions on completion, so a completed run certifies
\(\mathrm{MAS} = 543\) \emph{and} the closure of the level-\(\le 1\)
shell; the census identity of Appendix F is the completeness check.

\setcounter{table}{0}\renewcommand{\thetable}{F.\arabic{table}}\setcounter{figure}{0}\renewcommand{\thefigure}{F.\arabic{figure}}

\subsection{Appendix F. Paley(43): counts and verification
ledger}\label{appendix-f.-paley43-counts-and-verification-ledger}

This appendix documents the exhaustive verification behind
\(N(5) \le 43\) (Section 7): the structural counts of Paley(43), the
certified value \(\mathrm{MAS} = 543\), and the shell-and-screen search,
including the identity checks that confirm the search was complete.

\begin{longtable}[]{@{}lll@{}}
\caption{The Paley(43) verification ledger: structural counts and
certified quantities.}\tabularnewline*
\toprule\noalign{}
quantity & formula & value \\*
\midrule\noalign{}
\endfirsthead
\toprule\noalign{}
quantity & formula & value \\*
\midrule\noalign{}
\endhead
\bottomrule\noalign{}
\endlastfoot
arcs \(C\) & \(q(q-1)/2\) & 903 \\*
cyclic triangles \(N_{\Delta}\) & \((q^3 - q)/24\) & 3311 \\*
\(\lvert \mathrm{Aut}(P) \rvert\) & \(q(q-1)/2\) & 903 \\*
out-degree & \((q-1)/2\) & 21 \\*
triangles per arc & \((q+1)/4\) & 11 \\*
\(\mathrm{MAS}(P)\) & orbit-DP (Appendix E), certified & 543 \\*
\(\alpha^{*} = \mathrm{MAS}/C\) & & 181/301 \\*
slack \(5\,\mathrm{MAS} - 3C\) & & 6 \\*
forced top-two level & \(\lfloor \mathrm{slack}/4 \rfloor\) & 1 \\*
\end{longtable}

Shell census: level 0 = 19,651 orbits (17,744,853 orders); level 1 =
1,821,652 orbits (1,644,951,756 orders); total 1,841,303 orbits =
1,662,696,609 orders. Screen: 4,709,640 distinct razor restrictions;
5,092,111 candidate pairs; 678,686 dangerous restrictions; 347,694,990
dangerous pool orders; 4,376,325,129 (representative, pool) pairs
checked; no DB-disjoint pair found (0 of 4,376,325,129).

We conducted multiple sanity checks (Section 7.3): the MAS gauntlet
(\(q = 7, 11\) brute-force; \(q = 19/23/31\) reproduces known values
107/161/285); an exhaustive level-0 pair check over all
\(\approx 1.6 \times 10^{14}\) MAS-order pairs; a positive-detection
test -- replacing the exact double-back comparison by the razor
restriction alone turns \emph{every} candidate into a seed, and the seed
count then equals the checked count both at level 0 (333,809 = 333,809)
and at full scale (4,376,325,129 = 4,376,325,129); and the
\(\mathrm{Aut}\)-equivariance of double-back sets that we cross-checked
exhaustively over all 903 automorphisms (\(\mathrm{Aut}\) has 5 orbits
on triangles: 3 of size \(903\) and 2 of size \(301\)).

\setcounter{table}{0}\renewcommand{\thetable}{G.\arabic{table}}\setcounter{figure}{0}\renewcommand{\thefigure}{G.\arabic{figure}}

\subsection{\texorpdfstring{Appendix G. The \(n = 9\) obstacle
catalogue, the unit-margin boundary, and forced-arc reversal
(\(n \le 13\))}{Appendix G. The n = 9 obstacle catalogue, the unit-margin boundary, and forced-arc reversal (n \textbackslash le 13)}}\label{appendix-g.-the-n-9-obstacle-catalogue-the-unit-margin-boundary-and-forced-arc-reversal-n-le-13}

This appendix collects the detailed material behind the two-paragraph
summary of Section 5: the complete catalogue of minimal certificates
(G.1), the unit-margin boundary and its exact-coverage LP (G.2), the
triple-local reformulation that scales the census (G.3), the forced-arc
reversal dichotomy and its failures (G.4), and the census tables (G.5).
Census sizes are in Appendix H.

\subsubsection{G.1 The complete catalogue of minimal
certificates}\label{g.1-the-complete-catalogue-of-minimal-certificates}

Optimal duals of the inducibility LP are far from unique, so we
enumerated, for each of the 1,054 non-3-inducible 9-tournaments not
containing \(G_8\), \emph{all} inclusion-minimal supports of optimal
duals (a minimum-support MIP with no-good cuts; Appendix B.7). The
result was 1,748 minimal supports, \textbf{36\% of the tournaments carry
more than one minimal obstacle}, support sizes 23 to 28, and exactly
\textbf{40 isomorphism classes}. They are catalogued in Section 5. A
minimum set cover (Appendix B.4) of the 1,054 by the combined catalogue
needs 26 classes including \(G_8\) for the full census. Only 5 of the
obstacle classes of Section 5 are self-converse, but all are Eulerian,
and normalizing dual weights typically yields multi-digraphs with arc
multiplicity up to 4.

\subsubsection{G.2 The unit-margin
boundary}\label{g.2-the-unit-margin-boundary}

A modified LP requiring every arc covered \emph{exactly} \(\alpha\)
identifies 254 tournaments at \(n = 9\) whose value is 2/3 with the
relaxed condition but \(< 2/3\) with the exact one: precisely the
tournaments that are 3-inducible but \textbf{not with all margins
exactly 1}, i.e.~the witnesses to the strict inclusion
\(\mathcal{I}_{3,1} \subsetneq \mathcal{I}_{3,3}\) at \(n = 9\). For
these 254 the natural certificate comes from the \textbf{exact-coverage
variant} of the inducibility LP -- every arc covered \emph{exactly}
\(\alpha\) -- whose dual is \emph{free-signed}. Enumerating all minimal
supports of these signed duals (Appendix B.7) gives 51 classes under
unsigned support isomorphism. A minimum set cover of the 254 needs 25
classes (Appendix B.4). These tournaments have \(\alpha^{*} = 2/3\) and
are 3-inducible, as A(3) holds on \(n \le 9\) vertices. Since most
3-inducible tournaments admit unit-margin inductions, where fixing two
voters forces the third, this yields a practical search strategy for
3-inductions which we describe next.

\subsubsection{G.3 Scaling the census: 3-inducibility as a Constraint
Satisfaction
Problem}\label{g.3-scaling-the-census-3-inducibility-as-a-constraint-satisfaction-problem}

The \(n = 10\) census on all \(D_{10}\) tournaments rests on a
reformulation: for 3 voters, inducibility itself, not only the
unit-margin variant, is equivalent to a \emph{local} labelling problem
on vertex triples, with no orders ever constructed.

\textbf{Proposition G.1 (triple-local characterization).} \emph{\(T\) is
3-inducible if and only if its arcs admit a labelling by
\(\{0, 1, 2, 3\}\) (label \(0\) = unanimous; label \(i \ge 1\) = the
unique dissenting voter) such that (i) on every directed 3-cycle the
three labels are exactly \(\{1, 2, 3\}\); and (ii) on every transitive
triple with shortcut \(s\) and path arcs \(p_1, p_2\), no voter class
meets the triple in exactly \(\{s\}\), nor in exactly \(\{p_1, p_2\}\).
Moreover \(T\) is 3-inducible with all margins exactly 1 if and only if
the same holds with label \(0\) nowhere used.}

\emph{Proof.} Given an inducing profile, label each arc by its
dissenting voter if any ((P4a): at most one). Voter \(i\)'s class
\(B_i\) is the backward set of its order, so reversing \(B_i\) in \(T\)
gives a transitive tournament. Reversing arcs preserves completeness, so
the reversed digraph is again a \emph{tournament}, and a tournament is
acyclic if and only if it has no directed 3-cycle: in a shortest
directed cycle \(v_1 \to v_2 \to \cdots \to v_k \to v_1\) with
\(k \ge 4\), the arc between \(v_1\) and \(v_3\) either shortens the
cycle (if \(v_1 \to v_3\)) or closes the triangle \(v_1 v_2 v_3\) (if
\(v_3 \to v_1\)), both contradicting minimality. Acyclicity of the
reversal is therefore exactly the conjunction over triples of ``this
triple is not cyclic after reversing \(B_i\)'' -- no longer cycle can
escape the triple-by-triple check, because its chords are themselves
arcs of \(T\) and are themselves labelled. Enumerating the reversal
patterns of a single triple: a cyclic triple survives exactly when
\(B_i\) meets it in \(0\) or \(3\) arcs, and a transitive triple with
shortcut \(s\) and path arcs \(p_1, p_2\) turns cyclic exactly when
\(B_i\) meets it in \(\{s\}\) or in \(\{p_1, p_2\}\) -- precisely the
patterns excluded by (i)--(ii). Conversely, if the classes satisfy
(i)--(ii) then each \(B_i\) reverses to a transitive tournament, hence
is the backward set of a linear order \(\pi_i\); the profile
\((\pi_1, \pi_2, \pi_3)\) gives every arc labelled \(i\) margin 1 and
every arc labelled \(0\) margin 3, so it induces \(T\). In (i), the
rainbow pattern (all three labels distinct) is forced because the
classes are disjoint and each must meet every 3-cycle ((P3): no order
makes all three arcs of a cycle forward). \(\blacksquare\)

The resulting solver decides one \(n = 10\) tournament in
\(\approx 20\mu\)s; it reproduces the \(n \le 9\) censuses exactly and
agrees with an independent CP-SAT order-based model on all instances
checked. It extends the census to all \(D_{10}\) ten-vertex tournaments
and, at \(n = 11\)--\(13\), to the regular, semi-regular and
self-converse classes; the breakdown is tabulated in G.5 below, and the
finer stratifications of the middle class by forced-arc count and by the
number \(g\) of 3-cycles gained per forced-arc reversal can be
regenerated with the reproducibility package (Section 9).

\subsubsection{G.4 Forced-arc reversal: the dichotomy and its
failures}\label{g.4-forced-arc-reversal-the-dichotomy-and-its-failures}

The general lower bound that reversing a forced arc always creates at
least three cyclic triangles is Lemma 5.4 in the main text. We prove it
here, then record the census evidence around it.

\emph{Proof of Lemma 5.4.} Every inducing profile is unanimous on \(e\).
Among all inducing profiles choose \(R = (\pi_1, \pi_2, \pi_3)\)
minimizing, lexicographically after sorting, the vector of
between-counts \(d_i = |I_i|\), where
\(I_i = \{x : u \prec_i x \prec_i v\}\). Since \(e\) is forced it lies
in no directed 3-cycle ((P4b)), so no \(x\) has \(v \to x \to u\), and
every \(x \notin \{u, v\}\) is exactly one of: \textbf{\(P\)-type}
(\(u \to x \to v\)), \textbf{\(L\)-type} (\(x \to u\), \(x \to v\)), or
\textbf{\(R\)-type} (\(u \to x\), \(v \to x\)). Relative to \(\pi_i\),
each \(x\) occupies region \(A\) (\(x \prec_i u\)), \(B\)
(\(x \in I_i\)) or \(C\) (\(v \prec_i x\)); the majorities on \((u, x)\)
and \((x, v)\) force the region counts across the three voters:
\(L\)-type has \(\ge 2\) voters with \(x \in A\); \(R\)-type has
\(\ge 2\) with \(x \in C\); \(P\)-type has \(\ge 2\) with
\(x \in B \cup C\) and \(\ge 2\) with \(x \in A \cup B\).

\begin{enumerate}
\def\labelenumi{(\roman{enumi})}
\item
  \emph{Every \(I_i\) is nonempty.} If \(u, v\) were adjacent in
  \(\pi_i\), transposing them changes only the vote on \(e\) (from
  \(3{:}0\) to \(2{:}1\)), leaving an inducing profile that dissents on
  \(e\) -- contradicting forcedness.
\item
  \emph{Endpoint pivotality.} Let \(f_i\) be the successor of \(u\) in
  \(\pi_i\); by (i), \(f_i \in
  I_i\). Transposing \(u, f_i\) changes only the vote on \((u, f_i)\)
  and strictly lowers the sorted \(d\)-vector, so by minimality the new
  profile must fail to induce \(T\); since only \((u, f_i)\) changed,
  its majority must break. If \(f_i \to u \in T\), the pre-swap majority
  already required both other voters to rank \(f_i\) before \(u\), and
  the swap adds the third vote -- nothing breaks. Hence
  \(u \to f_i \in T\), and breakage forces the others to split:
  \textbf{exactly one other voter has \(f_i \in A\).} Dually, the
  predecessor \(l_i\) of \(v\) satisfies \(l_i \to v \in T\) with
  exactly one other voter placing \(l_i \in C\).
\item
  \emph{All endpoints lie in \(P\).} By (ii), \(f_i\) is \(P\)- or
  \(R\)-type. \(R\)-type requires two voters with \(f_i \in C\); but
  \(f_i \in B_i\), and by (ii) one of the two remaining voters has
  \(f_i \in A\) -- impossible. Dually \(l_i\) is \(P\)- or \(L\)-type,
  and \(L\)-type requires two voters with \(l_i \in A\), impossible
  against \(l_i \in B_i\) plus one \(C\). So \(f_i, l_i \in P\) for all
  \(i\); in particular \(P \neq \varnothing\).
\item
  \emph{Coincidence bounds.} The three firsts cannot all be one element
  \(x\): then \(x \in B_i\) for all \(i\) and no voter has \(x \in A\),
  violating (ii); dually for the lasts. If \(x = f_i =
  l_j\) with \(j \ne i\), then \(x \in B_i \cap B_j\) and the
  \(A\)-demand of (ii) at \(i\) and the \(C\)-demand at \(j\) must both
  be met by the single remaining voter -- impossible. So an element
  serving both as a first and as a last does so for a single voter \(i\)
  (its interval is \(\{x\}\)), with region pattern \(x \in B_i\),
  \(\in A\) for one other voter, \(\in C\) for the third -- in
  particular \(x\) lies in no other voter's interval.
\end{enumerate}

Suppose now \(|P| \le 2\). All six endpoints lie in \(P\) by (iii). If
\(|P| = 1\) the three firsts coincide, contradicting (iv). If
\(P = \{x, y\}\): by (iv) the firsts are not all equal, so both \(x\)
and \(y\) occur among the firsts, and likewise among the lasts; hence
each of \(x, y\) is a first-and-last, confined by (iv) to a single voter
-- distinct voters, since one voter's interval is a singleton. The third
voter's endpoints then lie in \(P\) but can equal neither \(x\) nor
\(y\) (neither lies in its interval), contradicting (iii). Hence
\(|P| \ge 3\). \(\blacksquare\)

\textbf{Theorem G.2 (forced-arc reversal; computational).} \emph{Let
\(T\) be a tournament on at most 9 vertices that is 3-inducible but not
with all margins exactly 1 (such \(T\) exist only at \(n = 9\): the 254;
the class is empty for \(n \le 8\), verified exhaustively). Call an arc
\(e\) \textbf{forced} if every 3-voter inducing profile of \(T\) makes
\(e\) unanimous. Then reversing any forced arc yields a non-3-inducible
tournament.} (All 244 instances: 240 tournaments carry exactly one
forced arc and 2 carry a forced pair, contributing \(2 \times 2 = 4\).)
The forcing hypothesis is necessary: among the 102 \textbf{avoidable}
saturations (arcs unanimous in some profile, at margin 1 in another), 48
reversals remain 3-inducible. Forcing can moreover be
\emph{disjunctive}: 12 of the 254 have no individually forced arc, yet
every profile saturates one of 3--4 candidate arcs. This settles the
question of whether reversing a unanimous arc destroys inducibility.

\textbf{Counterexample G.3 (forced-arc reversal fails at \(n = 10\)).}
Exactly 4 tournaments on 10 vertices, shown in Figure G.1 (vertices
labelled 1--10, the violating forced arc thick black), are 3-inducible
with a forced arc whose reversal remains 3-inducible, refuting the
extension of Theorem G.2 beyond \(n = 9\). This is the least such order.
In each case the forced arc gains exactly \(g = 3\) directed 3-cycles
upon reversal, the minimum possible by Lemma 5.4, so every inducing
profile of the reversed tournament has margin 1 on the reversed arc. The
tournaments and witnesses are in the reproducibility package (Section
9).

\begin{figure}[htbp]
\centering
\begin{minipage}[t]{0.42\textwidth}\centering
\includegraphics[width=\linewidth]{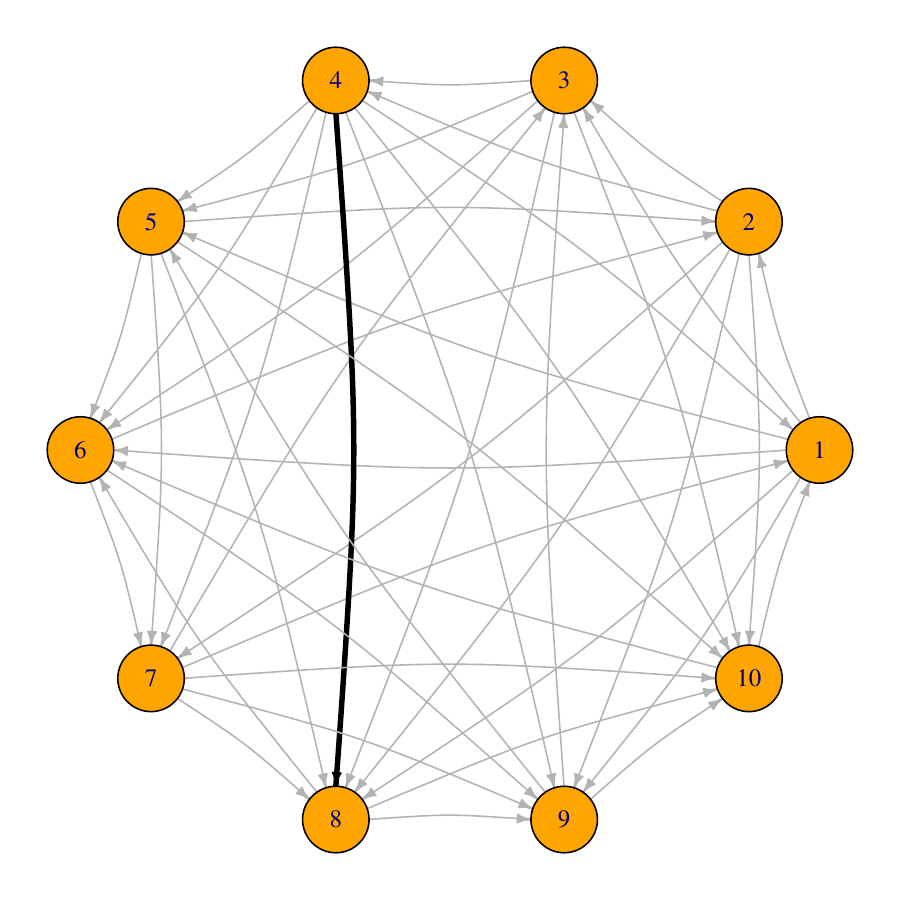}\par (a)
\end{minipage}\hfill
\begin{minipage}[t]{0.42\textwidth}\centering
\includegraphics[width=\linewidth]{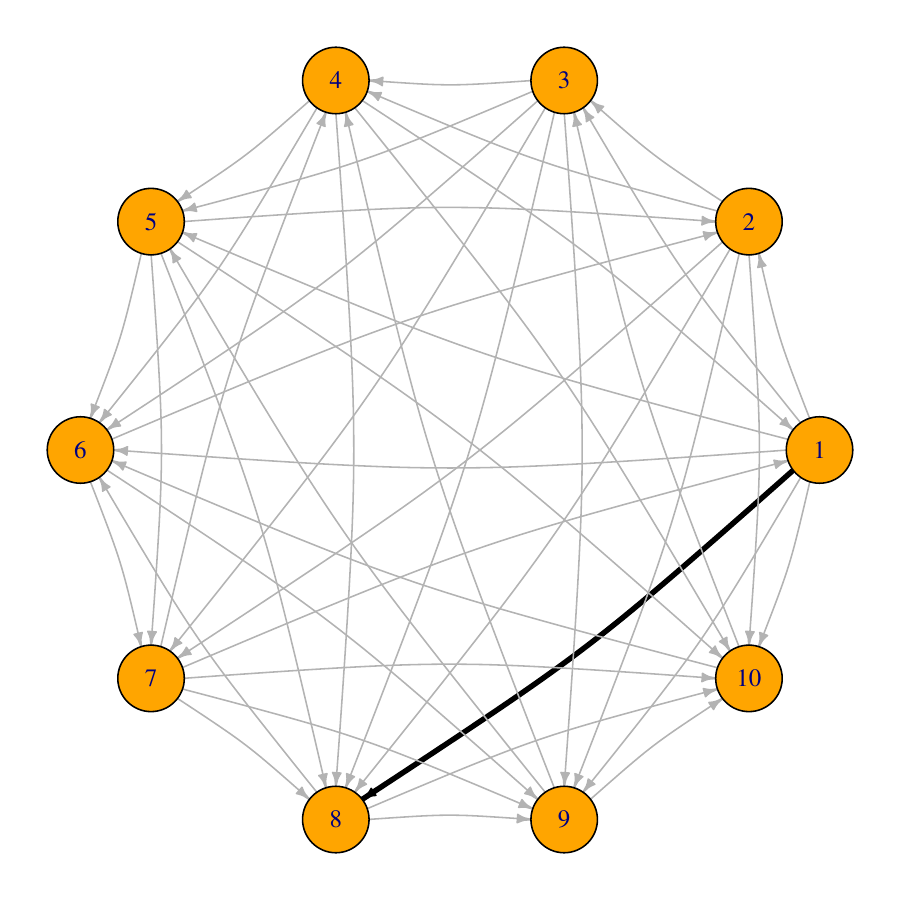}\par (b)
\end{minipage}\par\vspace{1ex}
\begin{minipage}[t]{0.42\textwidth}\centering
\includegraphics[width=\linewidth]{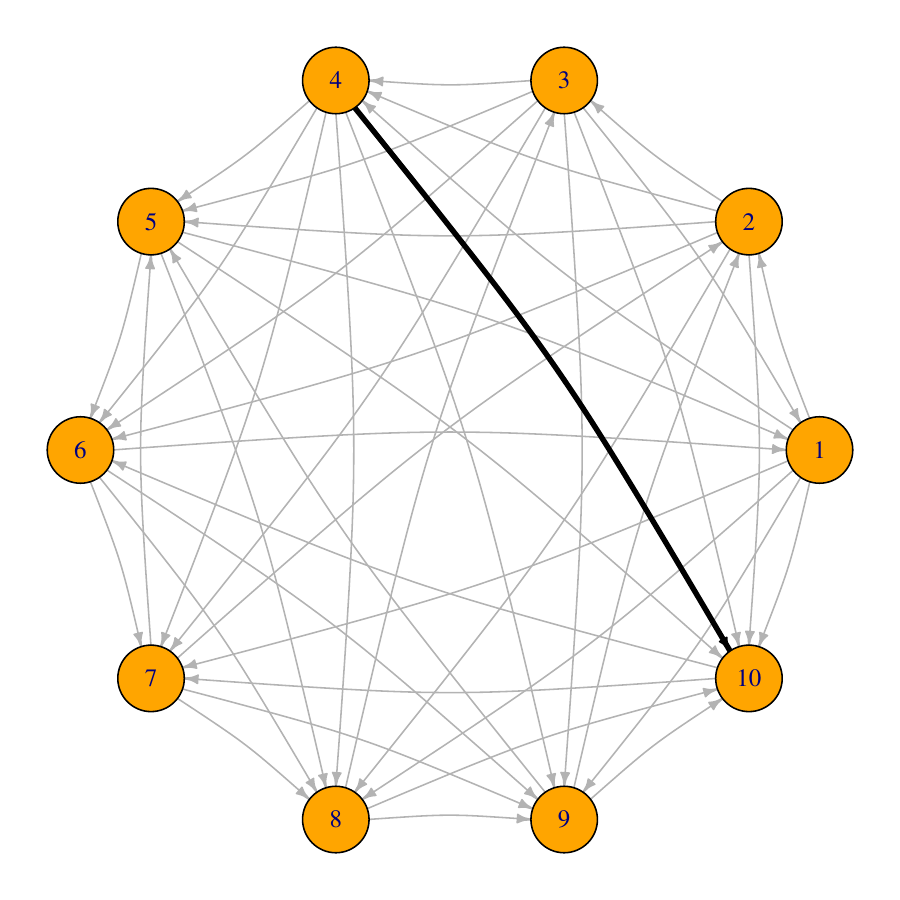}\par (c)
\end{minipage}\hfill
\begin{minipage}[t]{0.42\textwidth}\centering
\includegraphics[width=\linewidth]{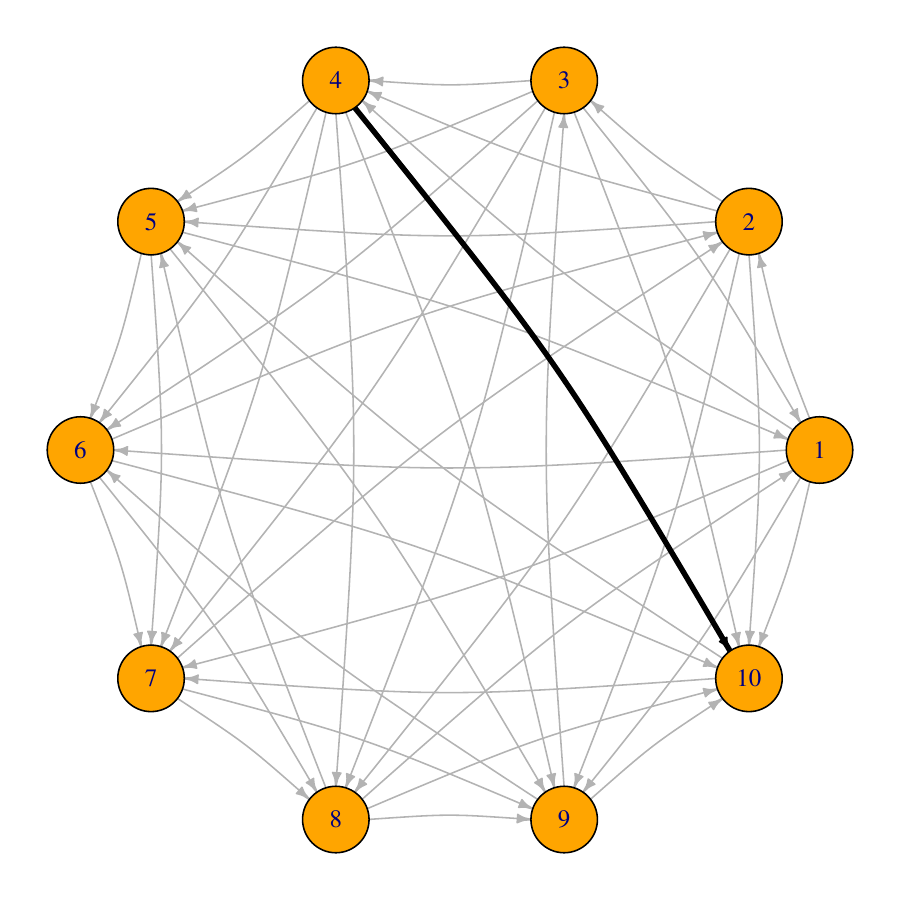}\par (d)
\end{minipage}
\caption{The four tournaments on 10 vertices with a forced arc (thick black) whose reversal
remains 3-inducible (Counterexample G.3).}
\end{figure}

\FloatBarrier

No repair by thresholding the ``triangle gain'' \(g\) survives, either.
The principle that a forced-arc reversal gaining \(g \ge 4\) directed
3-cycles destroys 3-inducibility holds at \(n \le 11\), but fails at
\(n = 12\) and collapses at \(n = 13\): among the \(S_{13}\) hits there
are 30 instances with \(g = 6\). Thus, the maximal failing \(g\) grows
\(3 \to 4 \to 6\) across \(n = 11, 12, 13\). Lemma 5.4, whose bound is
sharp, is the only statement about non-3-inducible tournaments that
remains valid as \(n\) grows.

\emph{Remark (regularity trivializes the census).} In a regular
tournament every arc is cyclic:
\(|N^{+}(v)| + |N^{-}(u)| = n - 1 > n - 2 = |V \setminus \{u, v\}|\)
forces a return 2-path through any arc \(u \to v\). By (P4a) every
3-inducing profile of a regular tournament therefore automatically has
unit margin, and the phenomena of this appendix occur strictly off the
regular locus. The semi-regular 12-vertex class is likewise empty of
hits (see G.5 below).

\subsubsection{G.5 The census table}\label{g.5-the-census-table}

Master census (Proposition G.1 solver, cross-validated as described in
G.3):

\begin{longtable}[]{@{}
  >{\raggedright\arraybackslash}p{(\linewidth - 8\tabcolsep) * \real{0.3288}}
  >{\raggedleft\arraybackslash}p{(\linewidth - 8\tabcolsep) * \real{0.1096}}
  >{\raggedleft\arraybackslash}p{(\linewidth - 8\tabcolsep) * \real{0.1370}}
  >{\raggedleft\arraybackslash}p{(\linewidth - 8\tabcolsep) * \real{0.1918}}
  >{\raggedleft\arraybackslash}p{(\linewidth - 8\tabcolsep) * \real{0.2329}}@{}}
\caption{Unit-margin, not-unit-margin, and not-3-inducible counts,
\(n = 8\)--\(13\).}\tabularnewline*
\toprule\noalign{}
\begin{minipage}[b]{\linewidth}\raggedright
collection
\end{minipage} & \begin{minipage}[b]{\linewidth}\raggedleft
total
\end{minipage} & \begin{minipage}[b]{\linewidth}\raggedleft
unit margin
\end{minipage} & \begin{minipage}[b]{\linewidth}\raggedleft
not unit margin
\end{minipage} & \begin{minipage}[b]{\linewidth}\raggedleft
not 3-inducible
\end{minipage} \\*
\midrule\noalign{}
\endfirsthead
\toprule\noalign{}
\begin{minipage}[b]{\linewidth}\raggedright
collection
\end{minipage} & \begin{minipage}[b]{\linewidth}\raggedleft
total
\end{minipage} & \begin{minipage}[b]{\linewidth}\raggedleft
unit margin
\end{minipage} & \begin{minipage}[b]{\linewidth}\raggedleft
not unit margin
\end{minipage} & \begin{minipage}[b]{\linewidth}\raggedleft
not 3-inducible
\end{minipage} \\*
\midrule\noalign{}
\endhead
\bottomrule\noalign{}
\endlastfoot
all, \(n = 8\) & \(D_8\) & 6,784 & 0 & 96 \\*
all, \(n = 9\) & \(D_9\) & 173,608 & \textbf{254} & 17,674 \\*
all, \(n = 10\) & \(D_{10}\) & 6,812,906 & \textbf{89,686} &
2,830,464 \\*
regular, \(n = 11\) & \(R_{11}\) & 48 & 0 & 1,175 \\*
self-converse, \(n = 11\) & \(S_{11}\) & 114,690 & \textbf{9,250} &
156,028 \\*
semi-regular, \(n = 12\) & \(R_{12}\) & 133,437 & 0 & 19,301,320 \\*
self-converse, \(n = 12\) & \(S_{12}\) & 549,607 & \textbf{38,815} &
903,866 \\*
self-converse, \(n = 13\) & \(S_{13}\) & 7,641,163 & \textbf{2,209,146}
& 85,608,251 \\*
\end{longtable}

\setcounter{table}{0}\renewcommand{\thetable}{H.\arabic{table}}\setcounter{figure}{0}\renewcommand{\thefigure}{H.\arabic{figure}}

\subsection{Appendix H. Census sizes}\label{appendix-h.-census-sizes}

The exact census sizes referred to in Sections 1--8 and Appendix G:
\(D_8 = 6{,}880\); \(D_9 = 191{,}536\); \(D_{10} = 9{,}733{,}056\);
\(D_{11} = 903{,}753{,}248\); \(R_{11} = 1{,}223\);
\(R_{12} = 19{,}434{,}757\) (semi-regular); \(S_{11} = 279{,}968\);
\(S_{12} = 1{,}492{,}288\); \(S_{13} = 95{,}458{,}560\); the
non-3-inducible 9-tournaments number 17,674, of which 16,620 contain
\(G_8\) (Eggermont et al.~2013). All tournament collections except the
Paley family come from McKay's digraph archive (McKay 2026). The
Paley(43) shell and screen counts are itemized in Appendix F.

\end{document}